Adam Mickiewicz University
Faculty of Physics
Poznań, Poland

**Master's Thesis**

# Numerical investigations of the Schwinger model and selected quantum spin models

Marcin Szyniszewski

**Supervisors:**

prof. UAM dr hab. Piotr Tomczak
dr Krzysztof Cichy

Quantum Physics Division,
Faculty of Physics, UAM

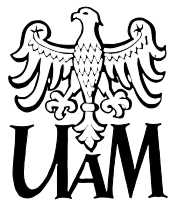

Poznań 2012

# Abstract

Numerical investigations of the XY model, the Heisenberg model and the $J - J'$ Heisenberg model are conducted, using the exact diagonalisation, the numerical renormalisation and the density matrix renormalisation group approach. The low-lying energy levels are obtained and finite size scaling is performed to estimate the bulk limit value. The results are found to be consistent with the exact values. The DMRG results are found to be most promising.

The Schwinger model is also studied using the exact diagonalisation and the strong coupling expansion. The massless, the massive model and the model with a background electric field are explored. Ground state energy, scalar and vector particle masses and order parameters are examined. The achieved values are observed to be consistent with previous results and theoretical predictions. Path to the future studies is outlined.

# Table of contents

# Introduction

The Schwinger model [1] is one-dimensional quantum electrodynamics, which was found to exhibit such interesting phenomena as confinement and chiral symmetry breaking [2]. Therefore, is it often used as a test ground for other, more complicated quantum field theories. This model can be solved exactly in two limits [3]: when the fermion mass is very heavy and very light. The behaviour in between those cases is a subject of intense studies [2,3,4].

A solution to this problem presents itself, especially nowadays, when we are experiencing a technological boom in computer sciences. The computer simulations of quantum models are the essential step in expanding our knowledge, particularly where exact calculations exceed our current mathematical aperture. A useful formalism, that will be used in this work, is the lattice representation of the models [5]. Finite lattice techniques are very successful tools in obtaining the results in the continuum and bulk limits.

Thus, our objective emerges – it is to use computer simulations to investigate several quantum spin-½ models and the Schwinger model in the lattice formulation. We will be interested in using many different simulation techniques, to observe which one is the most promising approach.

The outline of this work is as follows:

**Chapter 1** outlines the most important theoretical principles that are essential in understanding this project. Firstly, there is an overview of the basic concepts of quantum mechanics. Then, we summarise selected spin models that are studied in this work, i.e. the XY model, the Heisenberg model and the $J - J'$ Heisenberg model. Then follows a review of the Schwinger model, where we present the Hamiltonian for various cases of this model (massless and massive, how to incorporate the background electric field, etc.) in various representations (continuum and lattice cases).

In **Chapter 2** the reader is introduced to the methods used in the computer simulations of the mentioned models. We present the exact diagonalisation method and the strong coupling expansion of the Schwinger model. The renormalisation group approach is also delivered, with examples of the numerical renormalisation and the density matrix renormalisation group applied to the Heisenberg model.

The results of using the ED, NRG and DMRG methods on the spin models are presented in **Chapter 3**. Estimated values are compared to the theoretical predictions and previous results.

In **Chapter 4** we use the strong coupling expansion of the Schwinger model to simulate its Hamiltonian. We state our results and the analysis of the results to estimate the bulk and continuum limit values. In order to do this, we are using the finite size scaling methods and the extrapolants approach.

In **Chapter 5** we discuss the conclusions that can be derived from our results and sketch possible future improvements.

# Chapter 1
# Theoretical background

## 1.1 Basics of statistical physics and quantum mechanics

### 1.1.1 Mathematical tools of quantum mechanics

Since in quantum mechanics we deal with very abstract mathematical objects, like linear operators and wave functions that belong to the Hilbert space, it is important to make the reader familiar with these topics. The understanding of mathematical tools used in quantum mechanics is essential to interpret the physics behind them [6].

A field $\mathcal{F}$ is a set $\{a_1, a_2, a_3, \dots\} \equiv \{a_k\}$ of numbers (real or complex), that we will call scalars, with defined operations of multiplication and addition. *Linear vector space* $\mathcal{V}$ over a field $\mathcal{F}$ is defined as a set of vectors $\{v_1, v_2, v_3, \dots\} \equiv \{\vec{v}_k\}$ with two operations: vector addition and multiplication of vectors by scalars [6,7]. Vector addition is an abelian group (this means we have five axioms: closure, commutativity, associativity, existence of identity and inverse elements), while multiplication has properties: closure, distributivity with respect to additions of scalars and of vectors, compatibility with scalar multiplication, existence of identity and zero scalar elements.

| | Vector addition | | Multiplication of vectors by scalars | |
|---|---|---|---|---|
| Closure: | $(\vec{v}_k + \vec{v}_l) \in \mathcal{V}$ | | Closure: | $a_k \vec{v}_l \in \mathcal{V}$ |
| Commutativity: | $\vec{v}_k + \vec{v}_l = \vec{v}_l + \vec{v}_k$ | | Distributivity: | $a_k(\vec{v}_l + \vec{v}_m)$ $= a_k \vec{v}_l + a_k \vec{v}_m;$ $(a_k + a_l)\vec{v}_m$ $= a_k \vec{v}_m + a_l \vec{v}_m$ |
| Associativity: | $(\vec{v}_k + \vec{v}_l) + \vec{v}_m$ $= \vec{v}_k + (\vec{v}_l + \vec{v}_m)$ | | | |
| Identity element: | $\vec{0} + \vec{v}_k = \vec{v}_k$ | | Compatibility: | $a_k(a_l \vec{v}_m)$ $= (a_k a_l)\vec{v}_m$ |
| Inverse element: | $\vec{v}_k + (-\vec{v}_k) = \vec{0}$ | | Identity element: | $1\vec{v}_k = \vec{v}_k$ |
| | | | Zero element: | $0\vec{v}_k = \vec{0}$ |

$$\quad (1.1.1)$$

We define a **scalar product** (inner product) of vectors as follows [6,7]:
1) It is a map:
$$\vec{v}_k \cdot \vec{v}_l : \mathcal{V} \times \mathcal{V} \to \mathcal{F} \quad (1.1.2)$$
it takes two vectors as an argument and returns a scalar.
2) *Conjugate symmetry*:
$$\vec{v}_k \cdot \vec{v}_l = (\vec{v}_l \cdot \vec{v}_k)^* \quad (1.1.3)$$
where asterisk $*$ denotes complex conjugate.
3) It is *linear* with respect to the second factor:
$$\vec{v}_k \cdot (a_l \vec{v}_m + a_n \vec{v}_p) = a_l (\vec{v}_k \cdot \vec{v}_m) + a_n (\vec{v}_k \cdot \vec{v}_p)$$
It follows from (1.1.3) that it is antilinear with respect to the first factor:
$$(a_k \vec{v}_l + a_m \vec{v}_n) \cdot \vec{v}_p = a_k^* (\vec{v}_l \cdot \vec{v}_p) + a_m^* (\vec{v}_n \cdot \vec{v}_p)$$
4) The scalar product of a vector with itself is a *positive real number*:
$$\vec{v}_k \cdot \vec{v}_k = \|\vec{v}_k\|^2 \geq 0$$
where symbol $\|\vec{v}_k\|$ is called a **norm** of the vector and is thus associated to the scalar product by:
$$\|\vec{v}_k\| = \sqrt{\vec{v}_k \cdot \vec{v}_k} \quad (1.1.4)$$

A definition [8] of the norm says that it is a real positive number: $\|\vec{v}_k\| \geq 0$, is zero only for the zero vector: $\|\vec{v}_k\| = 0 \Leftrightarrow \vec{v}_k = \vec{0}$, it is linear with respect to scalars: $\|a_k \vec{v}_k\| = a_k \|\vec{v}_k\|$ and it obeys the triangle inequality: $\|\vec{v}_k + \vec{v}_l\| \leq \|\vec{v}_k\| + \|\vec{v}_l\|$.

A vector $\vec{v}_k$ is said to be *normalised* if:

$$\|\vec{v}_k\| = 1 \quad (1.1.5)$$

Two or more vectors are said to be *orthogonal* if for any pair of those vectors:

$$\vec{v}_k \cdot \vec{v}_l = 0 \quad (1.1.6)$$

Vectors $\{\vec{v}_1, \ldots, \vec{v}_n\}$ are *linearly independent* if the relation:

$$\sum_{k=1}^{n} a_k \vec{v}_k = \vec{0} \quad (1.1.7)$$

has one and only one solution: $a_1 = \cdots = a_n = 0$.

The *dimension* of the vector space $\dim \mathcal{V} = N$ is the maximum number of linearly independent vectors in this space. We say that the vector space $\mathcal{V}$ is then $N$-dimensional. If that number is infinite, then $\dim \mathcal{V} = +\infty$ and we say that the vector space is infinite dimensional.

Set $\mathcal{S}$ of $n$ vectors *spans* the vector space $\mathcal{V}$ if every vector in $\mathcal{V}$ can be written as a linear combination:

$$\vec{v}_k = \sum_{i=1}^{n} a_i \vec{e}_i, \qquad \vec{e}_i \in \mathcal{S} \tag{1.1.8}$$

If $\mathcal{S}$ is additionally to (1.1.8) a set of linearly independent (1.1.7) vectors, then we say that $\mathcal{S}$ is a **vector basis** of $\mathcal{V}$.

The basis $\mathcal{S}$ is called an *orthogonal basis* if all vectors of $\mathcal{S}$ are orthogonal (1.1.6). Additionally, if they are all normalised (1.1.5), the basis is called an **orthonormal basis**. If each vector of the basis is pointing in the direction of each axis of the Cartesian coordinates, and this basis is ordered and orthonormal, then it is called a **standard basis**.

Every vector can be represented by a one-column *matrix* [6,7], but we have to keep in mind that this representation depends on the choice of the coordinate system, i.e. choice of a standard basis used to represent a vector. For example, for a basis $\mathcal{S}$:

$$\vec{v}_k = \sum_{i=1}^{N} a_i \vec{e}_i = \begin{pmatrix} a_1 \\ a_2 \\ \vdots \\ a_N \end{pmatrix}, \qquad \vec{e}_i \in \mathcal{S}$$

It is now possible to introduce a *Hermitian adjoint*, sometimes called Hermitian conjugate, which is complex conjugate with transpose operation:

$$\vec{v}_k^\dagger = (\vec{v}_k^*)^T \tag{1.1.9}$$

In matrix representation, Hermitian adjoint of a vector is always one-row matrix, which is not a vector any more, but belongs to a dual vector space $\mathcal{V}^*$:

$$\vec{v}_k^\dagger = \begin{pmatrix} a_1^* & a_2^* & \cdots & a_N^* \end{pmatrix}, \qquad \vec{v}_k \in \mathcal{V}, \qquad \vec{v}_k^\dagger \in \mathcal{V}^*$$

**Hilbert space** $\mathcal{H}$ is a set of vectors $\{\vec{\psi}_k\}$ that satisfies the following rules [6,7,9]:
1) $\mathcal{H}$ is a *linear vector space* (1.1.1).
2) $\mathcal{H}$ has a defined *scalar product.*
3) $\mathcal{H}$ is *complete*, which means that if a series of vectors $\vec{\psi}_k \in \mathcal{H}$ converges absolutely [8], i.e.:

$$\sum_{k=0}^{\infty} \|\vec{\psi}_k\| < \infty \tag{1.1.10}$$

then it must converge in $\mathcal{H}$ (every partial sum of $\sum_{k=0}^{\infty} \vec{\psi}_k$ converges to an element of $\mathcal{H}$).

In quantum mechanics, we usually [9] talk about *separable* Hilbert spaces, which means that there exists a countable orthonormal basis in $\mathcal{H}$ [7]. Hilbert spaces in quantum mechanics are over a field of real or complex numbers ($\mathbb{R}$ or $\mathbb{C}$).

If a space is a linear vector space, has a defined norm and is complete, but the norm is not necessary associated to the scalar product like in equation (1.1.4), then such a space is called a *Banach space* [8]. Every Hilbert space is thus a Banach space, however not every Banach space is a Hilbert space.

**Dirac notation** is a special way of writing vectors and operations on vectors. This notation is an invaluable tool in quantum mechanics, that will be used in this work very often [6]. Most important is that the notation is free of coordinate system.

Every vector is denoted by symbol $|\cdot\rangle$ which is called a *ket*. A dual vector is denoted by symbol $\langle\cdot|$ which is called a *bra*. We can see that a scalar product of two vectors is then denoted by symbol $\langle\cdot|\cdot\rangle$ which is called a *bra-ket*.

$$\vec{\psi} \rightarrow |\psi\rangle$$
$$\vec{\psi}^\dagger \rightarrow \langle\psi|$$
$$\vec{\phi}\cdot\vec{\psi} \rightarrow \langle\phi|\psi\rangle$$

An **operator** $\hat{A}$ is an object that when applied to a ket transforms it into a ket of the same space, and when is applied to a bra transforms it into a bra of the same dual space [6]:

$$\hat{A}|\psi\rangle = |\psi'\rangle, \qquad \langle\phi|\hat{A} = \langle\phi'| \tag{1.1.11}$$

Most of the time we will deal with *linear operators*, that is with operators that commute with scalars and obey distributive law:

$$\hat{A}(a|\psi\rangle + b|\phi\rangle) = a\hat{A}|\psi\rangle + b\hat{A}|\phi\rangle \tag{1.1.12}$$

A *commutator* of two operators $\hat{A}$ and $\hat{B}$ is defined as:

$$[\hat{A},\hat{B}] = \hat{A}\hat{B} - \hat{B}\hat{A} \tag{1.1.13}$$

Similarly, an *anticommutator* is:

$$\{\hat{A},\hat{B}\} = \hat{A}\hat{B} + \hat{B}\hat{A} \tag{1.1.14}$$

We say that operators $\hat{A}$ and $\hat{B}$ commute if $[\hat{A},\hat{B}] = 0$ and they anticommute if $\{\hat{A},\hat{B}\} = 0$.

*Hermitian adjoint* of an operator $\hat{A}$ is defined by the relation:

$$\langle\psi|\hat{A}^\dagger|\phi\rangle = (\langle\phi|\hat{A}|\psi\rangle)^* \tag{1.1.15}$$

We say that operator $\hat{H}$ is *Hermitian* when it is equal to its Hermitian adjoint:

$$\hat{H}^\dagger = \hat{H} \tag{1.1.16}$$

Using Dirac notation we will denote an orthonormal basis $\mathcal{S} = \{|e_k\rangle\}$. Orthonormality of these vectors can be simply expressed by:

$$\langle e_k | e_l \rangle = \delta_{kl} = \begin{cases} 1 \text{ for } n = m \\ 0 \text{ for } n \neq m \end{cases} \quad (1.1.17)$$

where $\delta_{kl}$ is called a *Kronecker delta*. Closure of this basis is expressed by:

$$\sum_{k=1}^{\dim \mathcal{H}} |e_k\rangle\langle e_k| = \hat{I} \quad (1.1.18)$$

where $\hat{I}$ is an *identity* operator: $\hat{I}|\psi\rangle = |\psi\rangle$.

Matrix representation of vectors in a standard basis $\mathcal{S} = \{|e_k\rangle\}$ can be expressed using Dirac notation:

$$|\psi\rangle = \hat{I}|\psi\rangle = \sum_{k=1}^{\dim \mathcal{H}} |e_k\rangle \underbrace{\langle e_k|\psi\rangle}_{a_k} = \sum_{k=1}^{\dim \mathcal{H}} a_k |e_k\rangle = \begin{pmatrix} \langle e_1|\psi\rangle \\ \langle e_2|\psi\rangle \\ \langle e_3|\psi\rangle \\ \vdots \end{pmatrix} = \begin{pmatrix} a_1 \\ a_2 \\ a_3 \\ \vdots \end{pmatrix}$$

Operators are represented by:

$$\hat{A} = \hat{I}\hat{A}\hat{I} = \sum_{k=1}^{\dim \mathcal{H}} \sum_{l=1}^{\dim \mathcal{H}} |e_k\rangle \underbrace{\langle e_k|\hat{A}|e_l\rangle}_{A_{kl}} \langle e_l| = \sum_{kl} A_{kl} |e_k\rangle\langle e_l|$$

$$= \begin{pmatrix} \langle e_1|\hat{A}|e_1\rangle & \langle e_1|\hat{A}|e_2\rangle & \langle e_1|\hat{A}|e_3\rangle & \cdots \\ \langle e_2|\hat{A}|e_1\rangle & \langle e_2|\hat{A}|e_2\rangle & \langle e_2|\hat{A}|e_3\rangle & \cdots \\ \langle e_3|\hat{A}|e_1\rangle & \langle e_3|\hat{A}|e_2\rangle & \langle e_3|\hat{A}|e_3\rangle & \cdots \\ \vdots & \vdots & \vdots & \ddots \end{pmatrix} = \begin{pmatrix} A_{11} & A_{12} & A_{13} & \cdots \\ A_{21} & A_{22} & A_{23} & \cdots \\ A_{31} & A_{32} & A_{33} & \cdots \\ \vdots & \vdots & \vdots & \ddots \end{pmatrix}$$

To evaluate **eigensystem** of an operator is to find its *eigenvalues* and *eigenvectors*. An eigenvector $|\psi\rangle$ of operator $\hat{A}$ is said to satisfy the relation:

$$\hat{A}|\psi\rangle = a|\psi\rangle \quad (1.1.19)$$

where $a$ is a complex number called eigenvalue. Sometimes for a given eigenvalue there are more corresponding eigenvectors than just one – we say that this eigenvalue is *degenerate*. All eigenvalues of a Hermitian operator are real.

**Gram-Schmidt process** is a method of orthonormalising linearly independent vectors $\{|v_k\rangle\}$ and creating orthonormal basis of vectors $\{|e_k\rangle\}$ [7,8]. Define a projection operator $\mathbb{P}$:

$$\mathbb{P}_{|u\rangle}|v\rangle = \frac{\langle v|u\rangle}{\langle u|u\rangle}|u\rangle$$

Create orthogonal states $|u_k\rangle$ using:

$$|u_1\rangle = |v_1\rangle$$
$$|u_k\rangle = |v_k\rangle - \sum_{i=1}^{k-1} \mathbb{P}_{|u_i\rangle}|v_k\rangle \qquad (1.1.20a)$$

In the end, one can normalise the set $\{|u_k\rangle\}$ to get the orthonormal basis $\{|e_k\rangle\}$:

$$|e_k\rangle = \frac{|u_k\rangle}{\||u_k\|\|} \qquad (1.1.20b)$$

We will sometimes use a formalism of **square-integrable functions** [6]. A "vector" element is given there by a complex function $\psi(x)$, while the "dual vector" element is represented by its complex conjugate $\psi^*(x)$. Now, a scalar product is denoted by an integral over all possible values of $x$. It is important to note that index $x$ is continues and thus the space $\mathcal{X}$ of its all possible values is a range, not a set. So, a scalar product is:

$$\langle \psi | \phi \rangle = \int_{\mathcal{X}} \psi^*(x)\phi(x)\,dx \qquad (1.1.21)$$

If the scalar product of the function with itself exists (the integral does not diverge), then we say that this is a square-integrable function:

$$\langle \psi | \psi \rangle = \int_{\mathcal{X}} \psi^*(x)\psi(x)\,dx \text{ is finite.} \qquad (1.1.22)$$

In quantum mechanics square-integrable functions are called *wave functions*.

## 1.1.2 Basic concepts in quantum mechanics

First of all, we will introduce the reader to the quantities that arise in classical mechanics. One is the *Lagrangian L* (Lagrange's function) of the system, that is a function which describes all the mechanics of the system [10]. It is defined as:

$$L = T - V \qquad (1.1.23)$$

where $T -$ is the kinetic energy, $V -$ potential energy. The Lagrangian is a function of generalised coordinates $\vec{\varphi} = \{\varphi_i\}$, generalised velocities $\frac{d}{dt}\vec{\varphi} \equiv \dot{\vec{\varphi}}$ and time. We will often deal with the *Lagrangian density $\mathcal{L}$*, which is defined by an equation:

$$L = \iiint_V \mathcal{L}\,dx\,dy\,dz \qquad (1.1.24)$$

It is very easy to obtain the equations of motion from the Lagrangian. To do this we use the principle of stationary action (or least action), which is a very fundamental principle of physics and is used in almost every main area of modern physics. We define a quantity called the action:

$$S[\vec{\varphi}(t)] = \int_{t_1}^{t_2} L(\vec{\varphi}(t), \dot{\vec{\varphi}}(t), t) \, dt \qquad (1.1.25)$$

which is a functional – the function that takes elements of vector space as arguments (can be functions) and returns elements from the underlying scalar field (usually numbers). Stationary action principle says that the path $\vec{\varphi}(t)$ that is chosen for the physical system is such that the action is stationary (does not change) to the first order:

$$\delta S = 0 \qquad (1.1.26)$$

This leads to very important equations of classical mechanics, called the Euler-Lagrange equations:

$$\frac{\partial \mathcal{L}}{\partial \varphi_i} - \frac{d}{dt}\left(\frac{\partial \mathcal{L}}{\partial (\dot{\varphi}_i)}\right) = 0 \qquad (1.1.27)$$

These equations are in fact the equations of motion of the system described by the Lagrangian (density) $\mathcal{L}$. Knowing the Lagrangian gives us the full description of the system.

The other important quantity is the *Hamiltonian H* of the system, which describes the total energy of the system [10]:

$$H = T + V \qquad (1.1.28)$$

The Hamiltonian is a function of generalised coordinates $\vec{\varphi} = \{\varphi_i\}$, generalised linear momenta $\vec{p} = \{p_i\}$ and time.

Both the Lagrangian and the Hamiltonian are very similar and can be translated one into the other. For example Euler-Lagrange equations (1.1.27) are in the Hamiltonian formalism translated to Hamilton equations:

$$\dot{p}_i = -\frac{\partial H}{\partial \varphi_i}, \qquad \dot{\varphi}_i = +\frac{\partial H}{\partial p_i} \qquad (1.1.29)$$

Just like the Lagrangian, the Hamiltonian gives us the full description of the motion of the physical system it is describing.

Quantum mechanics has four basic **postulates** [6]:

1) *State vectors*:
    The state of a quantum-mechanical system is completely described by a state vector $|\psi\rangle$ which belongs to a Hilbert space. We will call this space a state space. We usually assume that the state vector is normalised.

2) *Observables*:

All physical quantities $A$ that can be measured and that can change in time (observables, dynamic variables) are described by Hermitian operators $\hat{A}$.

3) *Measurements*:

If one wants to measure an observable $A$, the only possible outcomes of this measurement are eigenvalues of $\hat{A}$. If the measurement is made, the system state immediately changes to the eigenstate $|\psi_n\rangle$ of $\hat{A}$ corresponding to the eigenvalue $a_n$ that was measured:

$$\hat{A}|\psi\rangle = a_n|\psi_n\rangle \qquad (1.1.30a)$$

Another way to represent the new state of the system is by using the projection operator on the direction of this new state, $\mathbb{P}_{|\psi_n\rangle} = |\psi_n\rangle\langle\psi_n|$:

$$|\psi_n\rangle = \frac{\mathbb{P}_{|\psi_n\rangle}|\psi\rangle}{\sqrt{P(a_n)}} \qquad (1.1.30b)$$

where $P(a_n)$ is the probability of measuring an eigenvalue $a_n$ and is given by:

$$P(a_n) = |\langle\psi_n|\psi\rangle|^2 = \langle\psi|\mathbb{P}_{|\psi_n\rangle}|\psi\rangle \qquad (1.1.30c)$$

This last equation is called the *Born's rule*.

4) *Time evolution*:

The state vector evolves in time as described by the Schrödinger equation:

$$i\hbar\frac{\partial}{\partial t}|\psi(t)\rangle = \hat{H}|\psi(t)\rangle \qquad (1.1.31)$$

where $\hat{H}$ is the Hamiltonian operator corresponding to the total energy of the system (1.1.28). We will often simply call this the Hamiltonian.

*Expectation value* $\langle\hat{A}\rangle$ of an operator $\hat{A}$ with respect to the state $|\psi\rangle$ is defined as:

$$\langle\hat{A}\rangle = \langle\psi|\hat{A}|\psi\rangle = \sum_k a_k P(a_k) \qquad (1.1.32)$$

which can be interpreted as an average of the outcome of many identical experiments.

**Angular momentum** in quantum mechanics [6,10] can be defined in a similar way as in classical physics, but using operators of position $\hat{\vec{r}}$ and momentum $\hat{\vec{p}}$:

$$\hat{\vec{L}} = \hat{\vec{r}} \times \hat{\vec{p}} \qquad (1.1.33)$$

where × denotes the cross product. $\hat{\vec{L}}$ is then called *orbital angular momentum* operator. As in classical mechanics, orbital angular momentum is conserved in systems with spherically symmetric or central potentials. It is important to note that every component $\hat{L}_x, \hat{L}_y, \hat{L}_z$ of this operator is Hermitian, which is also true for $\hat{L}^2 = \hat{\vec{L}}^2 = \hat{L}_x^2 + \hat{L}_y^2 + \hat{L}_z^2$. Commutation relations are:

$$[\hat{L}_i, \hat{L}_j] = i\hbar \sum_k \varepsilon_{ijk}\hat{L}_k, \qquad i,j,k \in \{x,y,z\} \tag{1.1.34}$$

where $\varepsilon_{ijk}$ is called Levi-Civita symbol (or fully antisymmetric symbol) defined as:

$$\varepsilon_{ijk} = \begin{cases} 1 \text{ if } i,j,k \text{ is an even permutation of } x,y,z \\ -1 \text{ if } i,j,k \text{ is an odd permutation of } x,y,z \\ 0 \text{ otherwise} \end{cases} \tag{1.1.35}$$

$\hat{L}^2$ commutes with any component of $\hat{\vec{L}}$:

$$[\hat{L}^2, \hat{L}_k] = 0 \tag{1.1.36}$$

Relation (1.1.36) means we can independently measure $\hat{L}^2$ and any of $\hat{L}_k$. We usually choose $\hat{L}^2$ and $\hat{L}_z$. Because they commute, we can find states that are simultaneously eigenstates of both of them. We will denote those eigenstates by two quantum numbers corresponding to the eigenvalues of $\hat{L}^2$ and $\hat{L}_z$: $l, m_l$. Similar to (1.1.19) following is true:

$$\begin{aligned}\hat{L}^2|l,m_l\rangle &= \hbar^2 l(l+1)|l,m_l\rangle \\ \hat{L}_z|l,m_l\rangle &= \hbar m_l|l,m_l\rangle\end{aligned} \tag{1.1.37}$$

States $|l, m_l\rangle$ form a complete, orthonormal basis. It is also useful to define two operators $\hat{L}_+$ and $\hat{L}_-$ as follows:

$$\hat{L}_\pm = \hat{L}_x \pm i\hat{L}_y \tag{1.1.38a}$$

Those operators act on states $|l, m_l\rangle$ in this way:

$$\hat{L}_\pm|l,m_l\rangle = \hbar\sqrt{l(l+1) - m_l(m_l \pm 1)}|l, m_l \pm 1\rangle \tag{1.1.38b}$$

which means that they rise and lower quantum number $m_l$. This is why they are called *rising and lowering operators.*

It is essential to add that, in contrary to classical mechanics, possible values of $\hat{L}^2$ and $\hat{L}_k$ are quantised. Measurement of those will give:

$$\begin{aligned}\hat{L}^2 &\to \hbar^2 l(l+1), & l &\in \{0,1,2,3,\dots\} \\ \hat{L}_k &\to \hbar m_l, & m_l &\in \{-l, (-l+1), \dots, (l-1), l\}\end{aligned} \tag{1.1.39}$$

so that quantum numbers $l, m_l$ have only *integer values*. Example in the graphical representation can be seen in Figure 1a.

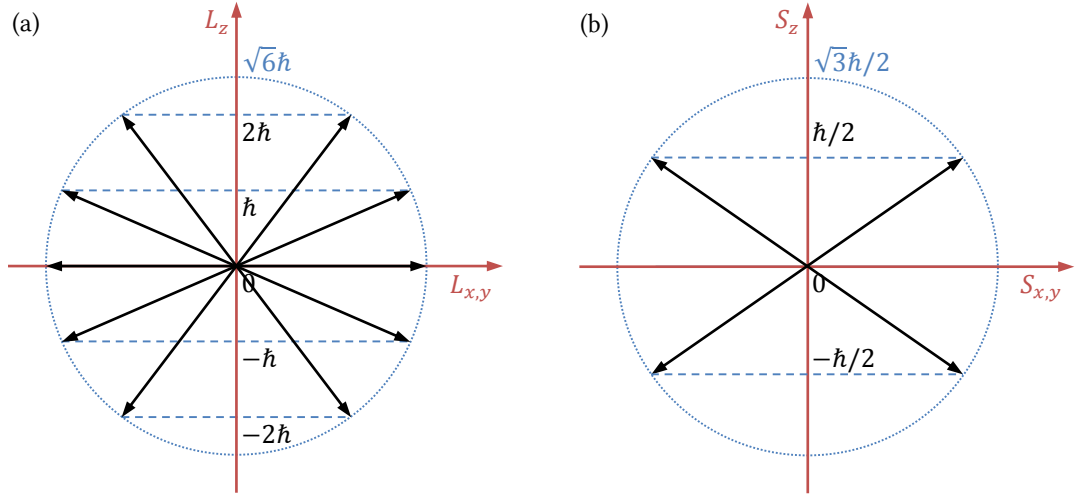

Figure 1. Graphical representation of angular momentum operator [6].
(a) Orbital angular momentum for $l = 2$. (b) Spin ½ angular momentum.

Apart from orbital angular momentum, in quantum mechanics we encounter **spin angular momentum**. It is often depicted as an internal spinning motion of a particle itself, but it is very misleading – in fact spin is essentially different from orbital angular momentum and is a fundamental property of every particle [6].

Mathematical properties of spin operator $\hat{\vec{S}}$ are almost the same as of $\hat{\vec{L}}$ (1.1.24-39) and there are similar quantum numbers $s, m_s$ describing the system. Important difference to (1.1.39) is that number $s$ can also have *half-integer values*:

$$\begin{aligned} \hat{S}^2 &\to \hbar^2 s(s+1), & s &\in \left\{0, \frac{1}{2}, 1, \frac{3}{2}, 2, \frac{5}{2}, 3, \dots\right\} \\ \hat{S}_k &\to \hbar m_s, & m_s &\in \{-s, (-s+1), \dots, (s-1), s\} \end{aligned} \qquad (1.1.40)$$

Example in graphical representation can be seen in Figure 1b. In this work we will mainly deal with particles of $s = ½$. In this case our orthonormal basis becomes:

$$\left|s = \frac{1}{2}, m_s = -\frac{1}{2}\right\rangle, \qquad \left|s = \frac{1}{2}, m_s = \frac{1}{2}\right\rangle \qquad (1.1.41a)$$

These two states are often described as "spin-up" and "spin-down", or $|\uparrow\rangle$ and $|\downarrow\rangle$. In matrix representation they are expressed in terms of two-element column matrices (called spinors):

$$|\uparrow\rangle = \begin{pmatrix} 1 \\ 0 \end{pmatrix}, \qquad |\downarrow\rangle = \begin{pmatrix} 0 \\ 1 \end{pmatrix} \qquad (1.1.41b)$$

Spin operator can be expressed as:

$$\hat{\vec{S}} = \frac{\hbar}{2} \hat{\vec{\sigma}} \qquad (1.1.42)$$

where $\hat{\vec{\sigma}}$ is a vector of **Pauli spin matrices**. From now on we will drop the hat $\hat{\phantom{\sigma}}$ from operator symbols when it will be clear from the context if we are talking about operator or not. So, vector of Pauli matrices is:

$$\vec{\sigma} = \begin{pmatrix} \sigma_1 \\ \sigma_2 \\ \sigma_3 \end{pmatrix} \tag{1.1.43a}$$

with Pauli spin matrices:

$$\begin{cases} \sigma_1 = \sigma_x = \begin{pmatrix} 0 & 1 \\ 1 & 0 \end{pmatrix} \\ \sigma_2 = \sigma_y = \begin{pmatrix} 0 & -i \\ i & 0 \end{pmatrix} \\ \sigma_3 = \sigma_z = \begin{pmatrix} 1 & 0 \\ 0 & -1 \end{pmatrix} \end{cases} \tag{1.1.43b}$$

We can check that the following relations are true:

$$\begin{array}{lll} \sigma_1|\uparrow\rangle = |\downarrow\rangle, & \sigma_2|\uparrow\rangle = i|\downarrow\rangle, & \sigma_3|\uparrow\rangle = |\uparrow\rangle \\ \sigma_1|\downarrow\rangle = |\uparrow\rangle, & \sigma_2|\downarrow\rangle = -i|\uparrow\rangle, & \sigma_3|\downarrow\rangle = -|\downarrow\rangle \end{array} \tag{1.1.43c}$$

Similarly to (1.1.38), lowering and rising operators $\sigma^\pm$ are defined as:

$$\sigma^\pm = \frac{1}{2}(\sigma_1 \pm i\sigma_2)$$
$$\sigma^+ = \begin{pmatrix} 0 & 1 \\ 0 & 0 \end{pmatrix}, \quad \sigma^- = \begin{pmatrix} 0 & 0 \\ 1 & 0 \end{pmatrix} \tag{1.1.44a}$$

And we can see that they act on states:

$$\begin{array}{ll} \sigma^+|\uparrow\rangle = 0, & \sigma^-|\uparrow\rangle = |\downarrow\rangle \\ \sigma^+|\downarrow\rangle = |\uparrow\rangle, & \sigma^-|\downarrow\rangle = 0 \end{array} \tag{1.1.44b}$$

### 1.1.3 Quantum mechanics of composite systems

In our work we will deal with systems that consist of many quantum subsystems. Such a large collection is called an **ensemble** [11,12]. If we deal with physically identical systems, then it is called a *pure* ensemble, otherwise it is *mixed*. In general we have a fraction $p_i$ that is in a state $|\psi_i\rangle$. To represent any ensemble, we use a statistical operator called **density matrix** $\hat{\rho}$ (or density operator):

$$\hat{\rho} = \sum_i p_i |\psi_i\rangle\langle\psi_i| \tag{1.1.45}$$

This operator is positive semidefinite, which means that it is Hermitian $\hat{\rho} = \hat{\rho}^\dagger$ and all of its eigenvalues are nonnegative [12].

Then, expectation value of operator $\hat{A}$ is:

$$\langle \hat{A} \rangle = \operatorname{tr} \hat{A}\hat{\rho} \tag{1.1.46}$$

where tr denotes a *trace* of an operator. In a given basis $\{|h_i\rangle\}$ trace is defined to be:

$$\operatorname{tr} \hat{A} = \sum_{i=1}^{\dim \mathcal{H}} \langle h_i|\hat{A}|h_i\rangle = \sum_{i=1}^{\dim \mathcal{H}} A_{ii} \tag{1.1.47}$$

so it is a sum of diagonal elements of $\hat{A}$. Normalisation of states (1.1.5) is now translated to:

$$\operatorname{tr} \hat{\rho} = 1 \tag{1.1.48}$$

When we know bases of each of the subsystems $\{|h_i^1\rangle\}, \{|h_i^2\rangle\}, \{|h_i^3\rangle\}, \ldots$ we can construct a basis of the ensemble by using the outer (tensor) product $\otimes$:

$$\text{basis} = \{|h_i^1\rangle\} \otimes \{|h_i^2\rangle\} \otimes \{|h_i^3\rangle\} \otimes \cdots \otimes \{|h_i^n\rangle\} = \bigotimes_{k=1}^{n} \{|h_i^k\rangle\} \tag{1.1.49}$$

For example, if we take two spin systems with spin ½ (1.1.41), basis (1.1.49) becomes:

$$\text{basis} = |\uparrow\rangle \otimes |\uparrow\rangle + |\uparrow\rangle \otimes |\downarrow\rangle + |\downarrow\rangle \otimes |\uparrow\rangle + |\downarrow\rangle \otimes |\downarrow\rangle$$

We will often use a short notation in which $|n\rangle \otimes |m\rangle = |nm\rangle$ or a notation involving names of subsystems: $|n\rangle \otimes |m\rangle = |n\rangle_A \otimes |m\rangle_B$.

If we have an ensemble consisting of two subsystems $A$ and $B$, we can define *reduced density matrices*:

$$\hat{\rho}_A = \operatorname{tr}_B \hat{\rho}, \qquad \hat{\rho}_B = \operatorname{tr}_A \hat{\rho} \tag{1.1.50}$$

where $\operatorname{tr}_X$ denotes a *partial trace* over subsystem $X$:

$$\operatorname{tr}_X \hat{A} = \sum_{i=1}^{M_X} \langle i|_X \hat{A} |i\rangle_X \tag{1.1.51}$$

$N_X$ denotes the dimension of subsystem $X$: $\dim X = M_X$, $\{|i\rangle_X\}$ is an orthogonal basis of $X$. Also, the density matrix (1.1.45) for two subsystems can be in general represented as:

$$\hat{\rho} = \sum_{i=1}^{M_A} \sum_{j=1}^{M_B} \sum_{k=1}^{M_A} \sum_{l=1}^{M_B} \alpha_{ij,kl} |i\rangle_A |j\rangle_B \langle k|_A \langle l|_B$$

We can see that using (1.1.51):

$$\operatorname{tr}_B \hat{\rho} = \sum_{i=1}^{M_B} \langle i|_B \hat{\rho} |i\rangle_B = \sum_{i=1}^{M_B} \langle i|_B |i\rangle_B$$

$$= \sum_{i=1}^{M_B} \langle i|_B \sum_{j=1}^{M_A} \sum_{k=1}^{M_B} \sum_{l=1}^{M_A} \sum_{m=1}^{M_B} \alpha_{jk,lm} |j\rangle_A \otimes |k\rangle_B \langle l|_A \otimes \langle m|_B |i\rangle_B$$

$$= \sum_{i=1}^{M_B} \sum_{j=1}^{M_A} \sum_{k=1}^{M_B} \sum_{l=1}^{M_A} \sum_{m=1}^{M_B} \alpha_{jk,lm} |j\rangle_A \delta_{ik} \langle l|_A \delta_{im} = \sum_{i=1}^{M_B} \sum_{j=1}^{M_A} \sum_{l=1}^{M_A} \alpha_{ji,li} |j\rangle_A \langle l|_A$$

$$= \sum_{j=1}^{M_A} \sum_{l=1}^{M_A} \left( \sum_{i=1}^{M_B} \alpha_{ji,li} \right) |j\rangle_A \langle l|_A$$

In short [11]:

$$(\mathrm{tr}_B \hat{\rho})_{ij} = \sum_{k=1}^{M_B} \alpha_{ik,lk} \tag{1.1.52a}$$

Similarly, we can calculate:

$$(\mathrm{tr}_A \hat{\rho})_{ij} = \sum_{k=1}^{M_A} \alpha_{ki,kj} \tag{1.1.52b}$$

As an example we can use a system with dimensions $M_A = 2, M_B = 3$:

$$\hat{\rho} = \begin{pmatrix} \alpha_{00,00} & \alpha_{00,01} & \alpha_{00,02} & \alpha_{00,10} & \alpha_{00,11} & \alpha_{00,12} \\ \alpha_{01,00} & \alpha_{01,01} & \alpha_{01,02} & \alpha_{01,10} & \alpha_{01,11} & \alpha_{01,12} \\ \alpha_{02,00} & \alpha_{02,01} & \alpha_{02,02} & \alpha_{02,10} & \alpha_{02,11} & \alpha_{02,12} \\ \alpha_{10,00} & \alpha_{10,01} & \alpha_{10,02} & \alpha_{10,10} & \alpha_{10,11} & \alpha_{10,12} \\ \alpha_{11,00} & \alpha_{11,01} & \alpha_{11,02} & \alpha_{11,10} & \alpha_{11,11} & \alpha_{11,12} \\ \alpha_{12,00} & \alpha_{12,01} & \alpha_{12,02} & \alpha_{12,10} & \alpha_{12,11} & \alpha_{12,12} \end{pmatrix}$$

$$\mathrm{tr}_B \hat{\rho} = \begin{pmatrix} \alpha_{00,00} + \alpha_{01,01} + \alpha_{02,02} & \alpha_{00,10} + \alpha_{01,11} + \alpha_{02,12} \\ \alpha_{10,00} + \alpha_{11,01} + \alpha_{12,02} & \alpha_{10,10} + \alpha_{11,11} + \alpha_{12,12} \end{pmatrix}$$

$$\mathrm{tr}_A \hat{\rho} = \begin{pmatrix} \alpha_{00,00} + \alpha_{10,10} & \alpha_{00,01} + \alpha_{10,11} & \alpha_{00,02} + \alpha_{10,12} \\ \alpha_{01,00} + \alpha_{11,10} & \alpha_{01,01} + \alpha_{11,11} & \alpha_{01,02} + \alpha_{11,12} \\ \alpha_{02,00} + \alpha_{12,10} & \alpha_{02,01} + \alpha_{12,11} & \alpha_{02,02} + \alpha_{12,12} \end{pmatrix} = C + D$$

We can see that $\mathrm{tr}_X$ is always of dimensions $M/M_X$.

### 1.1.4 Basic concepts of statistical physics

In statistical physics we are considering **physical models** of real systems. The main problem is that the reality is very complex and it is not easily described by mathematical equations. A model is a simplification that maintains specific physical properties of reality that we are interested in, and that can be used to formulate mathematical equations.

We are often interested in the **thermodynamic limit** of the model. This means that the system is of infinite size, we have infinite number of sites. At this point we have to distinguish between two cases: the first is when we add more and more sites to the system, but keeping the lattice spacing nonzero. This is the infinite volume limit or the *bulk limit* [2]. Another case is when we keep the volume of the system constant and add more sites by decreasing the lattice spacing. This way we can achieve the continuous space, or the *continuum limit* [2]. Those two limits are the same only when we don't consider the interactions in the system, because then there is no quantity that would change the physics of the system in different scales. However, if we include the interactions, we have to specify what limit we will use.

For the quantum spin models, we will be interested in the bulk limit, since the lattice spacing is a real physical quantity and should be non-zero. However, since the Schwinger model is a field theory, we will be always there interested in achieving the continuum limit and the bulk limit simultaneously.

Of course, we can't reach the bulk limit (neither the continuum limit) during the computer simulation, because in this way we can only investigate finite systems. But there are many ways to estimate this limit, for example we can simulate systems with different sizes $M$, but the same physics, and then assume that the quantity of interest is scaling like a function of $M$. Making a best fit to the data from the simulations, one can predict with a certain probability the value of the quantity in the limit $M \to \infty$.

Another very important aspect is how to describe the **boundary conditions** of a system with finite size. We often assume *periodic boundary conditions* in our models [13]. In one-dimensional chain of sites this means that the last site is connected to the first, i.e.:

$$i + M \equiv i \tag{1.1.53}$$

Thus, we have topology of the circle. Another case is the *non-periodic (hard-wall, open) boundary conditions*, where there is no link between the first and the last site. This removes the translation invariance of the lattice and we can also see that the physics is substantially different on the sites near the edges of chain, since the edge sites have less nearest neighbours than others. This is why periodic b.c. are usually preferred. Of course in the bulk limit, both cases will give the same result, but we expect periodic case to be more systematic in scaling with the number of sites $M$.

## 1.2 Overview of selected quantum spin systems

The reader will be now introduced to statistical models where we consider a set of magnetic dipoles. One can define a lattice where the dipoles are placed on its sites. We will only consider 1-dimensional lattice (i.e. chain). Each dipole has assigned spin that we will treat quantum mechanically (1.1.41), which means that it can be either up or down.

### 1.2.1 The XY model

In the XY model, we consider only two components of the spin vector $\vec{S}$ that are interacting. The third is arbitrarily set to zero or it is said that the interaction with it is zero, so we don't include it in the Hamiltonian [14]:

$$H = \sum_{i=1}^{M} \sum_{j=1}^{M} J_{i,j} \left( S_x(i) S_x(j) + S_y(i) S_y(j) \right) \tag{1.2.1}$$

where $J_{i,j}$ — interaction between two sites $i$ and $j$, $S_k(i)$ — $k$-th component of spin vector sitting on the site $i$. The interaction between two sites can be:

a) ferromagnetic, when $J_{i,j} < 0$, adjoint spins prefer to align in the same direction,
b) antiferromagnetic, when $J_{i,j} > 0$, adjoint spins prefer to align in opposite directions,
c) and there is no interaction if $J_{i,j} = 0$.

We usually consider an interaction between nearest neighbours (n.n.) only and set all other $J_{i,j}$ to zero. Also, we will introduce another simplification: interaction will be constant and equal to $J$. Using rising and lowering operators as defined by (1.1.38), the Hamiltonian becomes:

$$\frac{H}{J} = \frac{1}{2} \sum_{i=1}^{M} \left( S_+(i) S_-(i+1) + h.c. \right) \tag{1.2.2}$$

where $H/J$ means that we will treat the interaction constant as a measure of energy, i.e. the Hamiltonian and calculated energy levels will be in the units of $J$. To solve the antiferromagnetic XY model one can follow the method outlined in [14]. First of all, the vacuum state $|0\rangle$, i.e. all spins down, must be an eigenstate of $H$ and it has energy (eigenvalue) 0. We can determine the wave function of the state with only one spin up, by considering that $H$ is translationally invariant:

$$|\psi_{\vec{k}}\rangle = \frac{1}{\sqrt{M}} \sum_{i=1}^{M} e^{i\vec{k} \cdot \vec{r}_i} S_+(i) |0\rangle \tag{1.2.3}$$

where $\vec{r}_i$ — position vector of site $i$, $\vec{k}$ — wave vector.

The energy of this state is:

$$E_{\vec{k}} = \cos ka, \qquad |ka| < \pi \tag{1.2.4}$$

If we introduce periodic boundary conditions (1.1.53) as $e^{ikMa} = 1$, we will reach a very difficult situation. The wave function of the system with more than one spin up $|\psi_{\vec{k}_1,\vec{k}_2,\dots}\rangle$ changes discontinuously on the position $i = 0$. The system shouldn't behave like that, because it has a topology of a circle and the physics should be cyclic. Therefore, to solve this problem, we will use the antiperiodic boundary conditions, i.e.:

$$e^{ikMa} = -1 \tag{1.2.5a}$$

This equation is equivalent to saying that $k$ is a discrete set:

$$k = \frac{\pi(2p+1)}{Ma}, \qquad p = 0, \pm 1, \pm 2, \dots \tag{1.2.5b}$$

The wave function for any number of spins up, with wave vectors $\vec{k}_1, \vec{k}_2, \dots$ would be:

$$\left|\psi_{\vec{k}_1,\vec{k}_2,\dots}\right\rangle = C \sum_{i_1, i_2, \dots} F_{\vec{k}_1,\vec{k}_2,\dots}^{i_1, i_2, \dots} S_+(i_1) S_+(i_2) \dots |0\rangle \tag{1.2.6}$$

where $C$ — normalisation constant and $F_{\vec{k}_1,\vec{k}_2,\dots}^{i_1,i_2,\dots}$ — is in a form of determinant:

$$F_{\vec{k}_1,\vec{k}_2,\dots}^{i_1,i_2,\dots} = \varepsilon_{i_1,i_2,\dots} \begin{vmatrix} e^{ik_1 i_1} & e^{ik_2 i_1} & \cdots \\ e^{ik_1 i_2} & e^{ik_2 i_2} & \cdots \\ \vdots & \vdots & \ddots \end{vmatrix} \tag{1.2.7}$$

$\varepsilon_{i_1,i_2,\dots}$ is equal to $+1$ if $(i_1, i_2, \dots)$ is an even permutation of the natural order $i_k < i_m < i_l < \cdots$, equal to $-1$ if $(i_1, i_2, \dots)$ is an odd permutation, and equal to $0$ otherwise.

There is a certain distinction between odd and even number of spins up. If number of spins up is $n$ then:

$$\begin{aligned} n = \text{even} &\Rightarrow k = \frac{\pi}{Ma}(2p+1) \\ n = \text{odd} &\Rightarrow k = \frac{\pi}{Ma} \cdot 2p \end{aligned} \tag{1.2.8}$$

The energy of the state with $n$ spins up is therefore:

$$E_{k_1, k_2, \dots} = \sum_{i=1}^n \cos k_i a = \sum_{i=1}^n \cos\left(\frac{\pi}{M} \cdot \begin{cases} 2p_i \\ 2p_i + 1 \end{cases}\right) \tag{1.2.9}$$

The ground state is achieved by filling up only the negative energy states, as seen in Figure 2.

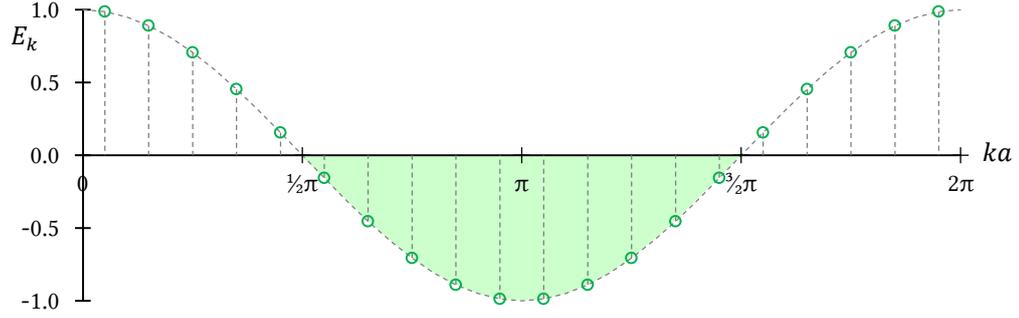

Figure 2. Example of energy levels for $M = 20$.
Green area shows which states are to be filled to achieve the ground state.

So, the energy of the ground state for finite number $M$ is:

$$E'_0 = \sum_{\frac{M}{4} \leq p \leq \frac{3M}{4}} \cos \frac{\pi}{M}(2p+1) \quad \text{and} \quad E''_0 = \sum_{\frac{M}{4} \leq p \leq \frac{3M}{4}} \cos \frac{\pi}{M}(2p+1) \quad (1.2.10)$$

Where $E'_0$ is for the even $M$ and $E''_0$ is for odd $M$. In the bulk limit it becomes (including the units of $J$):

$$\begin{aligned} \frac{E_0}{JM} &= \lim_{M \to \infty} \frac{1}{M} E'_0(M) = \lim_{M \to \infty} \frac{1}{M} E''_0(M) \\ &= -\frac{1}{\pi} \approx -0.3183098862 \end{aligned} \quad (1.2.11)$$

### 1.2.2 The Heisenberg model

In the Heisenberg model, we include interactions between all components of $\vec{S}$, so it is a more general model than the XY model, where spins are interacting only in $X$ and $Y$ directions. The Hamiltonian becomes [14]:

$$\begin{aligned} H &= \sum_{i=1}^{M} \sum_{j=1}^{M} J_{i,j} \vec{S}(i) \cdot \vec{S}(j) \\ &= \sum_{i=1}^{M} \sum_{j=1}^{M} J_{i,j} \left( S_x(i) S_x(j) + S_y(i) S_y(j) + S_z(i) S_z(j) \right) \end{aligned} \quad (1.2.12a)$$

where $J_{i,j}$ — interaction constant, $\vec{S}(i)$ — spin on site $i$, $M$ — number of sites. We will assume that there are only n.n. interactions with interaction constant $J_{i,j} \equiv J$. Using rising and lowering operators defined as in (1.1.38) we change the form of the Hamiltonian to:

$$H = J \sum_{i=1}^{M} \left( S_z(i) S_z(i+1) + \frac{1}{2} \left( S_+(i) S_-(i+1) + h.c. \right) \right) \quad (1.2.12b)$$

The energy of the ground state in the antiferromagnetic ($J > 0$) Heisenberg model can be calculated using the Bethe Ansatz approach [14]:

$$\frac{E_0}{JM} = -\ln 2 + \frac{1}{4} \approx -0.4431471805 \qquad (1.2.13)$$

where $+1/4$ is the ferromagnetic level.

### 1.2.3 The $J - J'$ Heisenberg model

The $J - J'$ Heisenberg model is the Heisenberg model with next nearest neighbours (n.n.n.) interactions. Its Hamiltonian is [15]:

$$H = J \sum_{i=1}^{M} \vec{S}(i) \cdot \vec{S}(i+1) + J' \sum_{i=1}^{M} \vec{S}(i) \cdot \vec{S}(i+2) \qquad (1.2.13a)$$

Or, in the dimensionless, expanded form:

$$\frac{H}{J} = \sum_{i=1}^{M} \left( S_z(i) S_z(i+1) + \frac{1}{2} \left( S_+(i) S_-(i+1) + h.c. \right) \right) \\ + \frac{J'}{J} \sum_{i=1}^{M} \left( S_z(i) S_z(i+2) + \frac{1}{2} \left( S_+(i) S_-(i+2) + h.c. \right) \right) \qquad (1.2.13b)$$

Interactions in the one dimensional $J - J'$ Heisenberg model are shown graphically in Figure 3. The model is sometimes called zigzag Heisenberg chain, or Heisenberg zigzag ladder [16], because it can be also represented as in Figure 3b).

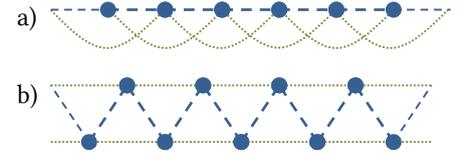

Figure 3. The Heisenberg zigzag chain. Blue dashed lines are the nearest neighbours $J$ interactions, green dotted lines are next nearest neighbours $J'$ interactions.

The signs of the interaction constants are essential [15]:

a) $J < 0$, $J' < 0$ — ferromagnetic interactions, all spins align in the same direction, ground state is stable,
b) $J > 0$, $J' < 0$ — n.n. align antiferromagnetically, n.n.n. align ferromagnetically, stable case,
c) $J < 0$, $J' > 0$ — n.n. align ferromagnetically, n.n.n. align antiferromagnetically, we expect that for large $J'$ ferromagnetic alignment will be destroyed,
d) $J > 0$, $J' > 0$ — n.n. tend to align in the opposite direction, but so are n.n.n. This causes a spin frustration [17].

For $J > 0$ the model exhibits a phase transition (see e.g. [18]) at the point:

$$(J'/J)_C = 0.241167 \pm 0.000005 \qquad (1.2.14)$$

For $J'/J$ lower than $(J'/J)_C$ the system is gapless (i.e. in the bulk limit there is no gap between the *ground singlet state* and the *lowest-lying triplet state*), while for $J'/J$ higher than $(J'/J)_C$ singlet and triplet states are separated by a finite gap [17]. In the finite size systems, this phase transition can be investigated by measuring the gap $\Delta$ between the first excited state with $S = 0$ and the lowest state with $S = 1$ [18].

$$\Delta = E_{S=1}^{(0)} - E_{S=0}^{(1)} \qquad (1.2.15)$$

Another phase transition occurs at the point (see e.g. [16,19]):

$$(J'/J)_{2C} = 0.5 \qquad (1.2.16)$$

where the system's symmetry is broken. For $J'/J$ above this value the ground state is a condensate of dimerised singlets of pairs of n.n. spins [19]. At the exact value of $J'/J = 0.5$ the model becomes the Majumdar-Ghosh model [19], which has the exact energy determined to be [20]:

$$\left.\frac{E}{JM}\right|_{\frac{J'}{J}=0.5} = -0.375 \qquad (1.2.17)$$

This equation also holds for the finite size systems with the even number of spins [15]. Another interesting property is that at this value of $J'/J$, the system has two degenerate ground states.

## 1.3 Introduction to the Schwinger model

The Schwinger model is one of the simplest quantum field theories and also the simplest gauge theory. It is a quantum electrodynamics (has the same Lagrangian as QED), but in one spatial and one temporal dimensions. It is often used as a "toy" model for other, more complicated models, because of its special properties. It shows fermion confinement [5] and other mechanisms present in (3+1)-dimensional quantum chromodynamics (QCD), and therefore it is a good test laboratory for Lattice QCD techniques [21].

### 1.3.1 Continuum formulation

The *Lagrangian density* (1.1.24) of the Schwinger model, in standard notation, is [4]:

$$\mathcal{L} = \bar{\psi}(i\slashed{D} - m)\psi - \frac{1}{4}F_{\mu\nu}F^{\mu\nu} \qquad (1.3.1)$$

We use here the Einstein summation convention, which implies summation over repeated indices. The possible values of the indices depend on the dimensions we are considering, so in the Schwinger model $\mu, \nu \in \{0; 1\}$. Slash over-imposed on a variable means the Feynman slash notation: $\slashed{D} \equiv \gamma^\mu D_\mu$ where $\gamma^\mu$ are $N$-dimensional gamma matrices satisfying $\gamma^\mu \gamma^\nu + \gamma^\nu \gamma^\mu = 2\eta^{\mu\nu} I_N$. $\eta^{\mu\nu}$ is the Minkowski metric (our convention):

$$\eta^{\mu\nu} = \begin{cases} 1 \text{ if } \mu = \nu \text{ describes temporal dimension} \\ -1 \text{ if } \mu = \nu \text{ describes spatial dimension} \\ 0 \text{ if } \mu \neq \nu \end{cases}$$

and $I_N$ is an identity matrix in $N$ dimensions. $\psi$ is a *two-component spinor field* [3]:

$$\psi = \begin{pmatrix} \psi_1 \\ \psi_2 \end{pmatrix} \tag{1.3.2}$$

"Psi-bar" $\bar{\psi}$ is a Dirac adjoint of this spinor field: $\bar{\psi} \equiv \psi^\dagger \gamma^0$.

$D_\mu$ is a gauge covariant derivative, necessary for the equation to be invariant under gauge transformations: $D_\mu \equiv \partial_\mu - igA_\mu$. $D_\mu$ behaves then like a true vector operator. $g$ is a coupling constant and has a dimension of mass $m$, so the theory is super-renormalisable. $A_\mu$ is the covariant potential of the electromagnetic field. $F_{\mu\nu} = \partial_\mu A_\nu - \partial_\nu A_\mu$ is the electromagnetic field tensor.

It is important to say that the equation (1.3.1) is also the Lagrangian density of quantum electrodynamics (QED). Of course, we have to change the dimensions of the model to $(3 + 1)$, i.e. three spatial and one temporal dimensions and the coupling constant becomes the elementary charge $e$. Then $F_{\mu\nu}$ is the electromagnetic field tensor and $A_\mu$ is electromagnetic four-potential.

The equations of motion for the Schwinger model are [4]:

$$(i\displaystyle{\not}D - m)\psi = 0 \qquad \text{(Dirac equation)} \tag{1.3.3a}$$

$$\partial_\mu F^{\mu\nu} = g\bar{\psi}\gamma^\nu\psi \qquad \text{(Maxwell's equations)} \tag{1.3.3b}$$

If we choose time-like axial gauge $A_0 = 0$ (Weyl gauge [2]), field tensor becomes $F^{10} = -\partial_0 A^1 = E$ and the *Hamiltonian* has a simple form:

$$H = \int \left( -i\bar{\psi}\gamma^1(\partial_1 + igA_1)\psi + m\bar{\psi}\psi + \frac{1}{2}E^2 \right) dx \tag{1.3.4}$$

The Schwinger model can be solved exactly in the limit of the massless fermions $m = 0$ [1]. In this case, the theory reduces to that of a non-interacting massive vector particle (boson).

### 1.3.2 Lattice formulation

The **lattice formulation** of the Schwinger model is very important in this work, simply because it is very useful for computer simulations. Lattice will consist of a set of $M$ sites (points) distributed with a distance $a$ (lattice spacing) to each other. In one spatial dimension, it is a simple chain, like shown in Figure 4.

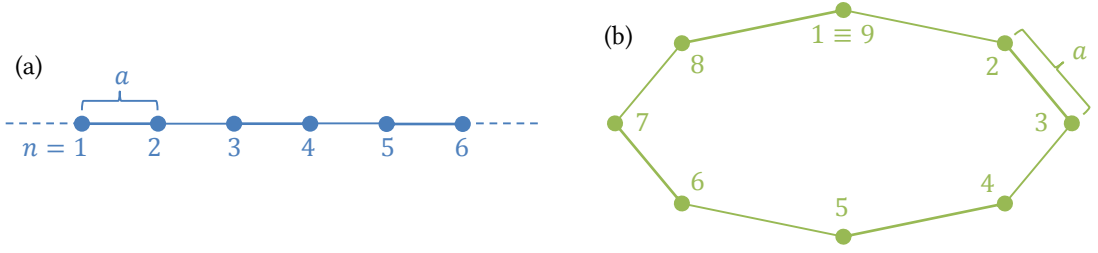

Figure 4. Simple chain of sites.
(a) Case of $M = 6$, open b.c. (b) Chain of $M = 8$ sites with periodic b.c. (1.1.53) forms a circle.

Kogut-Susskind formulation will be used, where we place *one-component Fermi field* $\phi(n)$ on each site $n$ of the lattice. To regain two-component spinor field (1.3.2), we assign [5]:

$$\begin{aligned} \psi_1 &= \frac{1}{\sqrt{a}}\phi(n), &&\text{for even } n \\ \psi_2 &= \frac{1}{\sqrt{a}}\phi(n), &&\text{for odd } n \end{aligned} \quad (1.3.5)$$

This means we deal with a **staggered lattice**, where fermions and antifermions are residing on *alternate* sites. Also, the unit cell of such a lattice consists of *two sites*. The field $\phi(n)$ is defined to obey the anticommutation relations:

$$\{\phi(n), \phi(m)\} = 0, \qquad \{\phi^\dagger(n), \phi(m)\} = \delta_{nm} \quad (1.3.6)$$

On each link between neighbouring sites we define a link variable [4]:

$$U(n; n+1) = e^{i\theta(n)} = e^{iagA^1(n)} \quad (1.3.7)$$

This is how the abelian gauge potential $A_\mu$ enters the lattice theory. We also define the operator $L(n)$, that generates cyclic translations in $\theta(n) = agA^1(n)$, so that the commutation relations between $\theta(n)$ and $L(n)$ are:

$$[\theta(n), L(m)] = i\delta_{nm} \quad (1.3.8)$$

It follows from (1.3.7) and (1.3.8) that the angular momentum operator $L(n)$ is [5]:

$$L(n) = \frac{1}{g}E(n)$$

So it is simply a multiple of electric field $E$.

The *Hamiltonian* for the lattice formulation, equivalent to (1.3.4), becomes [4]:

$$\begin{aligned} H = &-\frac{i}{2a}\sum_{n=1}^{M}\left(\phi^\dagger(n)e^{i\theta(n)}\phi(n+1) - h.c.\right) \\ &+ m\sum_{n=1}^{M}(-1)^n\phi^\dagger(n)\phi(n) + \frac{g^2 a}{2}\sum_{n=1}^{n}L^2(n) \end{aligned} \quad (1.3.9)$$

Now, we use the **Jordan-Wigner transformation** [5], so that we may write this Hamiltonian in spin variables. Firstly, we define Pauli spin matrices $\sigma_i(n)$ on each site $n$. The transformation is:

$$\phi(n) = \prod_{l<n} [i\sigma_3(l)] \sigma^-(n)$$
$$\phi^\dagger(n) = \prod_{l<n} [-i\sigma_3(l)] \sigma^+(n) \tag{1.3.10}$$

One can check how Dirac bilinears transform under (1.3.10), starting with charge density:

$$j^0 = \tfrac{1}{2}[\psi^\dagger, \psi] \leftrightarrow \tfrac{1}{2}[\phi^\dagger(n), \phi(n)] = \tfrac{1}{2}[\sigma^+(n), \sigma^-(n)] = \tfrac{1}{2}\sigma_3(n) \tag{1.3.11a}$$

scalar density becomes:

$$\bar{\psi}\psi = \psi^\dagger \gamma^0 \psi \leftrightarrow (-1)^n \phi^\dagger(n)\phi(n) = \tfrac{1}{2}(-1)^n(1+\sigma_3(n)) \tag{1.3.11b}$$

axial-vector (pseudovector) density, that is also the vector flux, becomes:

$$j_5^0 = \psi^\dagger \gamma_5 \psi \leftrightarrow \big(\phi^\dagger(n)\phi(n+1) + h.c.\big) = -i\big(\sigma^+(n)\sigma^-(n+1) - h.c.\big) \tag{1.3.11c}$$

and the pseudoscalar density becomes:

$$i\bar{\psi}\gamma_5\psi \leftrightarrow -i\big(\phi^\dagger(n)\phi(n+1) - h.c.\big) = \big(\sigma^+(n)\sigma^-(n+1) + h.c.\big) \tag{1.3.11d}$$

After performing transformation (1.3.10), the Hamiltonian (1.3.9) becomes [3]:

$$H = \frac{1}{2a}\sum_{n=1}^{M}\big(\sigma^+(n)e^{i\theta(n)}\sigma^-(n+1) + h.c.\big)$$
$$+ \frac{1}{2}m\sum_{n=1}^{M}\big(1+(-1)^n\sigma_3(n)\big) + \frac{ga^2}{2}\sum_{n=1}^{M} L^2(n) \tag{1.3.12}$$

What has to be noted is that while the Pauli matrices reside *on sites*, $e^{i\theta(n)}$ and $L(n)$ reside *on the links* between $n$-th and $(n+1)$-th sites, as suggested by equation (1.3.7).

### 1.3.3 Operators in the lattice formulation

To be able to simulate the Schwinger model we have to know how the operators that occur in the Hamiltonian (1.3.12) act on the states. Of course to do that we firstly have to know what is the basis (1.1.49) of the composite system that we are working in.

We already know that *spin operators* $\sigma_3, \sigma^\pm$ work in spin space $\{|\uparrow\rangle, |\downarrow\rangle\}$ (1.1.41) and that they act on states according to (1.1.43c) and (1.1.44b).

Because $\theta(n)$ and $A^1(n)$ only enter the theory through the operators $e^{\pm i\theta(n)}$, then the physically meaningful range of those is:

$$0 \leq \theta(n) \leq 2\pi \quad \Leftrightarrow \quad 0 \leq A^1(n) \leq \frac{2\pi}{ag} \tag{1.3.13}$$

It follows from equations (1.3.8) and (1.3.13) that $L(n)$ is quantised in steps of $0, \pm 1, \pm 2, \ldots$ and can be represented on a *ladder space* $\{|l(n)\rangle\}$ such that [5]:

$$L(n)|l(n)\rangle = l(n)|l(n)\rangle, \qquad l(n) = 0, \pm 1, \pm 2, \ldots \tag{1.3.14}$$

Now, to determine how operators $e^{\pm i\theta(n)}$ act on this ladder space $\{|l(n)\rangle\}$ we will consider one specific site $n$, so we will omit the $(n)$ index. Formula (1.3.8) becomes:

$$\theta L - L\theta = i \tag{1.3.15}$$

We start by acting with equation (1.3.15) on $|l\rangle$:

$$\theta L|l\rangle - L\theta|l\rangle = i|l\rangle$$

Now, using (1.3.14):

$$l\theta|l\rangle - L\theta|l\rangle = i|l\rangle$$

We act on this equation with $\theta$ over and over again, each time simplifying using the relation (1.3.15) in a form $\theta L = i + L\theta$:

$$l\theta^2|l\rangle - \theta L\theta|l\rangle = i\theta|l\rangle$$
$$l\theta^2|l\rangle - i\theta|l\rangle - L\theta^2|l\rangle = i\theta|l\rangle$$
$$l\theta^2|l\rangle - L\theta^2|l\rangle = 2i\theta|l\rangle$$

$$l\theta^3|l\rangle - \theta L\theta^2|l\rangle = 2i\theta^2|l\rangle$$
$$l\theta^3|l\rangle - i\theta^2|l\rangle - L\theta^3|l\rangle = 2i\theta^2|l\rangle$$
$$l\theta^3|l\rangle - L\theta^3|l\rangle = 3i\theta^2|l\rangle$$

$$\ldots$$

We can see that:

$$l\theta^m|l\rangle - L\theta^m|l\rangle = mi\theta^{m-1}|l\rangle$$

Multiplying by $\frac{(\pm i)^m}{m!}$ gives:

$$l\frac{(\pm i)^m}{m!}\theta^m|l\rangle - L\frac{(\pm i)^m}{m!}\theta^m|l\rangle = \frac{(\pm i)^m}{m!}mi\theta^{m-1}|l\rangle$$

We can sum over all possible values of $m$, that is $m = 1, 2, 3, \ldots$

$$l\sum_{m=1}^{\infty}\frac{(\pm i)^m}{m!}\theta^m|l\rangle - L\sum_{m=1}^{\infty}\frac{(\pm i)^m}{m!}\theta^m|l\rangle = \sum_{m=1}^{\infty}\frac{(\pm i)^m}{m!}mi\theta^{m-1}|l\rangle$$

In the first two sums we can add cases where $m = 0$, because those two expressions cancel each other out. Also, we can change $m \to m + 1$ in the third sum:

$$l \sum_{m=0}^{\infty} \frac{(\pm i)^m}{m!} \theta^m |l\rangle - L \sum_{m=0}^{\infty} \frac{(\pm i)^m}{m!} \theta^m |l\rangle = \sum_{m=0}^{\infty} \frac{(\pm i)^m (\pm i)}{m!(m+1)}(m+1)i\theta^m |l\rangle$$

$$l \sum_{m=0}^{\infty} \frac{(\pm i)^m}{m!} \theta^m |l\rangle - L \sum_{m=0}^{\infty} \frac{(\pm i)^m}{m!} \theta^m |l\rangle = \mp \sum_{m=0}^{\infty} \frac{(\pm i)^m}{m!} \theta^m |l\rangle$$

Using the definition of $e^{aA} = \sum_{m=0}^{\infty} \frac{a^m}{m!} A^m$ we obtain:

$$l e^{\pm i\theta} |l\rangle - L e^{\pm i\theta} |l\rangle = \mp e^{\pm i\theta} |l\rangle$$

$$l e^{\pm i\theta} |l\rangle \pm e^{\pm i\theta} |l\rangle = L e^{\pm i\theta} |l\rangle$$

$$(l \pm 1) e^{\pm i\theta} |l\rangle = L e^{\pm i\theta} |l\rangle$$

Because of relation (1.3.14) we can see that $e^{\pm i\theta} |l\rangle$ must be proportional to $|l \pm 1\rangle$:

$$L\left(e^{\pm i\theta} |l\rangle\right) = (l \pm 1)\left(e^{\pm i\theta} |l\rangle\right)$$

$$e^{\pm i\theta} |l\rangle = c |l \pm 1\rangle$$

We obtain the coefficient $c$ by multiplying $e^{\pm i\theta} |l\rangle$ with its Hermitian conjugate and assuming states $|l\rangle$ are normalized:

$$\langle l | \underbrace{e^{\mp i\theta} e^{\pm i\theta}}_{=1} | l \rangle = |c|^2 \langle l \pm 1 | l \pm 1 \rangle$$

$$|c|^2 = 1$$

So, we can just take $c = 1$:

$$e^{\pm i\theta} |l\rangle = |l \pm 1\rangle \tag{1.3.16}$$

We can see that $e^{\pm i\theta}$ are the *raising and lowering operators* of $|l\rangle$ [5].

Of course, spin operators act in a different space than the angular momentum operators $L(n)$ and $e^{\pm i\theta(n)}$. To create a **basis**, we can make outer product $\otimes$ of these spaces like in (1.1.49). So for each site we will have two spin states space times infinite ladder space of $L(n)$, resulting in a basis with infinite states:

$$\text{basis} = \bigotimes_{n=1}^{M} \left(\{|\uparrow\rangle, |\downarrow\rangle\} \otimes \{|l\rangle\}\right), \quad l \in \{0, \pm 1, \pm 2, \dots\} \tag{1.3.17}$$

It is worth to emphasise that we have an infinite basis on a finite lattice.

### 1.3.4 The free Schwinger model

Interesting case is the free Schwinger model, that is the Schwinger model with coupling constant $g = 0$. We can see that $\theta(n) = igA^1(n) = 0$ and rising/lowering operators of angular momentum become identities: $e^{\pm i\theta(n)} = \hat{1}$. The Hamiltonian (1.3.12) is simplified to:

$$H = \frac{1}{2a} \sum_{n=1}^{M} \left( \sigma^+(n) \, \sigma^-(n+1) + h.c. \right)$$
$$+ \frac{1}{2} m \sum_{n=1}^{M} \left( 1 + (-1)^n \sigma_3(n) \right) \quad (1.3.18)$$

As there is no angular momentum operator in the Hamiltonian, basis (1.3.17) is also simplified and it has now $2^M$ states:

$$\text{basis} = \bigotimes_{n=1}^{M} \{|\uparrow\rangle, |\downarrow\rangle\} \quad (1.3.19)$$

If, additionally, one considers the massless Schwinger model ($m = 0$), then the Hamiltonian becomes:

$$H = \frac{1}{2a} \sum_{n=1}^{M} \left( \sigma^+(n) \, \sigma^-(n+1) + h.c. \right) \quad (1.3.20)$$

This Hamiltonian is equivalent to the Hamiltonian of the XY model. Thus, we expect the massless free Schwinger model to be equivalent to the XY model (1.2.1). One can also show that the free Schwinger model with mass (1.3.18) is equivalent to the XY model in uniform transverse magnetic field.

### 1.3.5 The Schwinger model with background electric field

With the time-like gauge $A_0 = 0$, Maxwell's equations (1.3.3b) reduce to Gauss' law:

$$\frac{\partial E}{\partial x} = g\bar{\psi}\gamma^0\psi = gj^0 \quad (1.3.21)$$

where $j^0 = \bar{\psi}\gamma^0\psi$ is the charge density. We can integrate this equation to get:

$$E = g \int j^0(x)\, dx + F \quad (1.3.22)$$

We can see that the electric field $E$ is determined up to a constant "background field" $F$ [4,22]. The physics of the massive Schwinger model in a background electric field was particularly explored in a paper by Coleman [22]. One can imagine this additional field as it would be generated by a capacitor with plates on two ends of the one-dimensional system we're considering. There are two different processes that can happen: if $|F| > g/2$ then the charged pairs are produced and separate to infinity; if $|F| \leq g/2$ then the field is reduced and it lowers the electrostatic energy per unit length. So, it is obvious that our equations will be periodic in $F$, with period $g$. We define a dimensionless quantity:

$$\alpha = \frac{F}{g} \tag{1.3.23}$$

that will now denote the background field. The period now becomes 1, so we will work with $\alpha \in [0; 1)$

In the Hamiltonian (1.3.12) the electrostatic energy term is now modified [4] to:

$$\sum_{n=1}^{M} L^2(n) \longrightarrow \sum_{n=1}^{M} \left(L(n) + \alpha\right)^2 \tag{1.3.24}$$

and the Hamiltonian becomes:

$$\begin{aligned} H = &\frac{1}{2a} \sum_{n=1}^{M} \left(\sigma^+(n) e^{i\theta(n)} \sigma^-(n+1) + h.c.\right) \\ &+ \frac{1}{2} m \sum_{n=1}^{M} (1 + (-1)^n \sigma_3(n)) + \frac{ga^2}{2} \sum_{n=1}^{M} (L(n) + \alpha)^2 \end{aligned} \tag{1.3.25}$$

With an additional background field, in the bulk limit we will have a spontaneous symmetry breakdown. We will therefore be able to measure order parameters of the phase transition that occurs at $\alpha = 0.5$, where we have two degenerate ground states. In fact, there is a critical mass [4]:

$$(m/g)_c = 0.3335(2) \tag{1.3.26}$$

Above this critical mass, we observe the first order phase transition between the two degenerate vacuum states at $\alpha = 0.5$, while below the critical mass, we should see no transition [22].

Average electric field is given by [4]:

$$\Gamma^\alpha = \frac{1}{M} \left\langle \sum_n (L(n) + \alpha) \right\rangle_0 \tag{1.3.27}$$

Axial fermion density is defined as:

$$\Gamma^5 = \frac{1}{Mga} \left\langle \sum_n (-1)^n \left(\sigma^+(n) e^{i\theta(n)} \sigma^-(n+1) + h.c.\right) \right\rangle_0 \tag{1.3.28}$$

Both parameters should be equal to zero below the critical mass at $\alpha = 0.5$. In the limit of very large masses $m/g \to \infty$, $\Gamma^\alpha$ should be equal to $\pm 0.5$ (depending which ground state we are considering) [4].

String tension $T$ is defined as an increase in the vacuum energy because of the external field [2] and is given by:

$$T = \lim_{\substack{M \to \infty \\ a \to 0}} \frac{E_\alpha - E_0}{Ma} \qquad (1.3.29)$$

where $E_\alpha$ is ground state energy for system with background field $\alpha$. For $\alpha = 0.5$ we have a theoretical prediction:

$$T|_{\alpha=0.5} = \frac{e^\gamma}{\pi^{2/3}} mg \qquad (1.3.30)$$

where $e, \gamma, \pi$ are mathematical constants.

Finally, the chiral order parameter $\langle \bar{\psi}\psi \rangle_0$ or the chiral condensate is:

$$\langle \bar{\psi}\psi \rangle_0 = \lim_{\substack{M \to \infty \\ a \to 0}} \frac{1}{2Ma} \left\langle \sum_n (-1)^n \, \sigma_3(n) \right\rangle_0 \qquad (1.3.31)$$

For $m = 0$ there is a theoretical prediction of the chiral order parameter:

$$\langle \bar{\psi}\psi \rangle_0 = -\frac{e^\gamma}{2\pi^{3/2}} g \cos 2\pi\alpha \qquad (1.3.32)$$

# Chapter 2
# The details of used methods

## 2.1 Exact diagonalisation

The first method used in this work to obtain energy levels of a given model is Exact Diagonalisation (ED). This means, we create an exact Hamiltonian of a model and calculate its eigenvalues. Of course this method can be thus used only on finite basis models.

Firstly, we construct a basis. Here, we will talk about spin models, so we use basis (1.1.41) for every spin in an ensemble (1.1.49). So, every basis state can be written similarly to this:

$$|\downarrow\uparrow\downarrow\uparrow\downarrow\downarrow\cdots\rangle \tag{2.1.1}$$

This will give us $2^M$ basis states, where $M$ − number of sites.

In programming, we will translate this as follows: $|\uparrow\rangle = |1\rangle, |\downarrow\rangle = |0\rangle$ and we will number the sites from right to left [13]. Example state above (2.1.1) becomes:

$$|\cdots 001010\rangle \equiv |\mathbb{a}\rangle \tag{2.1.2}$$

If we treat this state as a binary number and convert it to integer $a$, then we can number every possible state starting with 0 up to $2^M - 1$.

Secondly, we compute the matrix representation of the Hamiltonian in this basis, which gives us $2^M \times 2^M$ matrix:

$$H_{ij} = \langle \psi_i | \hat{H} | \psi_j \rangle \tag{2.1.3a}$$

$$H = \begin{array}{c} \\ \langle \mathbb{0}| \\ \langle \mathbb{1}| \\ \langle \mathbb{2}| \\ \vdots \\ \langle \mathbb{m}| \end{array} \begin{pmatrix} |\mathbb{0}\rangle & |\mathbb{1}\rangle & |\mathbb{2}\rangle & \cdots & |\mathbb{m}\rangle \\ H_{11} & H_{12} & H_{13} & \cdots & H_{1n} \\ H_{21} & H_{22} & H_{23} & \cdots & H_{2n} \\ H_{31} & H_{32} & H_{33} & \cdots & H_{3n} \\ \vdots & \vdots & \vdots & \ddots & \vdots \\ H_{n1} & H_{n2} & H_{n3} & \cdots & H_{nn} \end{pmatrix}, \quad n = 2^M - 1 \tag{2.1.3b}$$

The biggest drawback is the memory problem – the number of states is $2^M$, so the Hamiltonian matrix has $2^{2M}$ elements. But it is possible to reduce the number of states by demanding the state to have a **given magnetisation** [13]. Of course this procedure will only give us certain energy levels, not the whole spectrum. For example, if we are looking for the ground state in the Heisenberg model without any background field, then we would look for magnetisation equal to zero (if $M$ is even). In spin models magnetisation $\hat{M}$ would be simply defined as:

$$\hat{M} = \sum_{i=1}^{M} S_z(i) \qquad (2.1.4)$$

By choosing only those states with certain magnetisation, we significantly reduce the dimensions of the Hamiltonian. For a given magnetisation every acceptable state has a certain number of spins up $n_\uparrow$ and spins down $n_\downarrow = M - n_\uparrow$. The number of those states is therefore calculated as a combination:

$$\mathcal{N} = C_M^{n_\uparrow} = C_M^{n_\downarrow} = \frac{M!}{n_\uparrow! \cdot n_\downarrow!} \qquad (2.1.5)$$

Finally, after creating the proper Hamiltonian, we calculate its eigenvalues. To do this we use the procedure of calculating eigenvalues of symmetric matrix included in LAPACK software package [23]. The method is to reduce the symmetric matrix to the tridiagonal form (matrix with elements only on the main diagonal, one diagonal above it and one below). Then the QL or QR algorithm[†] is used to calculate eigenvalues and (optionally) eigenvectors of this matrix.

However, sometimes we are interested only in the first lowest energy levels (eigenvalues of Hamiltonian). In that case, there are alternatives methods, for example one can use the Lanczos algorithm or Jacobi-Davidson method.

## 2.2 Strong coupling expansion

When we are considering the Schwinger model with $g \neq 0$, we face the problem of infinite number of the basis states. Strong coupling expansion is a method that will allow us to deal with this problem.

In computer simulations we can't use any units, we only use numbers. So, as the first step, we write the Hamiltonian (1.3.12) in a dimensionless form [3]:

$$W = \frac{2}{ag^2} H = W_0 - xV \qquad (2.2.1)$$

where

---

[†] If we want to calculate only the eigenvalues, the Pal-Walker-Kahan variant of the QL or QR algorithm is used. If we are also interested in calculating the eigenvectors, the implicit QL or QR method is used [23].

$$W_0 = \sum_{n=1}^{M} L^2(n) + \frac{1}{2}\mu \sum_{n=1}^{M} \left(1 + (-1)^n \sigma_3(n)\right) \tag{2.2.2}$$

$$V = \sum_{n=1}^{M} \left(\sigma^+(n) e^{i\theta(n)} \sigma^-(n+1) + h.c.\right) \tag{2.2.3}$$

$$\mu = \frac{2m}{g^2 a}, \qquad x = \frac{1}{g^2 a^2} \tag{2.2.4}$$

We can see that if we have a very strong coupling $g \to \infty$, then $x \to 0$. In this case we can use standard perturbation theory considering $V$ as perturbation operator [3].

### 2.2.1 "Scalar" states

The goal is to create a different basis than (1.3.17) by successive application of operator $V$. Of course, to make a complete basis, we would have to act with $V$ infinite times. In strong-coupling expansion of the Schwinger model previously created states are more important than those created later, if we are interested in the lowest energy levels. This is why we can limit the basis to the first $N$ applications of $V$, still preserving a good approximation of the results for the low-lying spectrum. The number $N$ is called a *perturbation order*. In the end we should obtain unperturbed eigenstates of $W_0$ and then calculate elements of $W$ between them [3].

Firstly, we create a ground state $|0\rangle$. To minimise the energy of $W_0$, one has to assume that for the ground state all $L(n)$ are zeros (we will call this state fluxless) and that $\sigma_3(n) = -1$ for all even sites and $\sigma_3(n) = +1$ for all odd sites. In short:

$$\sigma_3(n) = -(-1)^n \text{ and } L(n) = 0 \tag{2.2.5a}$$

We will use ket notation compatible with basis (1.3.17), for example:

$$|\cdots \uparrow \underbrace{\downarrow}_{n\text{-th site}} \uparrow \uparrow \cdots\rangle \otimes |\cdots 1 \underbrace{0}_{\substack{\text{link between}\\ n\text{-th and}\\ (n+1)\text{-th site}}} 0 \; 1 \cdots\rangle$$

Every state can be represented *diagrammatically*. Sites will be denoted by dots • and on every site there will be an arrow ↑, ↓ showing how the spin is flipped relative to the ground state (leave empty dot if not flipped). Gauge field excitations (sometimes called the flux lines [5] or fluxes for short) will be represented by squiggle arrows, while number of arrows will denote $|L(n)|$ and the direction will denote the sign (right arrow ⤳ for $L(n) > 0$, left arrow ⬳ for $L(n) < 0$) [3][‡].

Now, using equation (2.2.6a), the ground state is:

---

[‡] Note that [5] uses different notation, in which arrows on sites are equivalent to $\sigma_3(n)$.

$$|\mathbb{0}\rangle = |\uparrow\downarrow\uparrow\downarrow\uparrow\downarrow \cdots \rangle \otimes |000000 \cdots \rangle \qquad (2.2.5b)$$

Or diagrammatically:

$$\bullet \quad \bullet \quad \bullet \quad \bullet \quad \bullet \quad \bullet \quad \cdots \qquad (2.2.5c)$$

Secondly, we act sequentially with operator $V$ on the previously created state, starting with the ground state (2.2.6) and creating new states. If the initial state has order $K$, then the final states will be of order $(K + 1)$, but there will also appear states of order $(K - 1)$. We finish the procedure once we generate all $N$-th order states. Now, what is important, when we act with $V$, we don't necessary create only one state (see Figure 5). The states are mixed, so after acting with $V$, we have to separate them. Besides the states are not yet orthonormal. We will deal with these problems in the following way:

| Order | States created |
|---|---|
| 0 | • |
| 1 | • |
| 2 | • • • |
| 3 | • • • |
| 4 | • • • |
| 5 | • • |
| 6 | • • • |
| 7 | • • |
| 8 | • • |

Figure 5. Diagram showing the number of created states for $M = 8$.

1) From every new state we separate the states by setting the equivalence rules between states of the initial basis: the two states are equivalent if we transform one into the other using the operation of translation, "charge conjugation" and "helicity".
   a) *Translation*: Operation of moving the lattice by even number of sites (because the lattice is staggered). For example:
   $|\downarrow\uparrow\uparrow\uparrow\downarrow\downarrow\rangle \otimes |$-100100$\rangle$ is equivalent to $|\downarrow\downarrow\downarrow\uparrow\uparrow\uparrow\rangle \otimes |$000-1001$\rangle$
   Or, diagrammatically:

   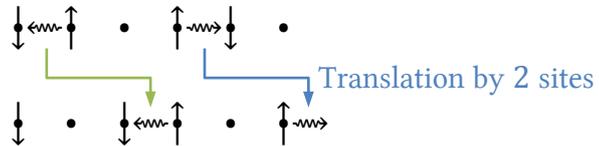

   In ket notation it means that we rotate spins and fluxes by even number of places. Of course, translation by $M$ number of sites is just an identity operation.
   b) *"Charge conjugation"*: Spins residing on sites in the Schwinger model are sometimes referred to as quarks and antiquarks. If two are connected by the gauge field excitation, then we can talk about a meson or an anti-meson (respectively for $L(n)$ positive and negative). In this operation we exchange mesons with antimesons and vice versa - $L(n)$ changes sign and all the spins are flipped. Now, because we have a staggered lattice, we have to additionally make a translation by an odd number of sites.
   For example:
   $|\downarrow\uparrow\uparrow\downarrow\uparrow\downarrow\rangle \otimes |$-100000$\rangle$ is equivalent to $|\uparrow\uparrow\downarrow\downarrow\uparrow\downarrow\rangle \otimes |$010000$\rangle$
   Or, diagrammatically:

   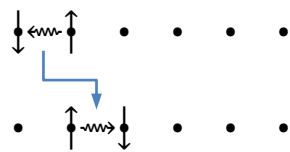

In ket notation, this means that we flip spins and change signs of fluxes and rotate spins by an odd number of places. We can see that two operations of "charge conjugation" result in a translation by an even number of sites.

c) *"Helicity"*: Operation of changing the numbering of spins to be in reversed order. Of course, we have to remember, that we're dealing with a staggered lattice.

For example:

$$|\uparrow\downarrow\uparrow\uparrow\downarrow\uparrow\downarrow\downarrow\downarrow\uparrow\rangle \otimes |00010100\text{-}10\rangle$$

is equivalent to $|\downarrow\uparrow\uparrow\uparrow\downarrow\uparrow\downarrow\downarrow\uparrow\downarrow\rangle \otimes |\text{-}1001010000\rangle$

Or, diagrammatically:

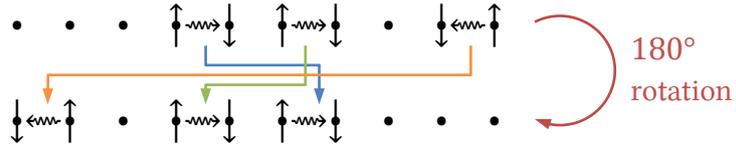

So, in the beginning we have two mesons followed by one empty site and then an antimeson on the right, while after the transformation an antimeson and one empty site is on the left of two mesons.

In ket notation, we have to reverse order of spins and flip them. Fluxes $L(n)$, $n \in \{1, \ldots, M-1\}$ are also in reversed order, but the one, $L(M)$, stays in place. Two "helicity" operations result of course in the identity operation.

This way, we will partition the new state into several groups of states of initial basis, called equivalence classes. We can add states in every class to create a new state of the new basis, resulting in as many states as there were equivalence classes. See Figure 6 for a concept diagram.

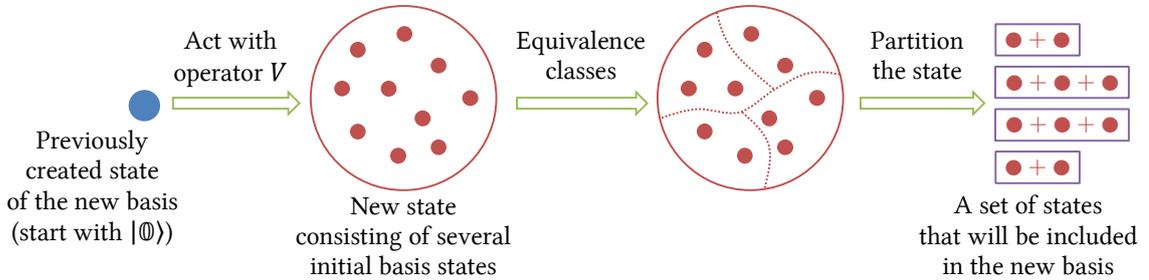

Figure 6. Diagram showing a process of creating new states.

2) We perform the Gram-Schmidt process (1.1.20), to orthonormalise the newly created states, also using previously calculated new basis states.
3) Using new basis states continue the process.

The result will be an orthonormal basis with $\mathcal{N}$ states, corresponding to perturbation order $N$. Number $\mathcal{N}$ depends on the order $N$ and the number of spins $M$ and $\mathcal{N} \geq N+1$. We will denote the new states by $|\Bbbk +\rangle$. Also, every state $|\Bbbk +\rangle$ in this basis is *even parity* (in this special case it means that every state transforms like a scalar under parity operation $x \to -x$; not true in general) and translationally invariant (it doesn't change under the translation by an even number of sites) [3]. We will call those **"scalar" states**. Unperturbed "energy" of the state is defined as an eigenvalue of the operator $W_0$, corresponding to this state. This of course immediately means that every state has to be an eigenstate of $W_0$. In fact, the separation of states described above ensures that every generated state is an eigenstate of $W_0$.

An example, for a two sites system $M = 2$ in order $N = 2$, where we have only three states in the new basis:

$$\begin{aligned}|\mathbb{0}\rangle &= |\uparrow\downarrow\rangle \otimes |00\rangle \\ |\mathbb{1}+\rangle &= \frac{1}{\sqrt{2}}\left(|\downarrow\uparrow\rangle \otimes |\text{-}10\rangle + |\downarrow\uparrow\rangle \otimes |01\rangle\right) \\ |\mathbb{2}+\rangle &= \frac{1}{\sqrt{2}}\left(|\uparrow\downarrow\rangle \otimes |\text{-}1\text{-}1\rangle + |\uparrow\downarrow\rangle \otimes |11\rangle\right)\end{aligned} \quad (2.2.6a)$$

Or diagrammatically:

$$\begin{aligned}|\mathbb{0}\rangle &= \bullet \quad \bullet \\ |\mathbb{1}+\rangle &= \downarrow\!\!\leftsquigarrow\!\!\uparrow \quad + \downarrow \quad \uparrow\!\!\rightsquigarrow \\ |\mathbb{2}+\rangle &= \bullet\!\!\leftsquigarrow\!\!\bullet\!\!\leftsquigarrow \quad + \bullet\!\!\rightsquigarrow\!\!\bullet\!\!\rightsquigarrow\end{aligned} \quad (2.2.6b)$$

Ground state has an unperturbed "energy" equal to 0, while $|\mathbb{1}+\rangle$ has $1 + 2\mu$ and $|\mathbb{2}+\rangle$ has 2. In state $|\mathbb{2}+\rangle$ we have the situation where there is a constant gauge field excitation $L(n)$ between every site. So, in state $|\mathbb{2}+\rangle$ we have a flux line going through the whole one-dimensional space, that we will call a *flux loop*. Of course, in higher order we will encounter states with more than one flux loops [3].

Finally, we create matrix $W$ using the new basis, i.e. find the representation of $W$ in this basis. To do this we use:

$$W_{kl} = \langle \mathbb{k}|W|\mathbb{l}\rangle = \langle \mathbb{k}|W_0|\mathbb{l}\rangle - x\langle \mathbb{k}|V|\mathbb{l}\rangle \quad (2.2.7)$$

For example, for states (2.2.6) operator $W$ connecting them is:

$$W = \begin{pmatrix} 0 & -\sqrt{2}x & 0 \\ -\sqrt{2}x & 1 + 2\mu & -x \\ 0 & -x & 2 \end{pmatrix}$$

It has to be noted that the strong coupling expansion significantly *decreases* the dimension of basis, still preserving a good approximation of the result. In exact diagonalisation method, we have $2^M$ states times the number of allowed values of $L(n)$. Here, the number of states depends on the order we will calculate. For example, in the exact diagonalisation method, if we limit the flux values $L(n) \in \{-5; -4; \cdots; 4; 5\}$, then the number of states is $65536 \times 11$. In strong coupling expansion, with $M = 16, N = 80$ (so that we also have $L(n) \in \{-5, \ldots, 5\}$), we have $\mathcal{N} = 4669$ states.

The quantity we will be interested in is the energy of the ground state. In the dimensionless form it becomes [3]:

$$f_0(x) = \frac{\omega_0}{2Mx} \tag{2.2.8}$$

where $\omega_0$ is the lowest eigenvalue of dimensionless Hamiltonian $W$. Because of the equivalence between the Schwinger model and the XY model we expect it to be given by (1.2.11) in the bulk limit.

Another quantity we can measure is the mass of the first excited state (scalar particle mass), which is proportional to the energy gap between the first excited state and the ground state [3]:

$$f_+(x) = \frac{\omega_+ - \omega_0}{2\sqrt{x}} - 2\frac{m}{g} \tag{2.2.9}$$

where $\omega_+$ is the second lowest eigenvalue of $W$. For $m \neq 0$, this function is called the scalar binding energy. In the $\{M \to \infty, x \to \infty\}$ limit it is calculated to be (for the massless Schwinger model $m = 0$) [1]:

$$\frac{m_+}{g} = \lim_{\substack{M \to \infty \\ x \to \infty}} f_+(x) = \frac{2}{\sqrt{\pi}} \approx 1.128379167 \tag{2.2.10}$$

The scalar binding energy dependence on the fermion mass $m$ in the limits $m \to 0$ and $m \to \infty$ is found to be [24,25]:

$$f_+(x) \underset{m/g \to 0}{\sim} 1.128 + 3.562\frac{m}{g} - 13.512\left(\frac{m}{g}\right)^2 \tag{2.2.11a}$$

$$f_+(x) \underset{m/g \to \infty}{\sim} 1.4729\left(\frac{m}{g}\right)^{-1/3} - \frac{1}{\pi}\left(\frac{m}{g}\right)^{-1} + 0.10847\left(\frac{m}{g}\right)^{-5/3} \tag{2.2.11b}$$

### 2.2.2 "Vector" states

There is also a possibility to create a so called **"vector" state** $|\mathbb{1}-\rangle$ from the ground state $|\mathbb{0}\rangle$, using the vector flux (1.3.11c) [3]:

$$|\mathbb{1}-\rangle = \frac{1}{\sqrt{M}} \sum_{n=1}^{M} \left(\sigma^+(n)e^{i\theta(n)}\sigma^-(n+1) - h.c.\right)|\mathbb{0}\rangle \tag{2.2.12}$$

Now, we can use the same procedure as described above, but using $|\mathbb{1}-\rangle$ in place of the ground state. This will result in orthonormal, *odd parity* (here it means that they transform as vectors under the parity operation; not true in general), translationally invariant basis $\{|\mathbb{k}-\rangle\}$ – we will call those: "vector" states.

Having the spectrum for "vector" states we can calculate vector particle mass [3], defined similarly to (2.2.9):

$$f_-(x) = \frac{\omega_- - \omega_0}{2\sqrt{x}} - 2\frac{m}{g} \qquad (2.2.13)$$

where $\omega_-$ is the first lowest energy of $W$ for "vector" states (ground state of states with vector quantum numbers) and $\omega_0$ is the vacuum energy (ground state for "scalar" states). For $m \neq 0$, $f_-(x)$ is called the vector binding energy. In the limit $\{M \to \infty, x \to \infty\}$ it is calculated to be (the $m = 0$ Schwinger model) [1]:

$$\frac{m_-}{g} = \lim_{\substack{M \to \infty \\ x \to \infty}} f_-(x) = \frac{1}{\sqrt{\pi}} \approx 0.5641895835 \qquad (2.2.14)$$

The vector binding energy dependence on the fermion mass $m$ in the limits $m \to 0$ and $m \to \infty$ is found to be [24,25]:

$$f_-(x) \underset{m/g \to 0}{\sim} 0.564 + 1.781\frac{m}{g} + 0.1907\left(\frac{m}{g}\right)^2 \qquad (2.2.15a)$$

$$f_-(x) \underset{m/g \to \infty}{\sim} 0.6418\left(\frac{m}{g}\right)^{-1/3} - \frac{1}{\pi}\left(\frac{m}{g}\right)^{-1} - 0.25208\left(\frac{m}{g}\right)^{-5/3} \qquad (2.2.15b)$$

We will also be interested in the ratio $f_+(x)/f_-(x)$ which in the $\{M \to \infty, x \to \infty\}$ limit will become ($m = 0$):

$$\lim_{\substack{M \to \infty \\ x \to \infty}} \frac{f_+(x)}{f_-(x)} = \frac{m_+}{m_-} = 2 \qquad (2.2.16)$$

Behaviour in the large and small $m$ is [24,26]:

$$\frac{f_+(x)}{f_-(x)} \underset{m/g \to 0}{\sim} 2 - 24.625\left(\frac{m}{g}\right)^2 \qquad (2.2.17a)$$

$$\frac{f_+(x)}{f_-(x)} \underset{m/g \to \infty}{\sim} 2.295 \qquad (2.2.17b)$$

### 2.2.3 Incorporating the constant background field

If we will include in our calculations **the constant background field** $\alpha$ and use the Hamiltonian (1.3.25), then the operator $W_0$ will be changed to:

$$W_0 = \sum_{n=1}^{M}\left(L(n) + \alpha\right)^2 + \frac{1}{2}\mu\sum_{n=1}^{M}\left(1 + (-1)^n\sigma_3(n)\right) \qquad (2.2.18)$$

Apart from this modification, it is also essential to note that for $\alpha = 0.5$ there exists an additional ground state [2,4], defined by:

$$\sigma_3(n) = -(-1)^n \text{ and } L(n) = 1 \qquad (2.2.19a)$$

In ket notation it will be:

$$|\mathbb{0}'\rangle = |\uparrow\downarrow\uparrow\downarrow\uparrow\downarrow\cdots\rangle \otimes |111111\cdots\rangle \tag{2.2.19b}$$

Diagrammatically, this state will be represented by one flux loop:

$$\bullet\text{-}\!\!\!\text{wv}\!\!\!\rightarrow\bullet\text{-}\!\!\!\text{wv}\!\!\!\rightarrow\bullet\text{-}\!\!\!\text{wv}\!\!\!\rightarrow\bullet\text{-}\!\!\!\text{wv}\!\!\!\rightarrow\bullet\text{-}\!\!\!\text{wv}\!\!\!\rightarrow\bullet\text{-}\!\!\!\text{wv}\!\!\!\rightarrow\cdots \tag{2.2.19c}$$

If $\alpha \neq 0.5$, then either $|\mathbb{0}\rangle$ or $|\mathbb{0}'\rangle$ will have slightly lower energy and will be the only ground state of the system. However, in the simulation it is essential to incorporate the other state, so that we will be able to create all relevant states firstly.

Now, to generate new states using previously described method, we will use both states (2.2.6) and (2.2.19). Also, while generating the states and separating the equivalence classes, we have to remember that because there is now a background electric field, then differently charged "particles" (mesons) will now behave in a different way. Therefore, the "charge conjugation" equivalence vanishes if we have the non-zero field.

We will be interested in measuring quantities described by equations (1.3.27)-(1.3.31). However, in computer simulations we have to use dimensionless values, so we will define dimensionless string tension (1.3.29):

$$\frac{2T}{g^2} = \frac{\omega_\alpha - \omega_0}{M} \tag{2.2.20}$$

where $\omega_\alpha$ is the ground state of $W_0$ for background field equal to $\alpha$. Chiral order parameter becomes:

$$\frac{\langle\bar{\psi}\psi\rangle_0}{g} = \frac{\sqrt{x}}{2M}\left\langle 0 \left| \sum_n (-1)^n\, \sigma_3(n) \right| 0 \right\rangle \tag{2.2.21}$$

As for $\Gamma^\alpha$ and $\Gamma^5$, on a finite lattice there is no spontaneous symmetry breakdown, and therefore expectation values remains zero. However, we can estimate them using an overlap matrix element:

$$\langle Q\rangle_0 \to \langle Q\rangle_{0,0'} = \langle 0|Q|0'\rangle \tag{2.2.22}$$

where $|0\rangle$ and $|0'\rangle$ are the two ground states of the Hamiltonian $W$. This is only true when both ground states become degenerate, i.e. when $\alpha = 0.5$.

## 2.3 Density matrix renormalisation group

### 2.3.1 Basic ideas of the renormalisation group approach

Renormalisation group (RG) can be used on systems where different scales are described by the same or similar equations. The main idea [27] is to divide the whole system or problem into smaller subsystems, which can be now described by simpler equations. We do this many times until we get a subsystem with such small size that it can be easily calculated exactly.

For example, there is a version of renormalisation group called block-spin transformation [27,28]. We form smaller blocks out of the whole system (see Figure 7a), replace those blocks with single spins, that have average value of the blocks (see Figure 7b) and finally restore the original scale by rescaling the lattice (see Figure 7c). One has to repeat those steps many times and for different configurations of spins, so that the probability of the configuration can be determined.

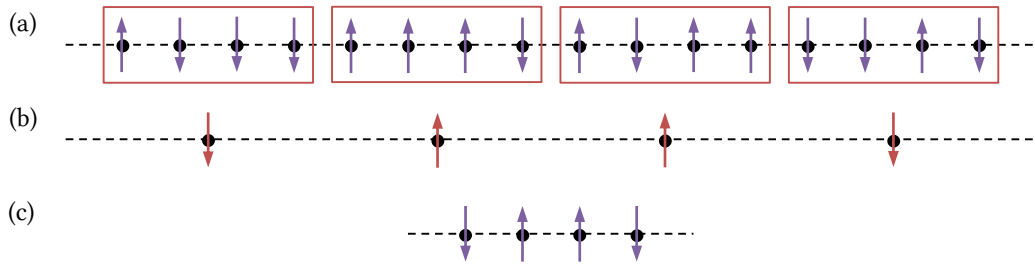

Figure 7. Basic idea of the renormalisation group approach.

In this work we will use a very similar approach, but in a reversed order – we will start from smaller scales of length and by adding more sites go to larger systems.

### 2.3.2   Numerical renormalisation group on the Heisenberg model

We consider a chain of $M$ spins with open boundary conditions (it is simpler that way). As an example we will use the Heisenberg model.

The algorithm is as follows:
1) Set up the exact Hamiltonian $H$ of the Heisenberg model for $M$ spins using (1.2.12b) – see Figure 8a.
2) Calculate eigenvalues $\{E_i\}$ and eigenvectors $\{|\psi_i\rangle\}$ of $H$.
3) Choose the relevant eigenvalues: keep only $2^{M-1}$ eigenvectors[§] that correspond to the smallest eigenstates and create a renormalisation matrix $O$:
$$O = \begin{pmatrix} |\psi_1\rangle & |\psi_2\rangle & \cdots & |\psi_{2^{M-1}}\rangle \end{pmatrix} \qquad (2.3.1)$$
"Relevant" or most "important" eigenvalues [28] are those for which the repeated RG iterations drive the results towards the true values in the smallest amount of time. In numerical renormalisation group we *guess* that those relevant eigenvalues are the smallest ones.
4) Use the renormalisation matrix (2.3.1) to create renormalised operators with dimensions $2^{M-1} \times 2^{M-1}$ (see Figure 8b):

---

[§] The number of kept eigenvectors is usually denoted by $l$ in NRG/DMRG literature [4]. In general, one can perform simulations with different values of $l$ and check the convergence of results when $l$ increases. In this work, we have only used values $l = 2^{M-1}$, with different initial $M$.

$$H_{Ren} = OHO^T$$
$$S_{x\,Ren} = O(I_{M-1} \otimes S_x)O^T$$
$$S_{y\,Ren} = O(I_{M-1} \otimes S_y)O^T \quad (2.3.2)$$
$$S_{z\,Ren} = O(I_{M-1} \otimes S_z)O^T$$

Here and in the following sections, we will use symbol $I_n$ to denote the identity matrix of dimensions $2^n$.

5) Create a new Hamiltonian $H$ for a system with one added site, using renormalised operators (2.3.2) – see Figure 8c. Initially:

$$H = J\sum_{i=1}^{N}\left(S_x(i)S_x(i+1) + S_y(i)S_y(i+1) + S_z(i)S_z(i+1)\right)$$
$$= H(N-1) + J\left(S_x(N)S_x(N+1) + S_y(N)S_y(N+1) + S_z(N)S_z(N+1)\right) \quad (2.3.3)$$

However, using renormalised operators:
$$H = H_{Ren} \otimes I_1 + S_{x\,Ren} \otimes S_x + S_{y\,Ren} \otimes S_y + S_{z\,Ren} \otimes S_z \quad (2.3.4)$$

This way, we have described the interaction between one added site and the renormalised block.

6) Return to 2).

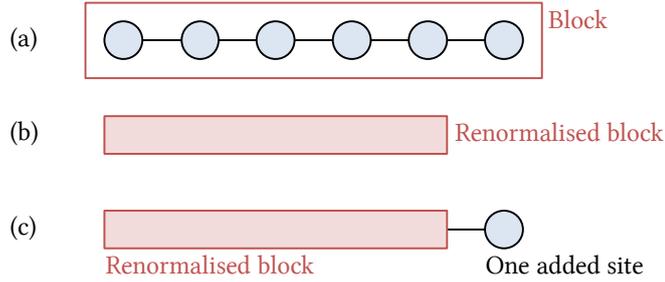

Figure 8. Diagram shows the basic idea of numerical renormalisation group for $M = 6$ sites.

## 2.3.3  DMRG on the Heisenberg model

In the numerical renormalisation group to create the renormalisation matrix $O$ we choose the states of the smallest eigenvalues. Unfortunately in composite systems, the states with smallest eigenvalues are not necessary the most "important" ones. This is the idea behind the density matrix renormalisation group (DMRG) – first we have to find out which states are relevant. To do this we use the density matrix (1.1.45) of the system. Because of its definition, if we calculate the eigensystem of $\rho$, the states that contribute in the most significant way are those with the biggest eigenvalues of $\rho$.

Now, we will consider not only the initial system $A$, consisting of $M$ sites, but also its mirror reflection $B$, so together they will form a system of $2M$ sites. The main idea behind it is as follows [29]: if we imagine the system as a "particle in a box" problem, then quantum mechanically we deal with plane waves that have nodes on the ends of the box (see Figure 9). In the procedure of renormalisation we have only considered one separate block, which means we neglected states without nods on the ends of the block (red states on the picture). Now, apart from our system block $A$ we will also include the environment block $B$. We will calculate the density matrix for the "superblock" $AB$, and then calculate the reduced density matrix (1.1.50) using partial trace over environment $B$ (1.1.52a). When we do the partial trace the ground state of the whole superblock $AB$ gets projected onto the reduced density matrix $\rho_A$ and thus we have improved the results and included the states analogic to the ones presented in Figure 9. So, the states that we are choosing now for the renormalisation matrix, are the ones that contribute the most to the entanglement between the system and the environment.

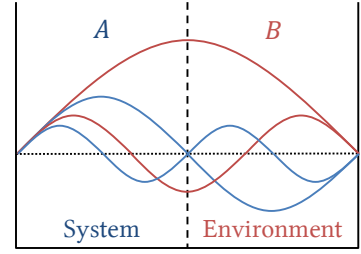

Figure 9. Single particle in the box. If we consider only system block, red waves are not included.

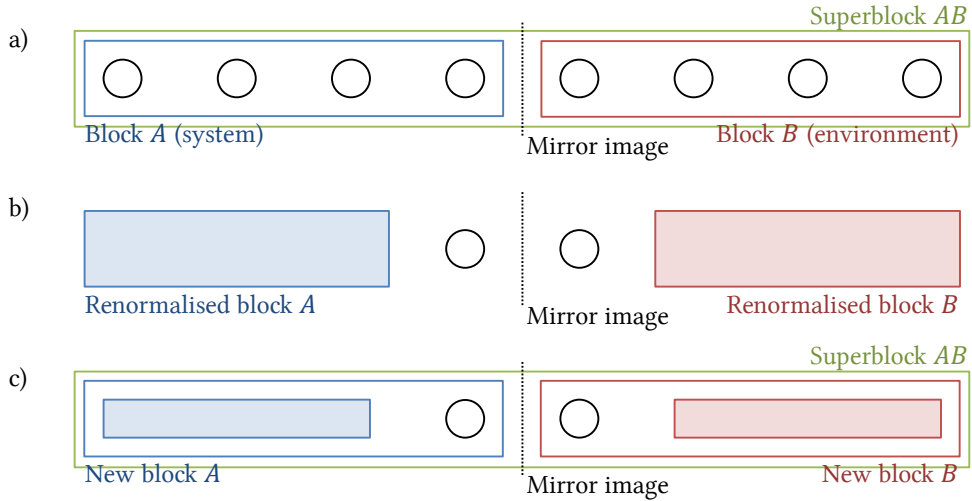

Figure 10. Diagram showing the DMRG procedure for $M = 4$ initial spins.

Algorithm of DMRG for the Heisenberg model is as follows [29]:
1) Set up the exact Hamiltonian $H$ of the Heisenberg model for $M$ spins using (1.2.12b). Denote it as subsystem (block) $A$ – see Figure 10a.
2) Create the Hamiltonian of an extended system – subsystem $A$ with subsystem $B$ (mirror image of subsystem $A$) – also see Figure 10a:
$$H_{AB} = H \otimes I_M + \underbrace{I_{M-1} \otimes S_x \otimes S_x \otimes I_{M-1}}_{\text{interaction between } A \text{ and } B} + I_M \otimes H \qquad (2.3.5)$$
3) Calculate eigenvalues $\{E_i^{AB}\}$ and eigenvectors $\{|\psi_i^{AB}\rangle\}$ of $H_{AB}$. Choose eigenvector which has the smallest eigenvalue (ground state) to be the target state $|\psi_{GS}\rangle$.
4) Create the density matrix of $AB$ system using the target state:
$$\rho = |\psi_{GS}\rangle\langle\psi_{GS}| \qquad (2.3.6)$$

5) Create the reduced density matrix for subsystem $A$. Ground state of superblock $AB$ gets projected onto this density matrix.
$$\rho_{Red} = \text{tr}_B \rho \qquad (2.3.7)$$

6) Choose relevant states: Determine eigenvalues $\{a_i\}$ and eigenvectors $\{|w_i\rangle\}$ of $\rho_{Red}$. Keep only those $2^{M-1}$ eigenvectors that correspond to the biggest eigenvalues. Create renormalisation matrix $O$ out of them:
$$O = \begin{pmatrix} |w_{2^{M-1}+1}\rangle & |w_{2^{M-1}+2}\rangle & \cdots & |w_{2^M}\rangle \end{pmatrix} \qquad (2.3.8)$$

7) Form renormalised operators (see Figure 10b):
$$\begin{aligned} H_{Ren} &= OHO^T \\ S_{x1\,Ren} &= O(I_{M-1} \otimes S_x)O^T \\ &\cdots \\ S_{x2\,Ren} &= O(S_x \otimes I_{M-1})O^T \\ &\cdots \end{aligned} \qquad (2.3.9)$$

8) Form a new Hamiltonian for subsystem $A$ using the renormalised operators (see Figure 10c):
$$H = H_{Ren} \otimes I_1 + S_{x1\,Ren} \otimes S_x + S_{y1\,Ren} \otimes S_y + S_{z1\,Ren} \otimes S_z \qquad (2.3.10)$$

9) Form a new Hamiltonian for system $AB$, using renormalised operators (also see Figure 10c):
$$\begin{aligned} H_{AB} = {} & H \otimes I_M + I_{M+1} \otimes H_{Ren} \\ & + I_{M-1} \otimes S_x \otimes S_x \otimes I_{M-1} + I_{M-1} \otimes S_y \otimes S_y \otimes I_{M-1} \\ & + I_{M-1} \otimes S_z \otimes S_z \otimes I_{M-1} \\ & + I_M \otimes S_x \otimes S_{x2\,Ren} + I_M \otimes S_x \otimes S_{x2\,Ren} + I_M \otimes S_x \otimes S_{x2\,Ren} \end{aligned} \qquad (2.3.11)$$

10) Return to step 3).

# Chapter 3
# Simulations of the spin models

## 3.1 Simulations of the XY model

### 3.1.1 Results and the analysis of the exact diagonalisation method

ED method was employed, using the Hamiltonian given by the equation (1.2.2), including the periodic boundary conditions. The sizes of the simulated systems were $M \in \{4; 5; ...; 13; 14\}$. We have to distinguish between systems with odd and even number of sites, because the equations for the energy are different for those two cases – see equation (1.2.9).

The results are shown in Figure 11 (even $M$) and Figure 12 (odd $M$). We have used the finite size scaling [30] to determine the values for the bulk limit**. The dashed lines are the polynomial fits, which were calculated using the least square fit method through the given points. The dotted lines mark the exact value of the ground state energy for the bulk limit, see equation (1.2.11).

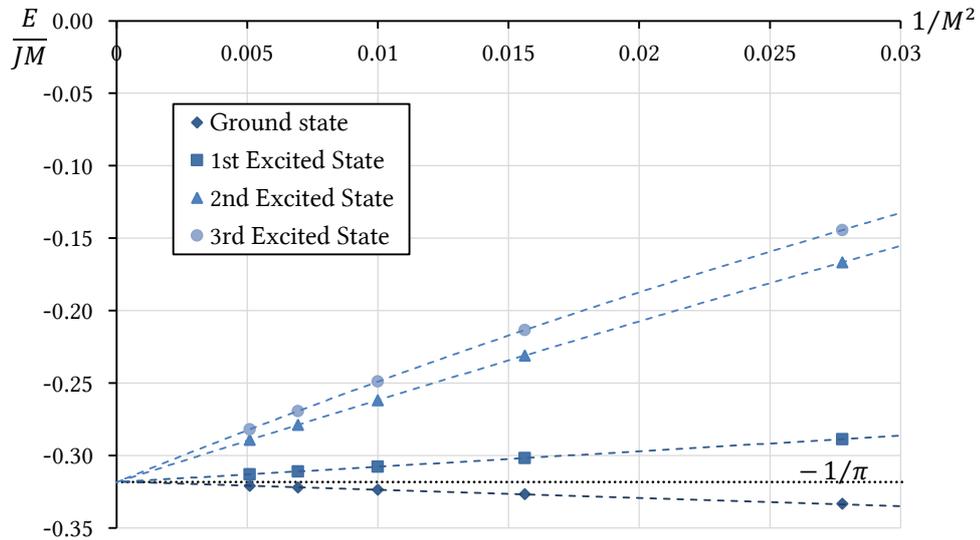

Figure 11. A plot showing first four energy levels of the XY models with sizes $M = 4, 6, 8, 10, 12, 14$. Dashed lines are the polynomial fits; dotted line is the bulk limit value.

---

** For one-dimensional systems, the energy density should scale similarly to the particle in a box problem, i.e. $E_0/L \sim 1/L^2$ (see e.g. [34]). In higher dimensions, this has to be translated, for example for 2D systems: $E_0/L^2 \sim 1/L^3$ (see e.g. [30]). However, for NRG and DMRG, the scaling will be different, because we don't include all the "modes". The scaling there strictly depends on how we create the renormalised Hamiltonian and cannot be easily determined. We will assume $\sim 1/M^\alpha$.

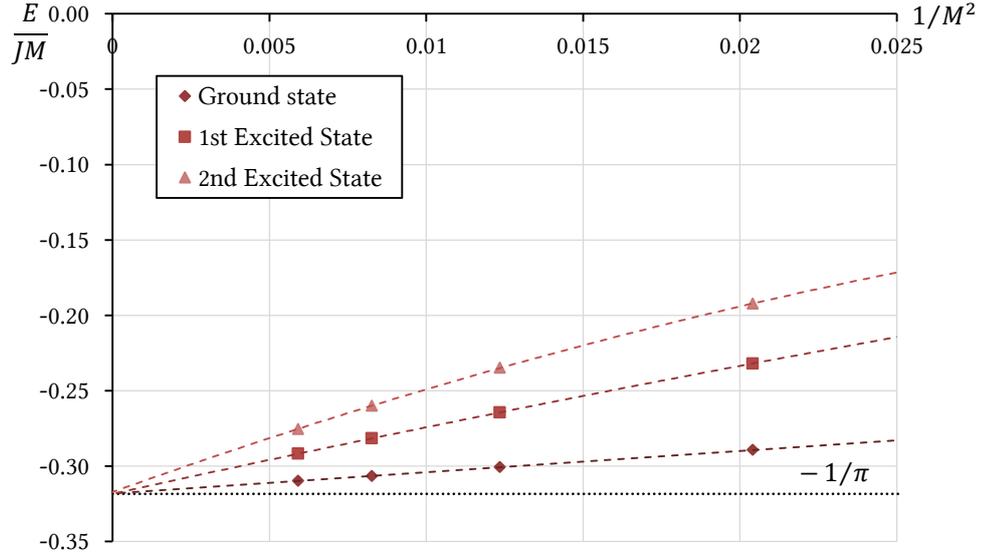

Figure 12. First three energy levels of the XY model with odd system sizes $M = 5, 7, 9, 11, 13$.

We can conclude from these plots that there is no energy gap between the ground state and the first excited state in the bulk limit for the XY model. Also, we can see that the series converge to the exact value $-1/\pi$.

To further estimate this number we use the following method. We fit the polynomial function of order $K$ to the data points using the least square fitting method [31]:

$$E\left(\frac{1}{M}\right) = a_0 + a_1\left(\frac{1}{M}\right) + a_2\left(\frac{1}{M}\right)^2 + \cdots + a_K\left(\frac{1}{M}\right)^K$$

Usually, we will use the linear fit ($K = 1$) or the quadratic fit ($K = 2$). The value at $M \to \infty$ (in the bulk limit) corresponds to $E(1/M \to 0) = a_0$. The error of determining this value includes the standard error obtained from the fitting and the uncertainty of choosing the right polynomial order $K$. The rough magnitude of the latter can be calculated as a discrepancy between bulk limit estimates for different orders $K$. We will usually use the difference between the linear and the quadratic fit. Otherwise, it will be specified.

Linear fit was used to approximate the value of the ground state energy for the XY model:

$$\frac{E_0}{JM} = \begin{cases} -0.3179 \pm 0.0004 & \text{(even } M\text{)} \\ -0.31806 \pm 0.00026 & \text{(odd } M\text{)} \end{cases} \qquad (3.1.1)$$

We can use the Student's $t$-test [31] approach to determine if this value is consistent with the exact value. To do this we use the $t$-value:

$$t = \frac{|X_{\text{theory}} - X_{\text{simulation}}|}{\sqrt{\sigma_{\text{theory}}^2 + \sigma_{\text{simulation}}^2}} \approx \begin{cases} 1.024 & \text{(even } M\text{)} \\ 0.961 & \text{(odd } M\text{)} \end{cases}$$

Now, we choose the significance level, usually $\alpha = 0.05$, and the number of degrees of freedom, that is $\nu = n - 1$, where $n$ is the number of points that were used to determine $X_{\text{simulation}}$. In our case $\nu = 5$ (even $M$), 4 (odd $M$). From the $t$ distribution table, we find the value $t_{\alpha,\nu} = 2.571$ (even $M$), $2.776$ (odd $M$). We test if $t < t_{\alpha,\nu}$. If it's not, we can lower the significance level, but usually not below $\alpha = 0.01$. Here, we find out that $\alpha = 0.05$ is sufficient. This means that the difference between $X_{\text{theory}}$ and $X_{\text{simulation}}$ is not statistically significant at the 5% level [31]. We conclude that there's no reason for rejecting the assumption that the number estimated from the simulation is the same as the exact value of the ground state energy for bulk limit.

We have also used the exact diagonalisation method with a given magnetisation, which allowed us to simulate even larger systems, $M \in \{4; 5; ...; 15; 16\}$. The results are shown in Figure 13 and Figure 14.

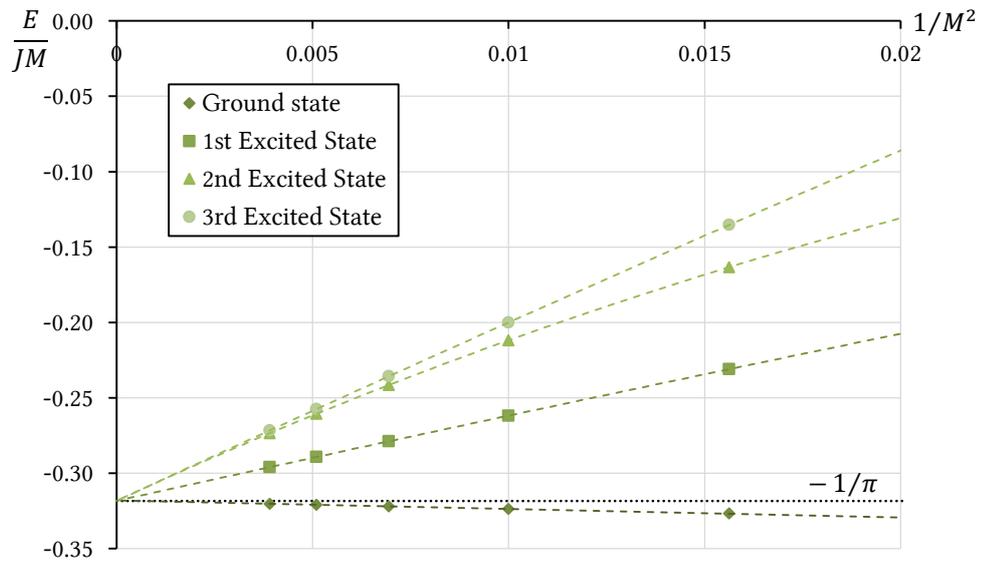

Figure 13. Simulation of the XY model with a given magnetisation $\mathcal{M} = 0$ for the even system sizes $M$.

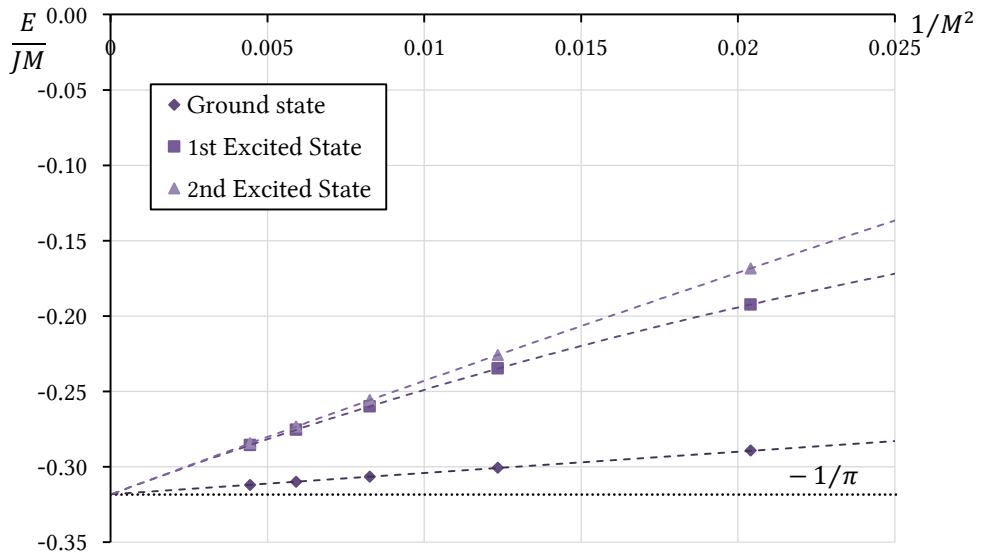

Figure 14. Simulation of the XY model with a given magnetisation $\mathcal{M} = 0.5$ for the odd system sizes $M$.

The finite size scaling gives the following results: there is no mass gap for the XY model and the ground state energy is estimated to be (linear fit):

$$\frac{E_0}{JM} = \begin{cases} -0.31797 \pm 0.00037 & \text{(even } M\text{)} \\ -0.31816 \pm 0.00021 & \text{(odd } M\text{)} \end{cases} \quad (3.1.2)$$

Employing Student's $t$-test, we have calculated: $t = 0.918$ (even $M$), $0.713$ (odd $M$). At the confidence level $\alpha = 0.05$, we can see that $t < t_{\alpha,\nu}$, and thus we conclude that the simulated values are no different from the exact value of the ground state energy.

### 3.1.2 Results and the analysis of the NRG method

We have used the numerical renormalisation group approach to simulate the ground state of the XY model using the Hamiltonian (1.2.1) with open boundary conditions. The simulations were set to have an initial system sizes $M_{\text{initial}} = 4, 5, 6$ and the target system size (the size at the end of the successive application of renormalisation step) was $100000$. The ground state energy was measured for every $M$ that was a multiple of $1000$, which has given us $100$ measure points.

Figure 15 presents the results. The finite size scaling was used and we can see that the numerical values converge in the bulk limit $M \to \infty$. Dashed lines show the linear fits. The estimated results are shown below:

| $M_{\text{init}}$ | $E_0/JM$ |
|---|---|
| 4 | $-0.2984358 \pm 0.0000021$ |
| 5 | $-0.3068305 \pm 0.0000012$ |
| 6 | $-0.308131 \pm 0.000005$ |

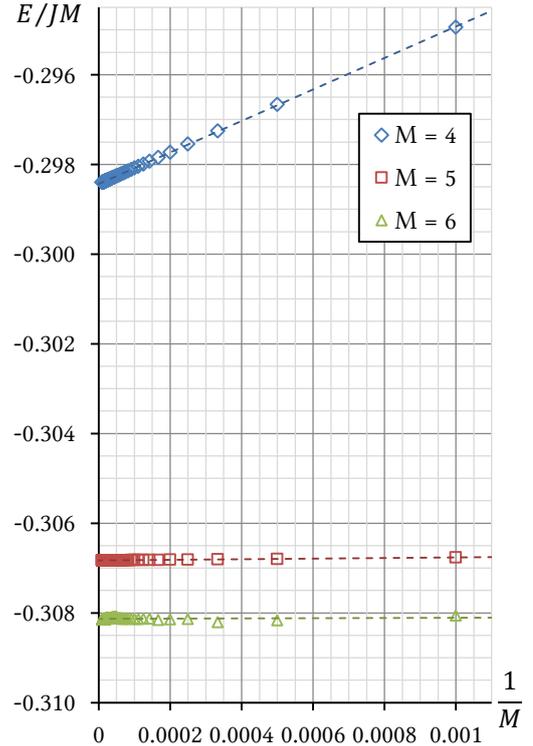

Figure 15. The finite size scaling in the NRG method used on the XY model.

To get an estimate for $M_{\text{initial}} \to \infty$, we have employed the finite size scaling against $1/M_{\text{initial}}$, as presented in Figure 16. The exponential fit was made and the estimated value is:

$$\frac{E_0}{JM} = -0.308761 \pm 0.000006 \quad (3.1.3)$$

We can see that the achieved value is not consistent with the exact value (1.2.11).

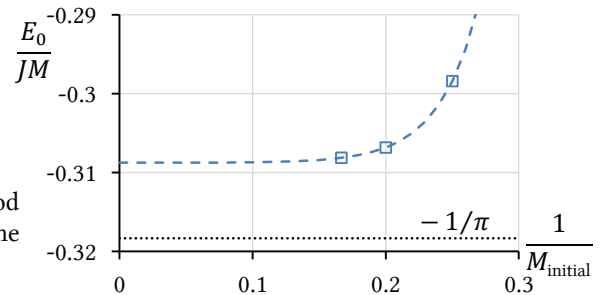

Figure 16. Finite size scaling of the NRG method results, that should give us an approximate idea of the true value.

This was expected since the numerical renormalisation group approach gives only approximate results and has some flaws, as discussed in Chapter 2.3. Therefore, increasing the value of $M_{\text{initial}}$ would not necessary lead us to the exact value. NRG is helpful for estimating the general properties for the system, but is still not a proper device to achieve good values.

### 3.1.3 Results and the analysis of the DMRG method

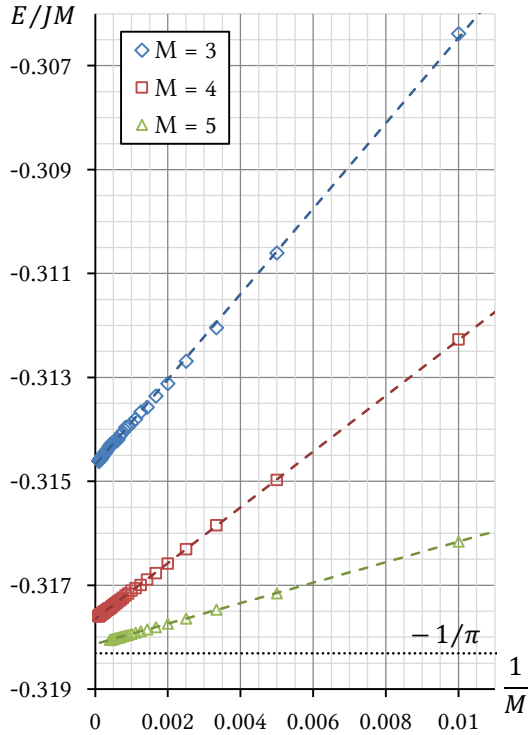

Figure 17. DMRG results for the XY model, using the initial system sizes $M_{\text{initial}} = 3, 4, 5$.

We have employed the DMRG method for the Hamiltonian (1.2.2) with open boundary conditions. Initial system sizes: $M_{\text{initial}} = 3, 4, 5$ and the target system sizes: $10000, 10000, 2500$. The measurements were done every time $M$ was a multiple of $100$, which gives us $100, 100$ and $25$ points.

Finite size scaling of the ground state energy is shown in Figure 17. The numbers converge rapidly for the bulk limit and give result that is close to the exact value (1.2.11). The estimates from the linear fits are:

| $M_{\text{init}}$ | $E_0/JM$ |
| --- | --- |
| 3 | $-0.314682 \pm 0.000010$ |
| 4 | $-0.317650 \pm 0.000016$ |
| 5 | $-0.318133 \pm 0.000011$ |

Approximate value for the very large $M_{\text{initial}}$ was estimated (see ):

$$\frac{E_0}{JM} = -0.31844 \pm 0.00005 \qquad (3.1.4)$$

The value can be tested against the exact value, and the Student's $t$-value was found to be $t = 2.602$. At the confidence level $\alpha = 0.01$, $t < t_{\alpha,\nu} = 2.797$, and we deduce that there is no reason to reject the hypothesis of equality between the value estimated in the simulation and the exact value of the ground state energy. To be able to come to the same conclusion but using a higher confidence level $\alpha$, we can simulate this system using bigger initial size in the future.

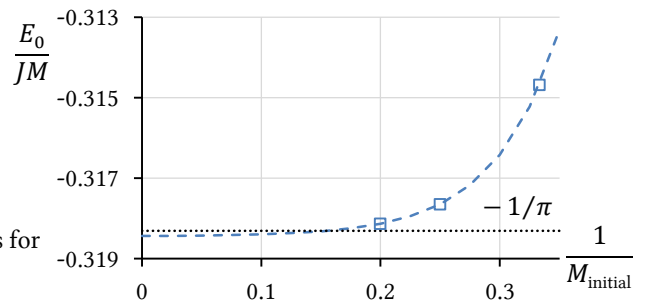

Figure 18. Finite size scaling of the DMRG results for the XY model.

## 3.2 Simulations of the Heisenberg model

### 3.2.1 Results and the analysis of the exact diagonalisation method

We have used the ED method to simulate the Heisenberg model given by the Hamiltonian (1.2.12b) with periodic boundary conditions. The simulated system sizes were $M \in \{4; 5; ...; 13; 14\}$. Similarly to the approach used in the Ch. 3.1, we have to distinguish between systems with odd and even sizes.

The results are shown in Figure 19 and Figure 20. Finite size scaling was performed to estimate the values at the bulk limit $M \to \infty$. The dashed lines are the polynomial fits and the dotted lines are the exact values given by equation (1.2.13).

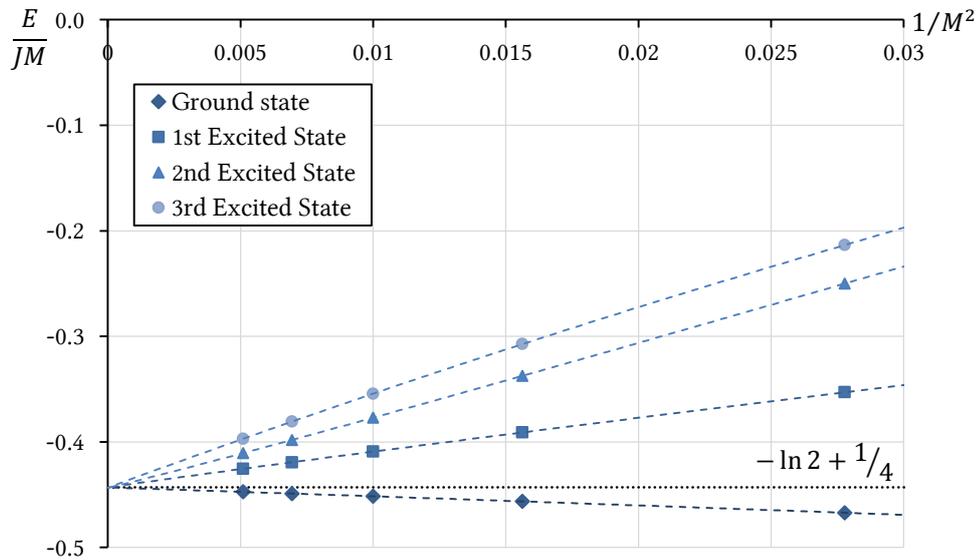

Figure 19. Finite size scaling for the Heisenberg model low-lying energy levels for even system sizes $M = 4, 6, 8, 10, 12, 14$.

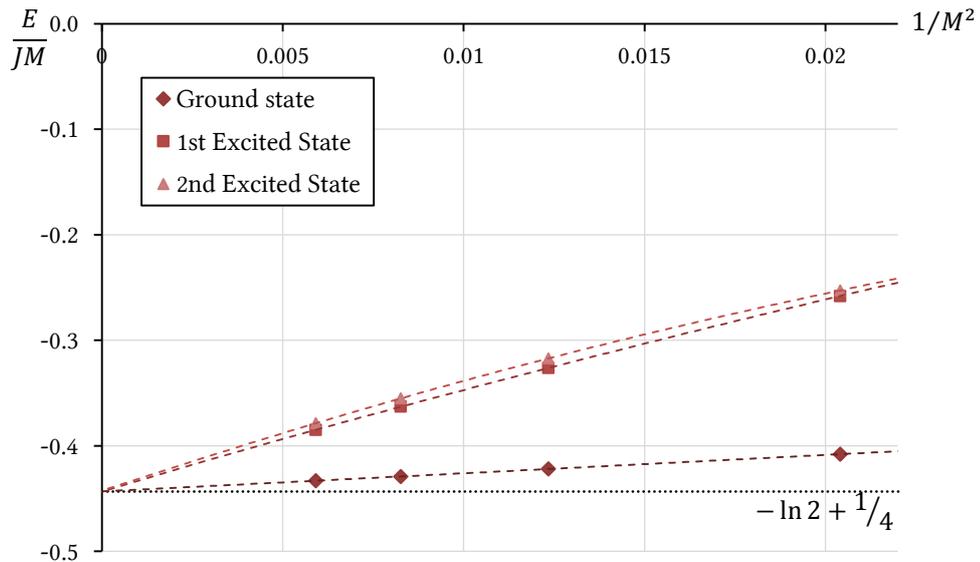

Figure 20. Finite size scaling for the Heisenberg model low-lying energy levels for odd system sizes $M = 5, 7, 9, 11, 13$.

Using the results we can conclude that the energy gap between the ground state energy and the first excited state energy is equal to zero in the bulk limit. Besides, we have used the linear fits to estimate the value of the ground state energy in the $M \to \infty$ limit:

$$\frac{E_0}{JM} = \begin{cases} -0.4423 \pm 0.0009 & \text{(even } M\text{)} \\ -0.44342 \pm 0.00015 & \text{(odd } M\text{)} \end{cases} \quad (3.2.1)$$

Employing the same approach as before, we can calculate Student's $t$-value, which is $t = 0.947$ (even $M$), $1.850$ (odd $M$). Setting the confidence level to $\alpha = 0.05$ and using $\nu = 5$ (even $M$), $4$ (odd $M$), we can conclude that for both cases, $t < t_{\alpha,\nu}$, which means we have very high confidence that the value of the ground state energy estimated in the simulation is the same as the exact value.

Exact diagonalisation with a given magnetisation was also used. The simulated system sizes are $M \in \{4; 5; \ldots; 15; 16\}$. The results are presented in Figure 21 and Figure 22.

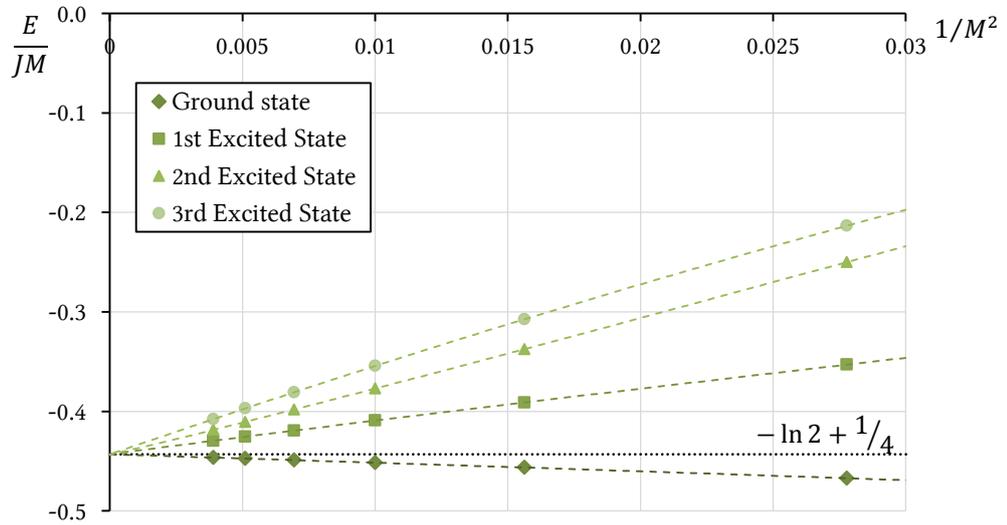

Figure 21. ED method result for the Heisenberg model low-lying energy levels with even number of sites. Given magnetisation is $\mathcal{M} = 0$.

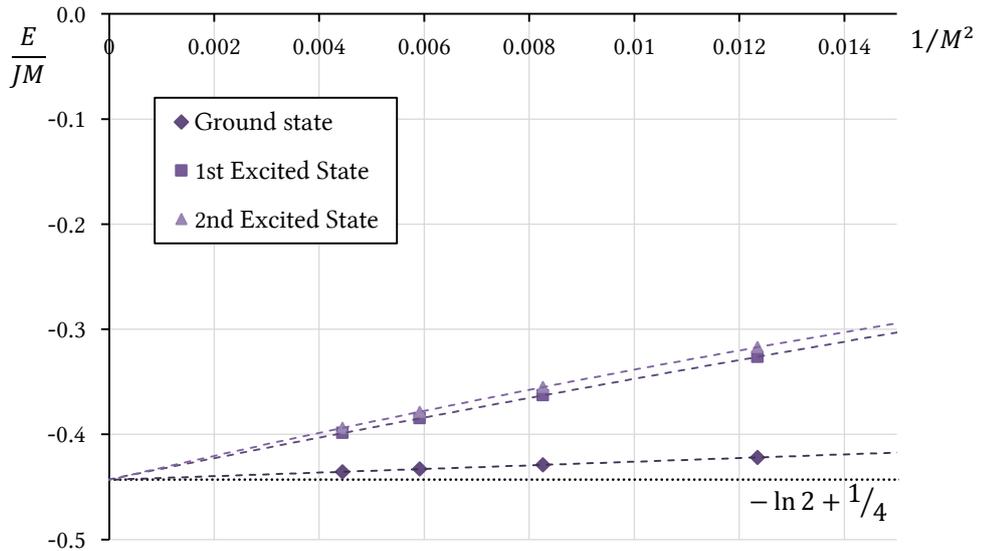

Figure 22. ED method result for the Heisenberg model low-lying energy levels with odd number of sites. Given magnetisation is $\mathcal{M} = 0.5$.

Like before, we can see that in the bulk limit there is no energy gap. The estimated values (linear fits) of the ground state energy are:

$$\frac{E_0}{JM} = \begin{cases} -0.4425 \pm 0.0007 & \text{(even } M\text{)} \\ -0.44335 \pm 0.00009 & \text{(odd } M\text{)} \end{cases} \quad (3.2.2)$$

The Student's $t$-values are $t = 0.947$ (even $M$), $2.144$ (odd $M$). Using $\alpha = 0.05, \nu = 6$ (even $M$), $\nu = 5$ (odd $M$), we can see that $t < t_{\alpha,\nu}$ and we deduce that the estimated value is consistent with the theoretical calculation. Still, there is a possibility of doing the additional simulations with bigger system sizes (if one has the appropriate memory space for the simulation), which would give us better approximation.

### 3.2.2 Results and the analysis of the NRG method

We have used the numerical renormalisation for the Heisenberg model (1.2.12b) with open boundary conditions and with initial sizes $M_{\text{initial}} = 4, 5, 6$. The target size was $100000$, while the measurements were performed every time $M$ was a multiple of $1000$, which gives us $100$ measure points.

The results are presented in Figure 23. The numbers converge rapidly in the bulk limit, and we have employed the finite size scaling to estimate the values for $M \to \infty$. We have used the linear fits to determine the estimates:

| $M_{\text{init}}$ | $E_0/JM$ |
|---|---|
| 4 | $-0.423073875 \pm 0.000000028$ |
| 5 | $-0.43317388 \pm 0.00000009$ |
| 6 | $-0.43707876 \pm 0.00000036$ |

We can see that those points differ significantly from the exact value given by equation (1.2.13). However, we can estimate the correct value by scaling the energy against $1/M_{\text{initial}}$, as shown in Figure 24:

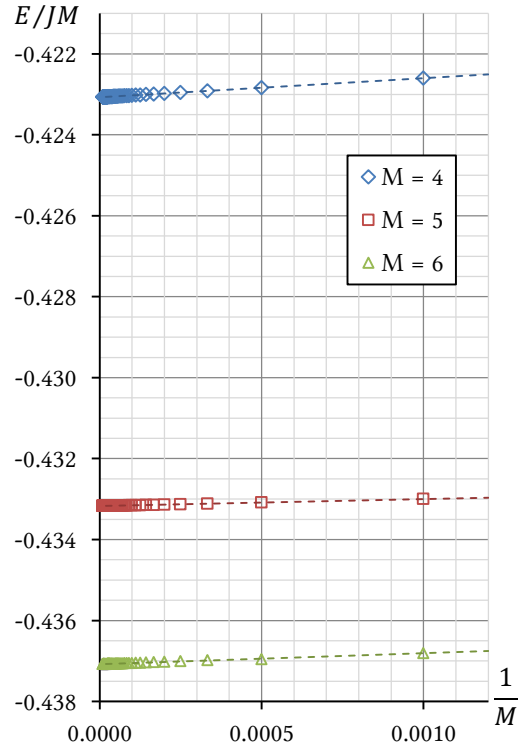

Figure 23. NRG results for the Heisenberg model showing the ground state energy.

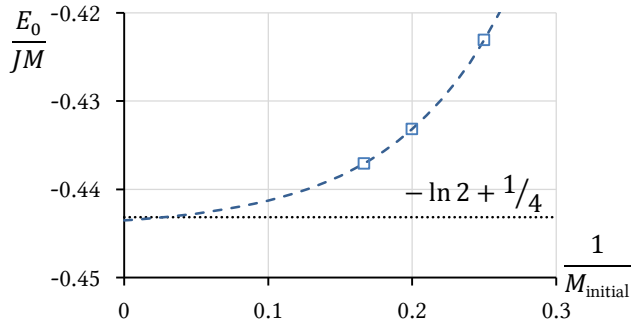

Figure 24 (on the left). Scaling of the ground state energy against $1/M_{\text{initial}}$. The errorbars are too small to be seen on the plot.

Afterwards, we have used the exponential fit to determine the approximate value of the ground state energy:

$$\frac{E_0}{JM} = -0.44438 \pm 0.00011 \tag{3.2.3}$$

It has to be noted that the method of scaling against $1/M_{\text{initial}}$ should be done with more simulations for different $M_{\text{initial}}$, so that the true dependence of the ground state energy on $M_{\text{initial}}$ can be determined. In our case it was not possible due to the limited computer memory resources. Thus, we consider this result to have a bigger error than it is calculated using the fitting.

To test the result, we have determined Student's $t$-value to be $t \approx 11$, which means that the estimated value is not consistent with the exact result. Still, as one can see in Figure 24, the value is very near the expected number. Numerical renormalisation is therefore suitable for calculating the approximation, but for better and consistent results, we need to start the simulation with bigger values of $M_{\text{initial}}$.

### 3.2.3   Results and the analysis of the DMRG method

The density matrix renormalisation group approach was used to study the ground state energy of the Heisenberg model, given by equation (1.2.12b), with open boundary conditions. The initial system sizes were $M_{\text{initial}} = 3, 4, 5$ and target sizes were respectively $10000, 10000, 2500$.

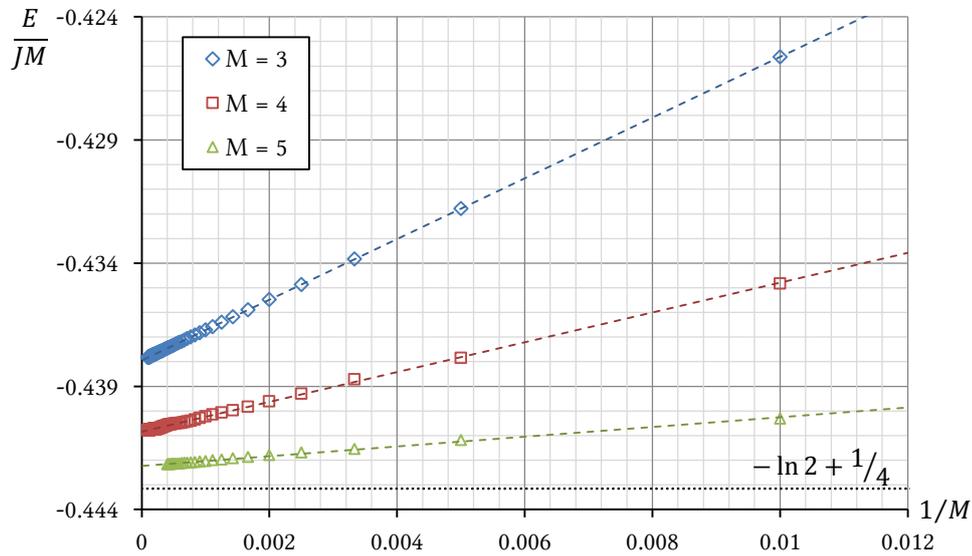

Figure 25. Ground state energy of the Heisenberg model using DMRG approach.

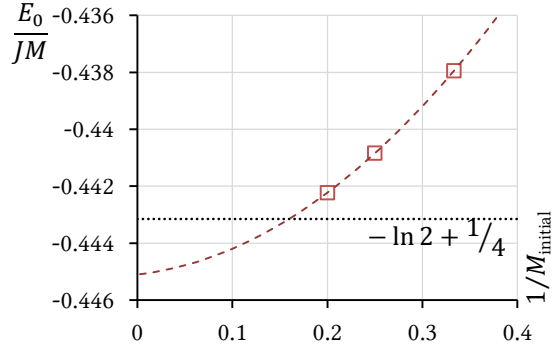

Figure 26. Scaling of the ground state energy for different initial system sizes.

The results are shown in Figure 25. One can see that the numbers converge in the infinite volume limit, but the linear estimates for the ground state energy differ from the exact value:

| $M_{init}$ | $E_0/JM$ |
|---|---|
| 3 | $-0.437945899 \pm 0.000000024$ |
| 4 | $-0.4408444 \pm 0.0000030$ |
| 5 | $-0.442228 \pm 0.000007$ |

We have performed the same method as in Chapter 3.2.2, i.e. we have plotted the ground state energy versus $1/M_{initial}$, as seen in Figure 26. After making the exponential fit (dashed lines), we have estimated the value of the ground state energy for very large $M_{initial}$:

$$\frac{E_0}{JM} = -0.4451 \pm 0.0009 \qquad (3.2.4)$$

Performing the Student's $t$-test, gives: $t = 2.169$, which compared to the critical $t$ value for $\alpha = 0.02$ and $\nu$ being a very large number ($\to \infty$), yields: $t < t_{\alpha,\nu} = 2.326$. As we can see there is no reason to discard the hypothesis that the ground state energy determined in the simulation is no different from the theoretical value. However, to further compensate the uncertainty, we suggest simulating systems starting with larger system sizes.

## 3.3 Simulations of the $J - J'$ Heisenberg model

In the computer simulation, we will assume $J > 0$ and investigate the range $J'/J \in [0; 1]$, so that both phase transitions at points described by equations (1.2.14) and (1.2.16) can be examined. The interaction constant has been sampled with a step $0.01$.

### 3.3.1 Results and the analysis of the exact diagonalisation method

The simulation has been performed for finite systems with sizes $M = 10, 12, 14$, using the Hamiltonian (1.2.13b) with periodic boundary conditions. We have chosen the systems with an even number of sites, because it was suggested by the findings of [15] that this gives better results. The following three plots show the lowest energy levels for each of these systems.

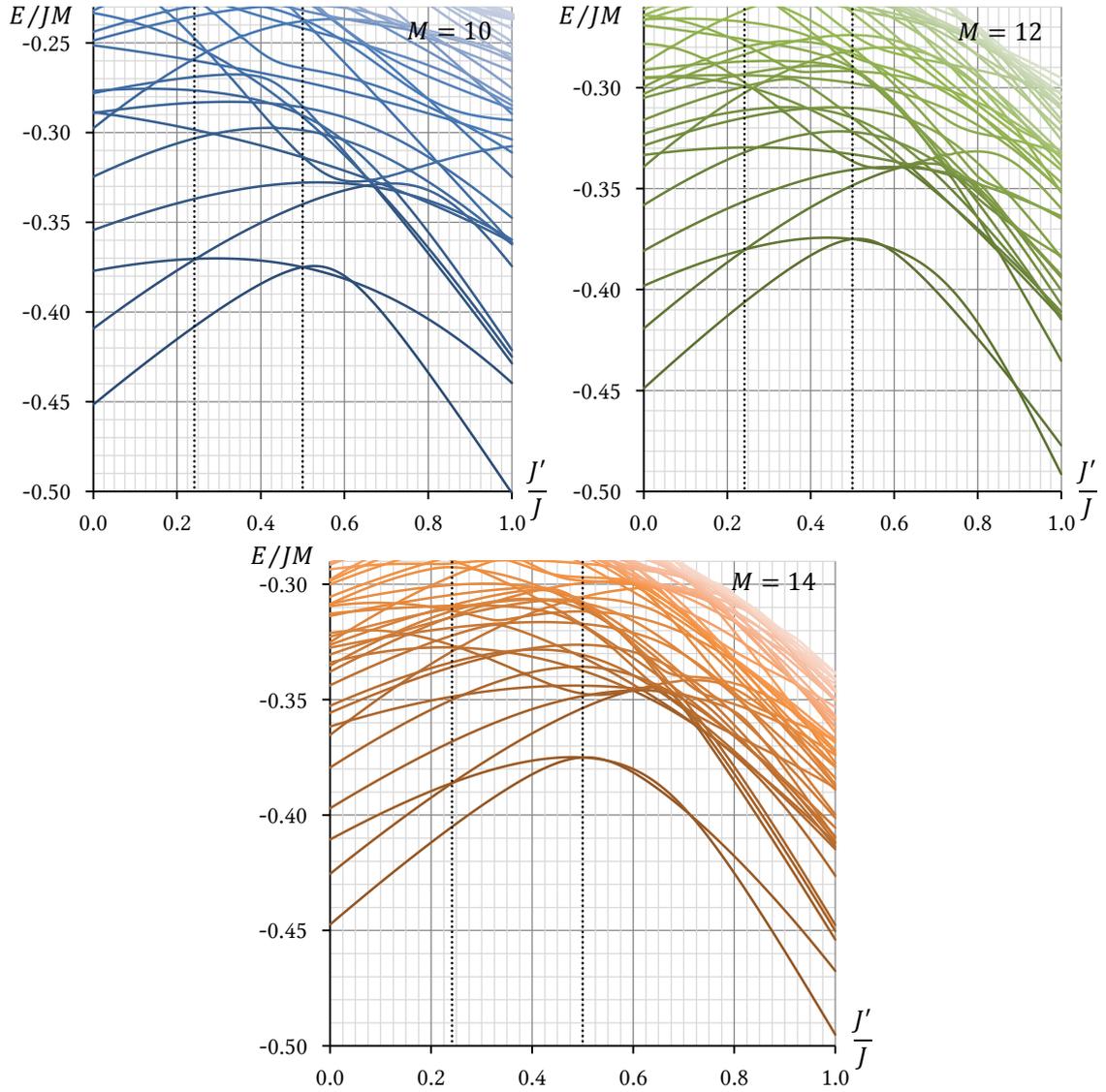

Figure 27. Plots showing the low-lying energy levels of the systems with sizes $M = 10$ (blue), $M = 12$ (green), $M = 14$ (orange). Dotted lines indicate the phase transitions at $J'/J = \{0.2411; 0.5\}$.

In Figure 27 we can see that when we change the ratio of the interaction constants, some states may become degenerate and afterwards exchange their positions in the energy ladder. In particular, at the phase transition point $(J'/J)_{2C} = 0.5$ we have two degenerate ground states, which is consistent with the theory (see Ch. 1.2.3). Also, the energy at this point is *exactly* equal to $E/JM = -0.375$, as predicted by (1.2.17). Another interesting aspect, is that the first and the second excited states become degenerated at the second phase transition point $(J'/J)_C \approx 0.24$.

From the overall shape of the ground state level, we can forecast that in the bulk limit, at the point $(J'/J)_{2C} = 0.5$ there is a (local?) maximum of the energy.

The lowest-lying energy level of the triplet $S = 1$ state was also calculated, using the exact diagonalisation method with a given magnetisation $\mathcal{M} = 1$. Hence, we have determined the gap $\Delta$ using equation (1.2.15).

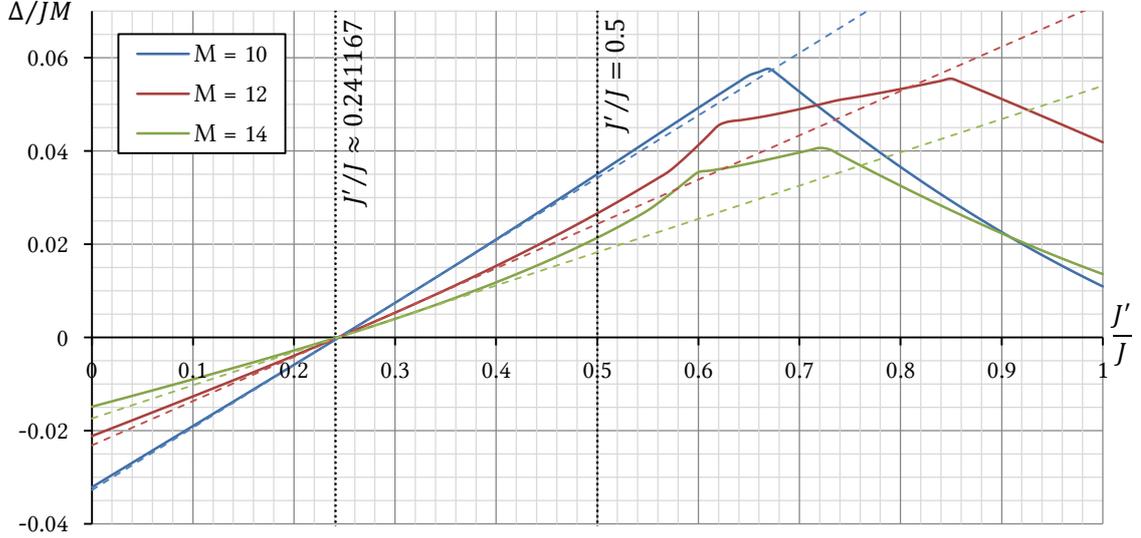

Figure 28. The energy gap $\Delta$ for the $N = 10, 12, 14$ finite systems.
The dashed lines are the linear fits for $J'/J$ near the phase transition.

Figure 28 shows how the energy gap depends on the interaction constant. Below the transition point $(J'/J)_C$ the system has a small gap and we expect it to become gapless in the bulk limit, as discussed in Ch. 1.2.3. We have used the linear fit for a few points near the phase transition, to determine the correct value of $(J'/J)_C$ for the system with a given number of spins. This occurs when the gap $\Delta$ becomes zero. Finally, we can use finite size scaling, as suggested in [29] to obtain the estimate for the bulk limit $M \to \infty$, as shown in Figure 29.

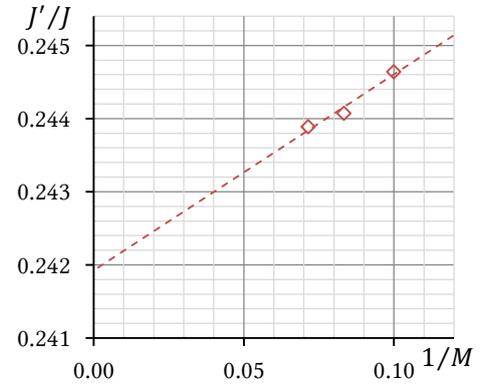

Figure 29. Finite size scaling for $J'/J$.

The estimated value is:

$$(J'/J)_C = 0.2419 \pm 0.0005 \qquad (3.3.1)$$

Using the calculated before value (1.2.14), we can examine our results with the Student's $t$-test:

$$t = \frac{|0.2419 - 0.241167|}{\sqrt{0.0005^2 + 0.000005^2}} \approx 1.47$$

On the significance level of $\alpha = 0.05$, $u_\alpha = 1.95$ and we can see that $t < u_\alpha$. Therefore, there is no reason to discard the hypothesis that the result is no different from the expected value.

### 3.3.2  Results and the analysis of the NRG method

Our objective will be to determine the dependence of the ground state energy on the interaction constants ratio $J'/J$. We expect it to have a maximum around the point $(J'/J)_{2C}$. In the simulation we have used the Hamiltonian (1.2.13b) with open boundary conditions and $M_{\text{initial}} = 4, 5, 6$ system sizes, i.e. we started with $3, 4$ or $5$ sites and added an additional one site and then renormalised the system. The goal was to achieve the system of size $M = 100000$.

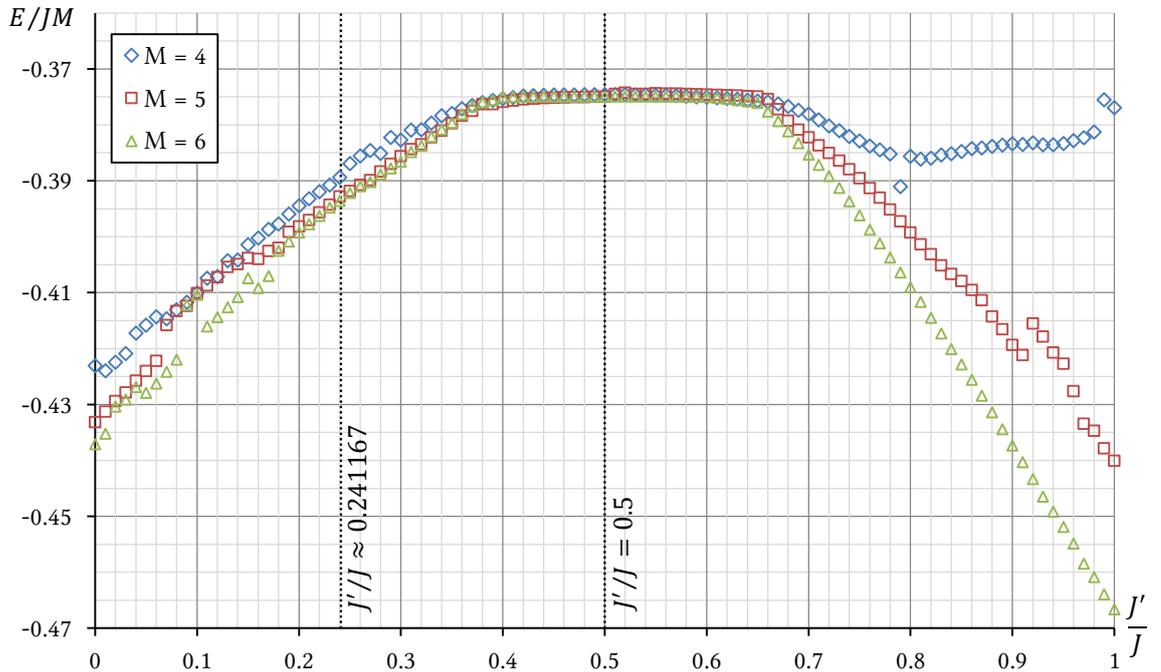

Figure 30. Ground state energy dependence on the interaction constants ratio (NRG).

The results of the numerical renormalisation group approach on the $J - J'$ Heisenberg model are shown in Figure 30. We can see that the method failed to show us the phase transition that occurs at $(J'/J)_{2C} = 0.5$. It was expected, since NRG has many flaws, as discussed in Chapter 2.3.

As a test for this method, we can use the finite size scaling during the renormalisation of the system, with a specific value of $J'/J$. We will use $J'/J = 0.5$, since the exact value of the energy is known for this point – equation (1.2.17). Also, this finite size scaling will be performed for the largest simulated system ($M_{\text{initial}} = 6$).

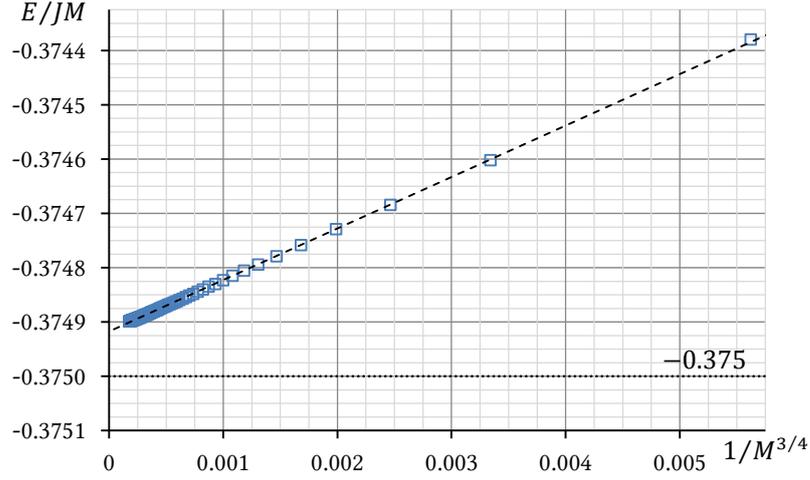

Figure 31. Finite size scaling of the ground state energy for $J'/J = 0.5$.

The linear fit shown in Figure 31 gives the result:

$$\left.\frac{E_0}{JM}\right|_{\frac{J'}{J}=0.5} = -0.3749175 \pm 0.0000012 \qquad (3.3.2)$$

Though the relative error of this value is very small, there's no doubt that the result is wrong – we can conclude this from the Figure 31. Using Student's $t$-test:

$$t = \frac{|0.375 - 0.3749175|}{0.0000012} \approx 69$$

The value of $t$ is so big that we have to reject the hypothesis that the result is equal to the exact value. As we can see, the numerical renormalisation group approach gives fairly accurate results and can be used for determining the approximate behaviour of the system, but it is still far from giving us the exact results.

### 3.3.3 Results and the analysis of the DMRG method

In the DMRG simulation, we have used the following parameters: the target system size is $10000$, when we begin with $M = 3$ and $M = 4$ sites and is $2500$ when we begin with $M = 5$ sites. The objective is the same as in the previous chapter: to determine the dependence of the ground state energy on the interaction constants ratio. The results are shown in Figure 32.

We can see that there is a clear maximum in ground state energy level at the critical point $(J'/J)_{2C} = 0.5$. The overall shape of the result is very smooth (especially for the biggest initial system size $M_{\text{initial}} = 5$), as it was expected from the exact diagonalisation plots (Figure 27).

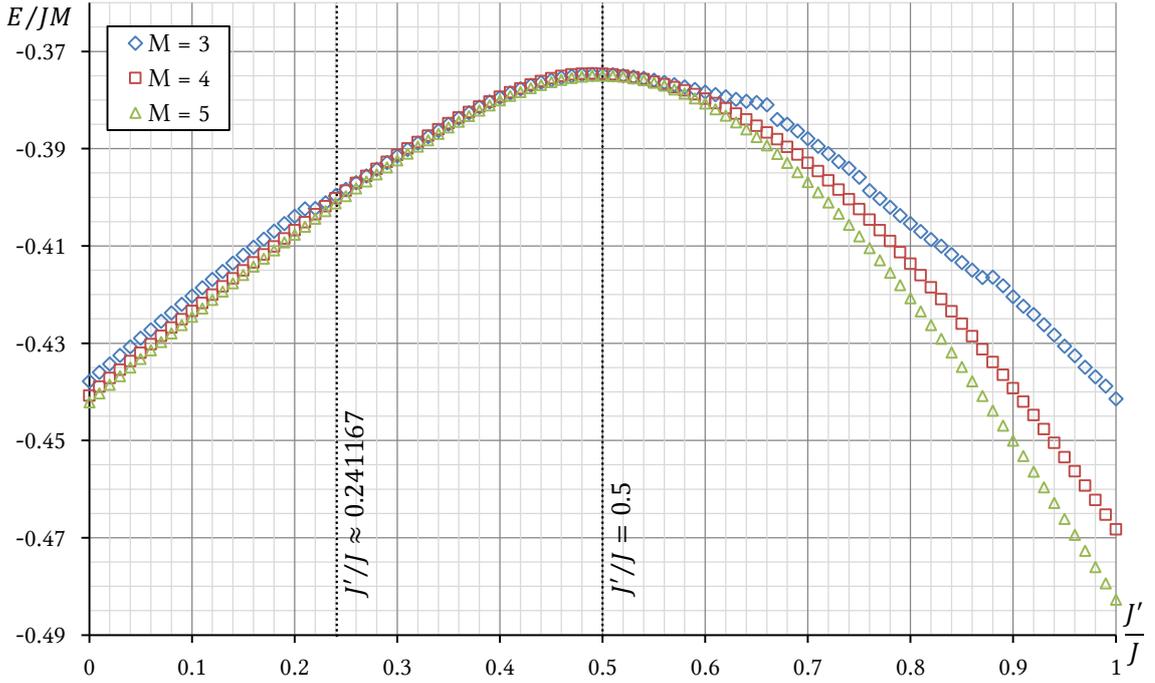

Figure 32. DMRG results for the initial system sizes $M_{\text{initial}} = 3, 4, 5$.

Like in the Chapter 3.3.2, we can perform the finite size scaling for the critical value $(J'/J)_{2C}$. This time we expect the result to be very precise.

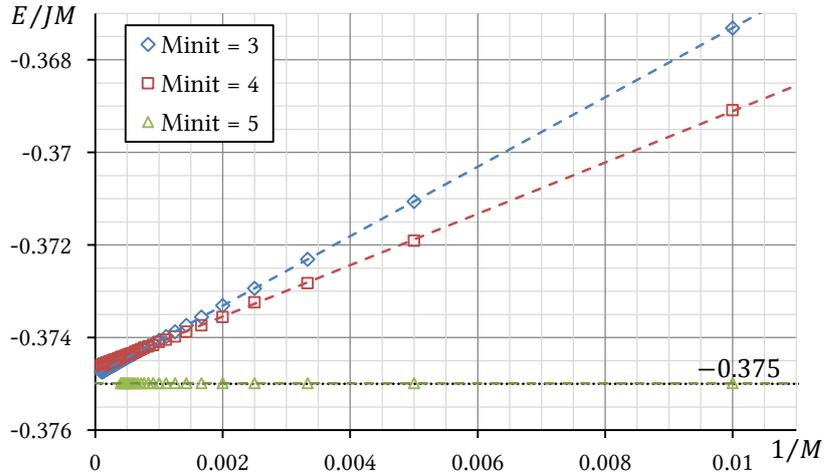

Figure 33. The finite size scaling for the ground state energy at $(J'/J)_{2C} = 0.5$.

Figure 33 shows that for $M_{\text{initial}} = 5$, the ground state energy is indeed very close the exact value. Linear fit produces the estimate:

$$\left.\frac{E_0}{JM}\right|_{\frac{J'}{J}=0.5} = -0.374985 \pm 0.000008 \tag{3.3.3}$$

Using the Student's $t$-test, we determine that the $t$-value is:

$$t = \frac{|0.375 - 0.374985|}{0.000008} \approx 1.87$$

The significance level $\alpha = 0.05$ is chosen and the number of the degrees of freedom is $\nu = 24$. Since $t < t_{\alpha,\nu} = 2.064$, we deduce that the value achieved in the simulation is consistent with the exact result given by equation (1.2.17).

We can see that the DMRG method is quite accurate and gives very consistent results. We have successfully used this method to determine the dependence of the ground state energy on the interaction constants ratio in the Heisenberg zigzag model and succeeded in obtaining a very precise result for the critical point $(J'/J)_{2c}$.

# Chapter 4
# Simulations of the Schwinger model

## 4.1 Exact diagonalisation of the free Schwinger model

Our goal was to use the ED method to determine the first free energy levels of the free Schwinger model in the bulk limit. As mentioned in Chapter 1.3.4, the **free massless Schwinger model** is equivalent to the XY model. We have successfully simulated the XY model energy levels in Ch. 3.1.1. The summary of the results is as follows:

- In the bulk limit the ground state energy of the model was determined to be (from equation (3.1.2) ):
$$\frac{E_0}{JM} = \begin{cases} -0.31797 \pm 0.00037 & \text{(even } M\text{)} \\ -0.31816 \pm 0.00021 & \text{(odd } M\text{)} \end{cases}$$
which is consistent with the exact value $-1/\pi$.

- There is no mass gap (energy gap between the ground state and the first excited state) in the bulk limit, which can be seen in Figure 11 and Figure 12.

We will now move on to the **free Schwinger model with mass** $m$. As discussed in Chapter 1.3.4, this model is equivalent to the antiferromagnetic XY model with a uniform transverse magnetic field.

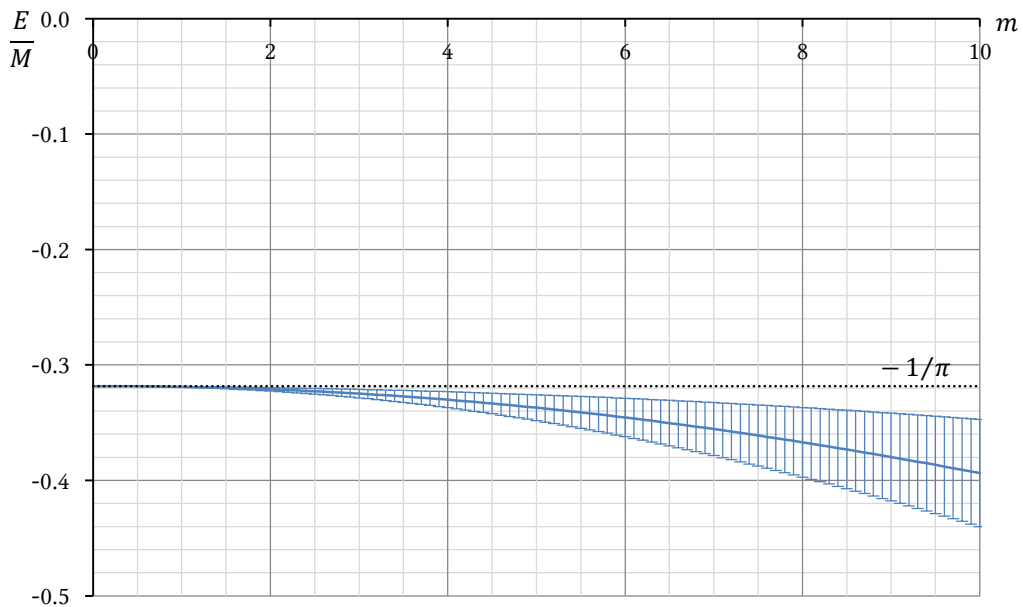

Figure 34. Ground state energy for the massive Schwinger model in the bulk limit.

Hamiltonian (1.3.18) was used to simulate system sizes $M \in \{4; 6; 8; 10; 12; 14\}$ and to obtain the ground state energy for each system. The examined range of mass was $m \in [0; 10]$ with sampling 0.1. After the simulation was done, the finite size scaling ($1/M^2$) was performed for every value of $m$. Quadratic fits were used to obtain the estimates of $E_0/M$ in the bulk limit. Errors were estimated as a difference between the values of quadratic and order 4 polynomial fits. The final results with errorbars are presented in Figure 34.

The theoretical prediction [5,3] is that the ground state energy should be independent of $m$. The fitting parameter errors shown in Figure 34 are found to be larger for larger values of $m$. This indicates, that when we go to higher $m$, the small size systems exhibit such high finite size effects, that we should use even higher order polynomials to estimate the true value of $E_0$. Of course, to employ such polynomials we have to also simulate bigger system sizes in the future.

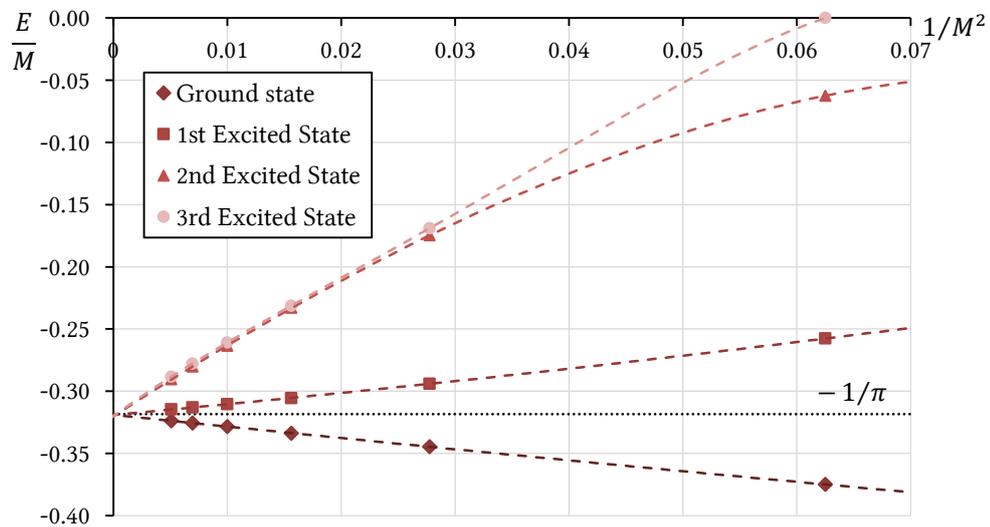

Figure 35. Low-lying energy levels of the massive Schwinger model with $m = 1.0$.

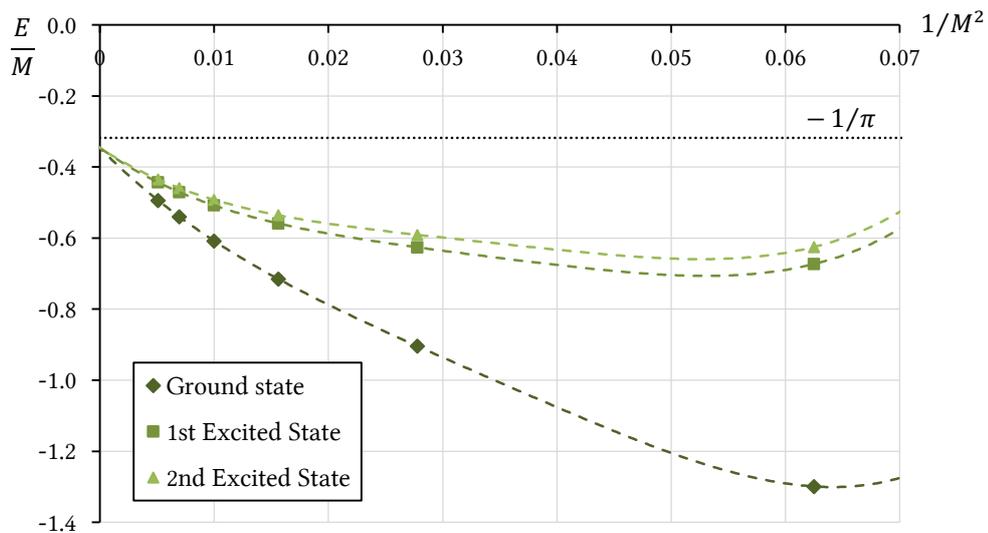

Figure 36. Low-lying energy levels of the massive Schwinger model with $m = 10.0$.

The simulation of low-lying energy levels was also performed for $m = 1.0$ and $m = 10.0$ using the same system sizes as before. The objective was to determine if the mass gap is present for the free Schwinger model with non-zero mass. Figure 35 and Figure 36 show the results. The finite size scaling with polynomial fits (dashed lines) was used to determine the values in the bulk limit. It should be emphasized that though we are dealing with the massive Schwinger model, we observe no mass gap in the infinite volume limit.

Besides, closer investigation of the plots ensures us that much bigger system sizes are needed for a good estimate of the ground state energy. We can see that even 4$^{\text{th}}$ order polynomial fit in Figure 36 was unable to provide us with a good answer. Conclusion is that to accurately determine $E_0$ in the massive Schwinger model with the exact diagonalisation method, we need to simulate very large systems.

## 4.2 Using the strong coupling expansion method for the Schwinger model

We have employed the strong coupling expansion (see Chapter 2.2) to simulate the Hamiltonian of the Schwinger model.

First of all, we have to discuss what order $N$ is suitable for a given number of sites $M$. Generally, for any perturbation method it is not certain that increasing the order $N$ will lead us to better results, i.e. we don't know if the results will converge in the limit $N \to \infty$ [6]. As it was found by [3], the most useful choice is $N = M - 2$, and we have used this relation in most cases. However, we were also interested in the question, if by increasing $N$, the results will diverge or not.

Secondly, it has to be stressed that we will be interested in the measured eigenvalues of the Hamiltonian in the two limits: the bulk limit $M \to \infty$ and the continuum limit $a \to 0 \Leftrightarrow x \to \infty$ at the same time (see equations (2.2.10) and (2.2.14) as an example). To do this we can follow two solutions proposed by [3]:

1. Increase $M$ and $N = M - 2$: the results should converge rapidly for any finite value of $x$ and the eigenvalue for the bulk limit will be "mapped out" to very large $x$.
2. Use different sizes $M$ and very large $N$: by increasing $N$ eigenvalue for the finite lattice will be mapped out to very large $x$. Use the finite size scaling relation by Fisher and Barber [32]:

$$f^\infty(x) - f^M(x) \underset{M \to \infty}{\sim} \alpha e^{-\beta M} \tag{4.2.1}$$

to obtain the bulk limit $M \to \infty$. $f^M(x)$ – the eigenvalue at size $M$, $f^\infty(x)$ – the eigenvalue in the bulk limit, $\alpha, \beta$ – constants. Such a fitting was found to be not working near $x \to \infty$ [33], but we can use it to determine the eigenvalues in the bulk limit for $x \approx 1$ and lower.

In the end we will use linear or quadratic extrapolants to estimate the value in the continuum limit $x \to \infty$. To do this, we will plot studied functions against $1/\sqrt{x}$, so that the continuum limit will be on the plot in $1/\sqrt{x} = 0$. Now, for specific values of $x$, the linear and the quadratic fits in $1/\sqrt{x}$ will be made and the interceptions with the axis $1/\sqrt{x} = 0$ will be observed – we will call those values the extrapolants $G$. Errors will be estimated as a discrepancy between the linear and quadratic extrapolants.

### 4.2.1 The massless Schwinger model

The Schwinger model with $m = 0$ was studied for different values of $x$. We were interested in obtaining the ground state energy (vacuum energy) function $f_0(x)$ described by equation (2.2.8). The quantity $1/\sqrt{x}$ was investigated in the range $1/\sqrt{x} \in [0.025; 2.0]$ with sampling $0.025$. The measurements were done for $M = \{6; 8; …; 16; 18\}$ and $N = M - 2$. We will use a symbol $[N, M]$ to denote such a measurement.

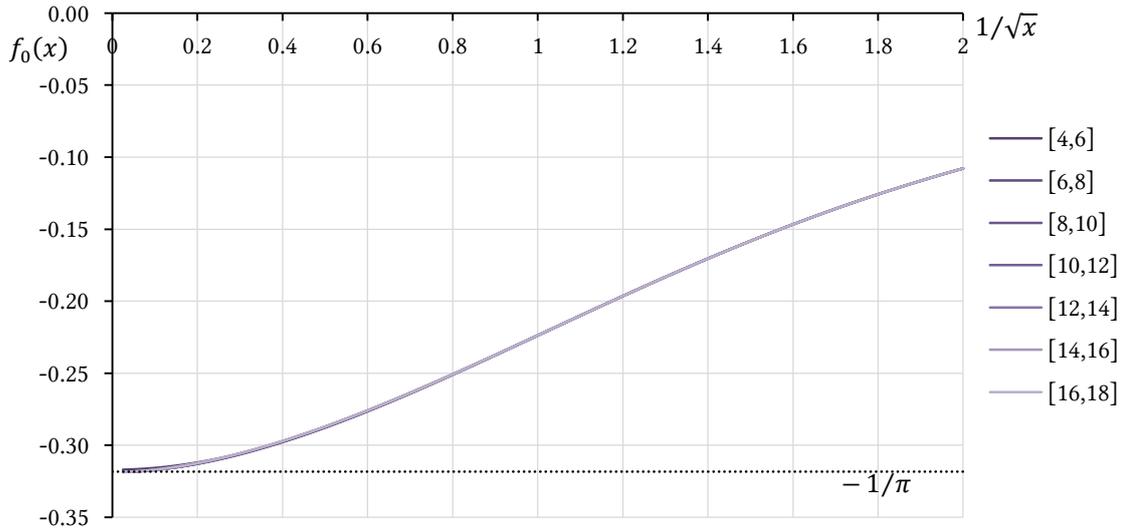

Figure 37. Vacuum energy $f_0(x)$ of the massless Schwinger model for different values of $[N, M]$.
Dotted line is the exact value for the continuum limit $x \to \infty$.

Interestingly, function $f_0(x)$ converges so fast, that all curves for different $[N, M]$ in Figure 37 are nearly overlapping. Also, the dotted line shows the exact value in the continuum limit, as given by (1.2.11), and we can see on the plot that near $1/\sqrt{x} \to 0$, this value is approximated very well. Using the linear and quadratic fits for the first few points, we can calculate the estimated value of $f_0(x \to \infty)$:

$$f_0(x)|_{x\to\infty} = -0.31888 \pm 0.00038 \quad (4.2.2)$$

To test this result, Student's $t$-test was employed, giving $t = 1.5$. With $\alpha = 0.05$ and $\nu = 4$, we can calculate that $t < t_{\alpha,\nu} = 2.776$ which means that the result is consistent with the exact value.

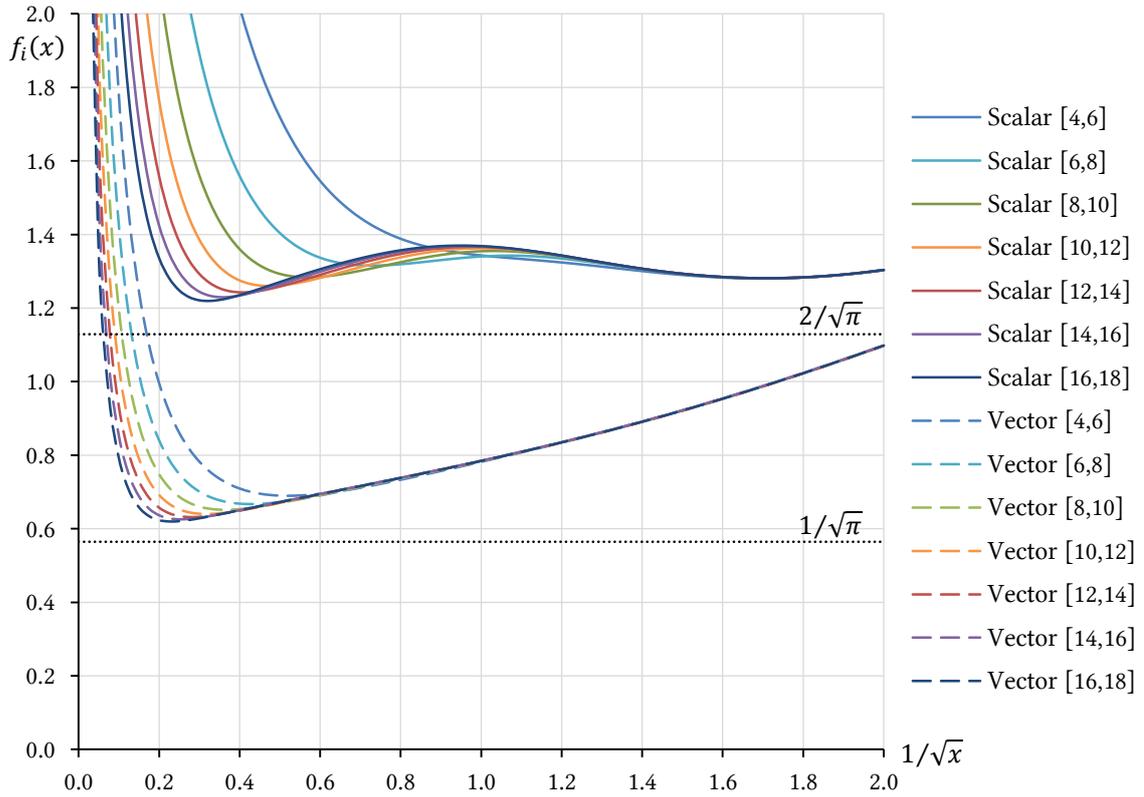

Figure 38. Scalar and vector particle masses $f_i(x), i \in \{+, -\}$ for the massless Schwinger model.

Scalar and vector particle masses given by equations (2.2.9) and (2.2.13) were also studied. The results are shown in Figure 38. The dotted lines are the exact results for the scalar and vector particle masses, given by equations (2.2.10) and (2.2.14). One can see that by increasing $[N, M]$, the curves become closer to the exact results in the bulk and continuum limits. The results are converging very fast for the vector state and relatively fast for the scalar state.

We will now use the linear and quadratic extrapolants [3] to estimate the values for this limit. For example, if we want to know the minimum values of $f_i(x)$, that can be approximated from our results, we would look for the values of extrapolants at the inflexion points of $f_i(x)$, like in Figure 39.

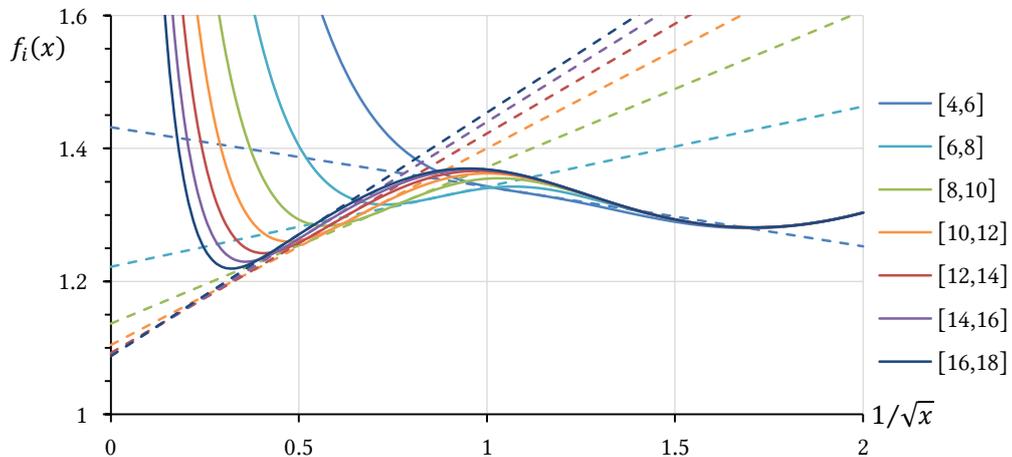

Figure 39. Finding the extrapolants at the inflexion points for scalar particle mass curve.

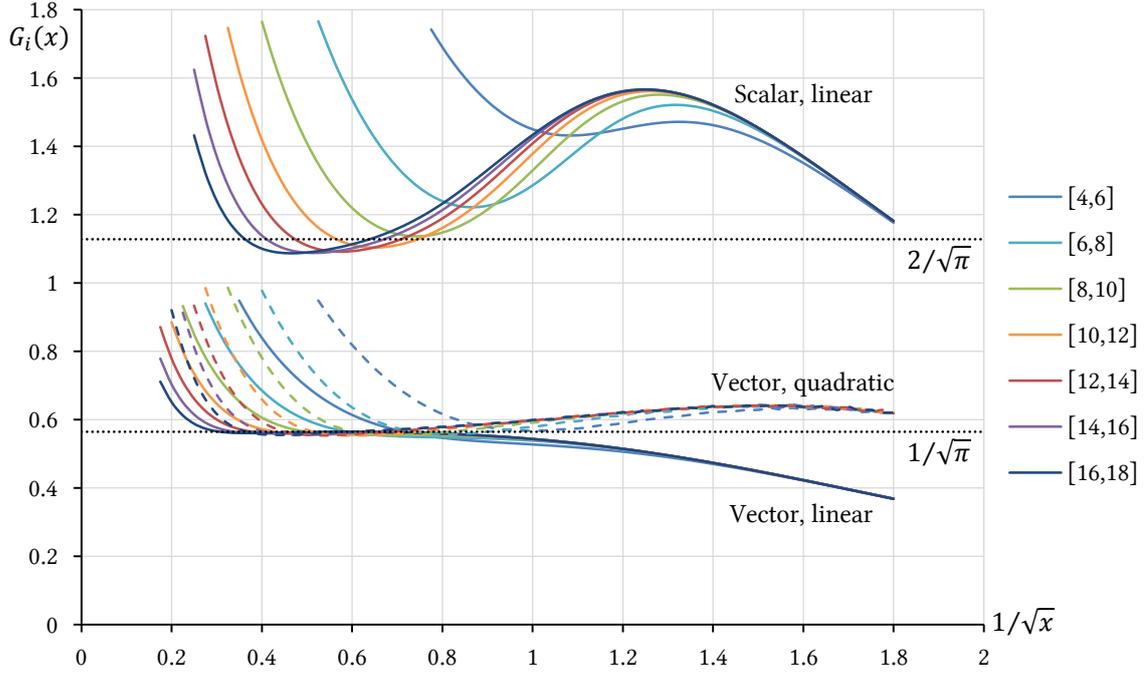

Figure 40. Linear and quadratic extrapolants for the scalar and vector particle masses in the massless Schwinger model.

Extrapolants $G_i(x), i \in \{+, -\}$ are shown in Figure 40. We have not included the quadratic extrapolants for a scalar state because they are generally not giving very good results [3]. From the overall shape of the extrapolants, we can see that for a vector state, $G_-(x)$ gives results that are very close to the exact value, especially in the range $1/\sqrt{x} \in [0.4; 0.7]$. After close inspection, one can find that in this range the linear $G_-(x)$ has a minimum and a maximum. Results for the maximum are very close to the exact value in the continuum limit (see Table 1).

If we observe the first turning point of the scalar state, we find out that the estimated values are close but noticeably lower than the exact ones, as one increases $[N, M]$. This aspect was not yet found in [3] since smaller system sizes were used there. We expect this to be due to the improper extrapolation from too high values of $1/\sqrt{x}$. In order to improve this result, we should use bigger $x$ together with higher perturbation orders $N$.

| $[N, M]$ | $f_+(x)\|_{x \to \infty}$ | $f_-(x)\|_{x \to \infty}$ |
|---|---|---|
| [4,6]   | 1.431976 | –        |
| [6,8]   | 1.221849 | –        |
| [8,10]  | 1.136521 | 0.555434 |
| [10,12] | 1.104152 | 0.559699 |
| [12,14] | 1.092068 | 0.562208 |
| [14,16] | 1.088228 | 0.563617 |
| [16,18] | 1.087434 | 0.564485 |
| Exact   | 1.128379 | 0.564189 |

Table 1. Estimated values of $f_i(x)$ using extrapolants.

We were also interested in the ratio $f_+(x)/f_-(x)$, which in the continuum limit would become $m_+/m_- = 2$ – equation (2.2.16). The results are shown in Figure 41.

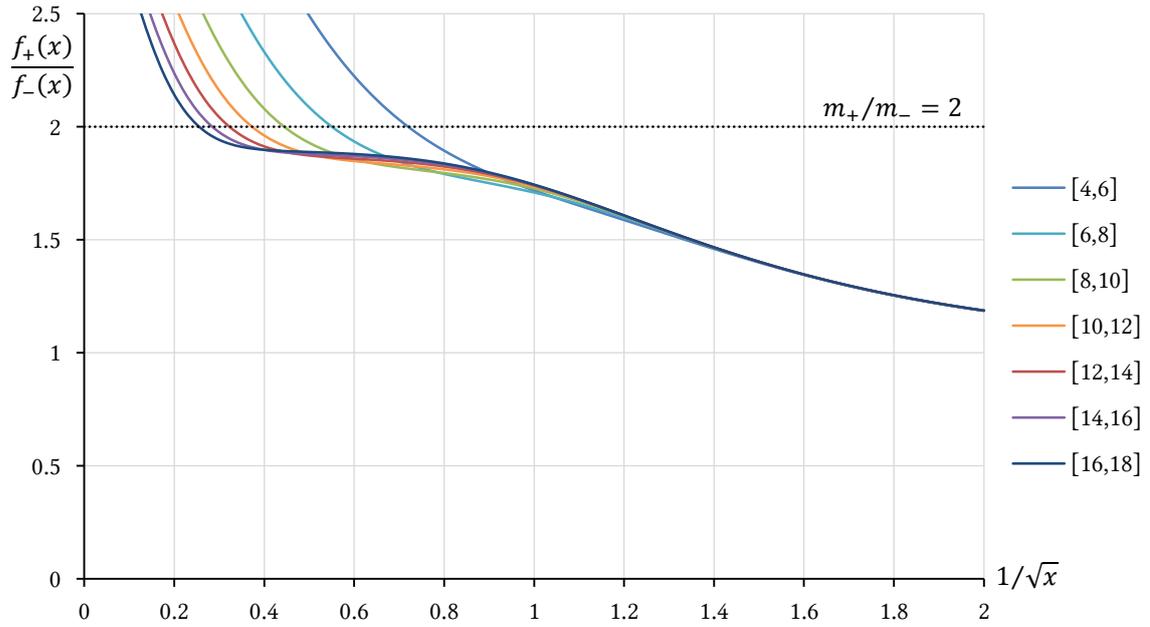

Figure 41. Ratio of the scalar and vector particle masses $f_+(x)/f_-(x)$ for the massless Schwinger model.

We can see that the result converge as we go to the higher $[N, M]$, however the estimation of the exact value in the limit $x \to \infty$ needs even larger systems.

As mentioned before, we have also used the finite size scaling of Fisher and Barber [32]. Strong coupling expansion with high order $N$ was used to do that. It has to be stressed that our results seem to converge as $N \to \infty$, which is not guaranteed when using this perturbation theory method. We have used such $N$ that our results were not changing with the accuracy of 6 decimal places when going to even higher orders.

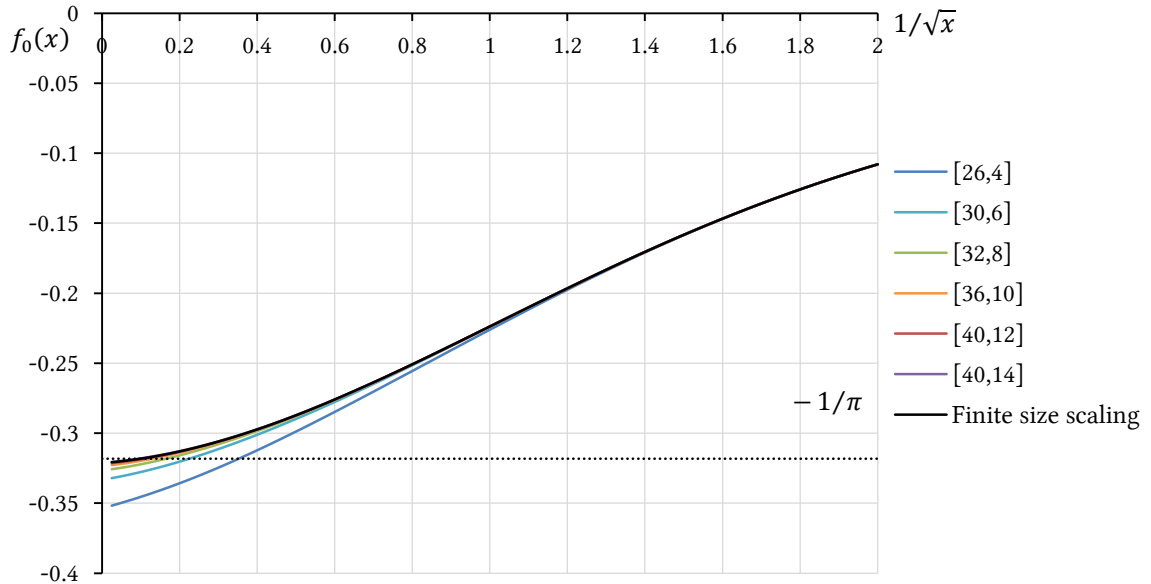

Figure 42. Vacuum energy $f_0(x)$ of the massless Schwinger model for very high values of perturbation order $N$. The bulk limit value was estimated using finite size scaling.

The ground state energy $f_0(x)$ was presented in Figure 42. The black solid line is the estimated value in the bulk limit, using the finite size scaling equation (4.2.1). One can see that very near $1/\sqrt{x} \to 0$, the approximated bulk limit results are dropping lower than the exact value $-1/\pi$. This is consistent with the findings of Hamer et al. [33], that when $x \to \infty$, we should expect a power law scaling, rather than the exponential one, because this limit is a critical point of the lattice model.

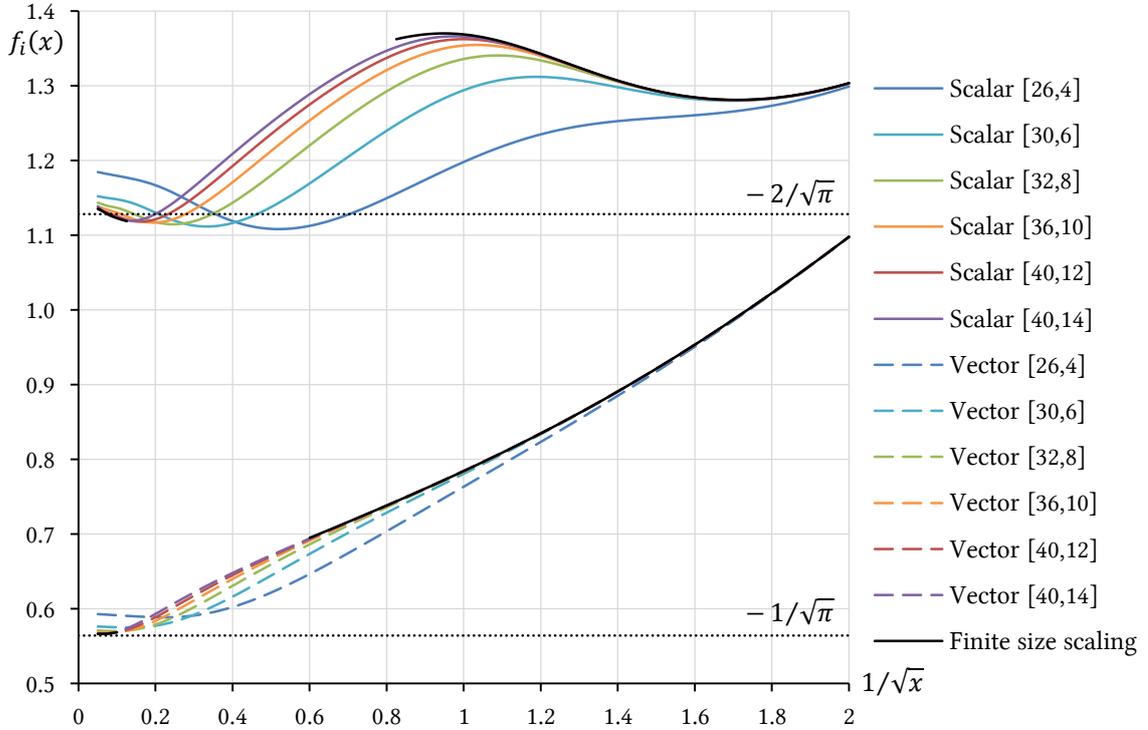

Figure 43. Scalar and vector particles masses $f_i(x)$ of the massless Schwinger model for very high perturbation order $N$. Bulk limit values estimated using exponential finite size scaling.

Particle masses $f_+(x), f_-(x)$ were also investigated and the results are shown in Figure 43. The finite size scaling could not be performed for range of $1/\sqrt{x}$ where the plots were overlapping. Below the overlapping region, one can see that the scaling for the vector particle mass is generally consistent with the exact result, while the scaling for the scalar particle mass is not a very good approximation.

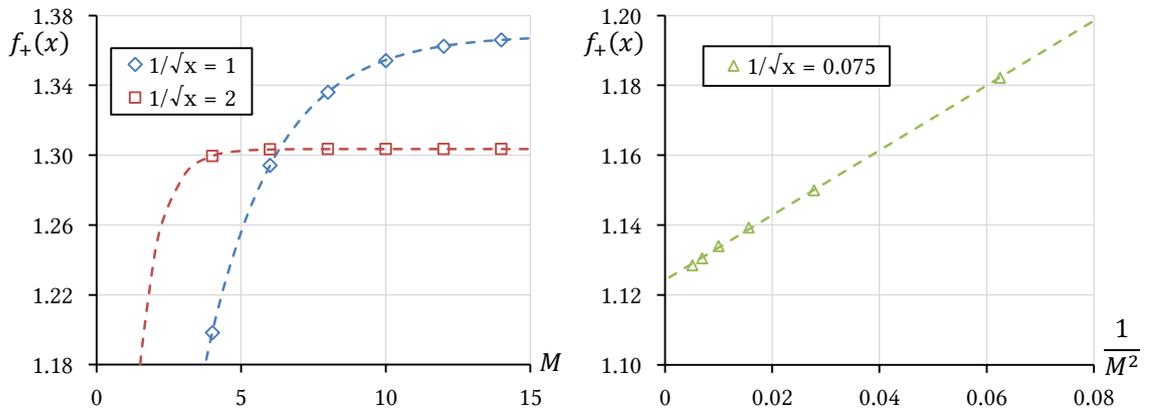

Figure 44. Scaling of the scalar particle mass for large values of $1/\sqrt{x}$ (left) and small value of $1/\sqrt{x}$ (right).

Figure 44 shows the examples of scaling of the scalar mass particle $f_+(x)$. We can see that for values $1/\sqrt{x} = 1, 2$ there is an exponential scaling, while near the continuum limit, for $1/\sqrt{x} = 0.075$ power-law scaling applies.

### 4.2.2 The massive Schwinger model

The massive Schwinger model was also studied using the strong coupling expansion. First of all, we were interested in the vacuum energy (2.2.8) dependence on the fermion mass $m$. We have studied masses $m/g \in \{0.0; 0.2; ...; 1.4; 1.6\}$ and the range of $1/\sqrt{x}$ was, like before, $[0.025; 2.0]$ with sampling $0.025$. The used system sizes were $[N, M] = [10,12], [12,14]$, and $[14,16]$, but since they almost overlap, we will present only the plots for $[14,16]$. The results are shown in Figure 45.

Increasing the fermion mass, also increases the value of $f_0(x)$ for a given $x$. However, using the quadratic fits (see the close-up), we can see that in the limit $x \to \infty$ all curves approximately converge to the same value $-1/\pi$. This is in agreement with the results of Hamer et al. [2,3], that the value of $f_0(x)$ in the continuum limit should be independent of the fermion mass $m$.

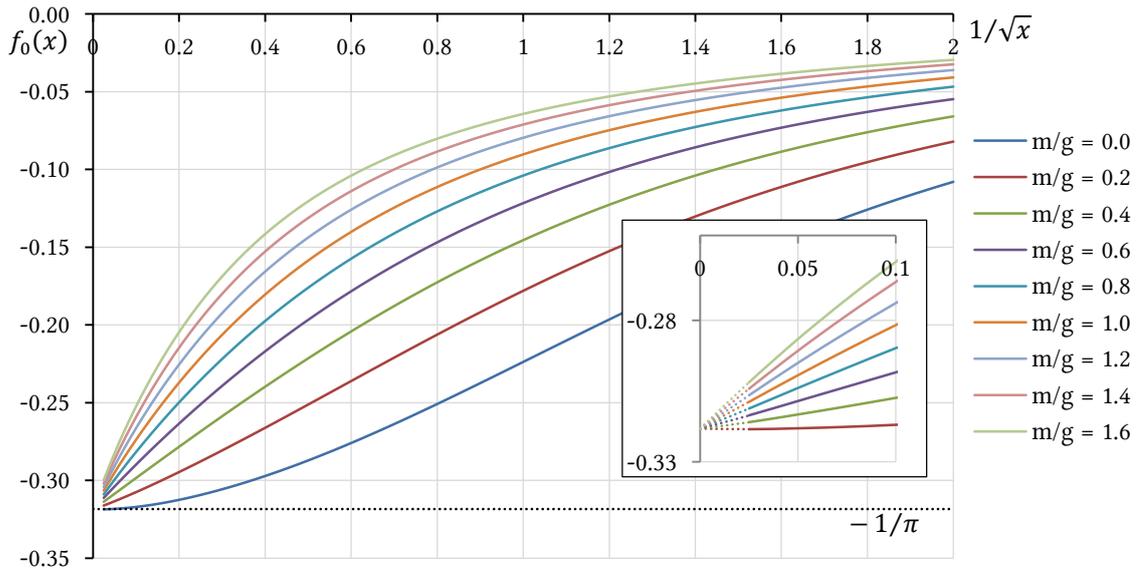

Figure 45. Vacuum energy function $f_0(x)$ of the massive Schwinger model for various masses $m/g$.

| Mass $m/g$ | Estimated $f_0(x)$ | Error | Student's $t$ |
|---|---|---|---|
| 0.0 | −0.31839 | 0.00007 | 1.144 |
| 0.2 | −0.31840 | 0.00008 | 1.126 |
| 0.4 | −0.31840 | 0.00008 | 1.126 |
| 0.6 | −0.31839 | 0.00007 | 1.144 |
| 0.8 | −0.31836 | 0.00005 | 1.002 |
| 1.0 | −0.31833 | 0.00004 | 0.503 |
| 1.2 | −0.31828 | 0.00006 | 0.498 |
| 1.4 | −0.31823 | 0.00010 | 0.799 |
| 1.6 | −0.31817 | 0.00016 | 0.874 |
| Exact | −0.31831 | | |

Table 2. Estimated values of $f_0(x)$ for different fermion masses.

Student's $t$-test was used to judge if the results are indeed converging. Inspection of Table 2 shows that the $t$-value is always less than the critical value $t_{\alpha,\nu=6} = 1.440$ even for the confidence level as high as $\alpha = 0.2$, which means that the numbers are consistent with the exact result.

The scalar and vector binding energies given by equations (2.2.9) and (2.2.13) were also studied. An example plot for $m/g = 10$ is shown in Figure 46. We can see that the curves converge fast when increasing $[N,M]$, especially for $f_-(x)$. The largest system size [16,18] was used to estimate the continuum limit value. Linear extrapolants approach was employed (black dashed lines), where we looked for the first turning point of scalar extrapolants and second turning point for vector extrapolants (compare to Figure 40 discussion).

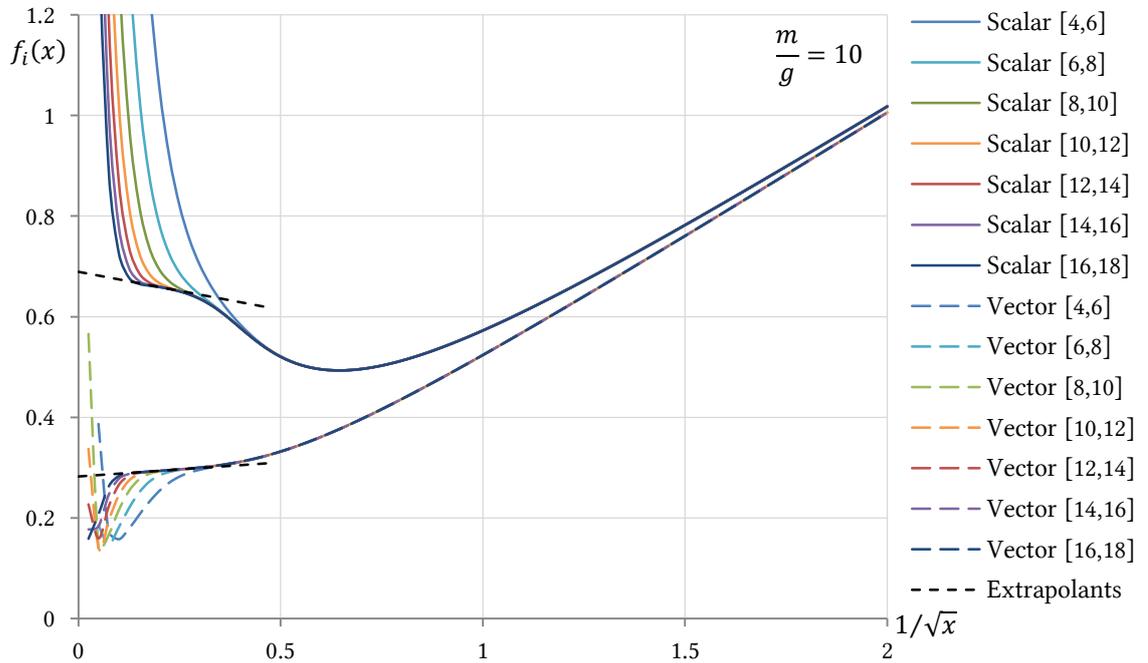

Figure 46. Scalar and vector binding energies for the massive Schwinger model with $m/g = 10$.

This method of finding values of $f_i(x)$ in the limit $x \to \infty$ was applied to systems with $m/g = 10^k$, $k \in \{-2.0; -1.9; ...; 1.9; 2.0\}$ and $[N,M] = [14,16]$. The goal was to obtain the dependence of binding energies on the fermion mass in the Schwinger model. As suggested in [3], we have used the mean difference between linear and quadratic extrapolants to find the error of the measured value $f_i(x)$. Figure 47 presents the results and the exact values for very large and small mass limits (black lines), as specified by equations (2.2.11) and (2.2.15).

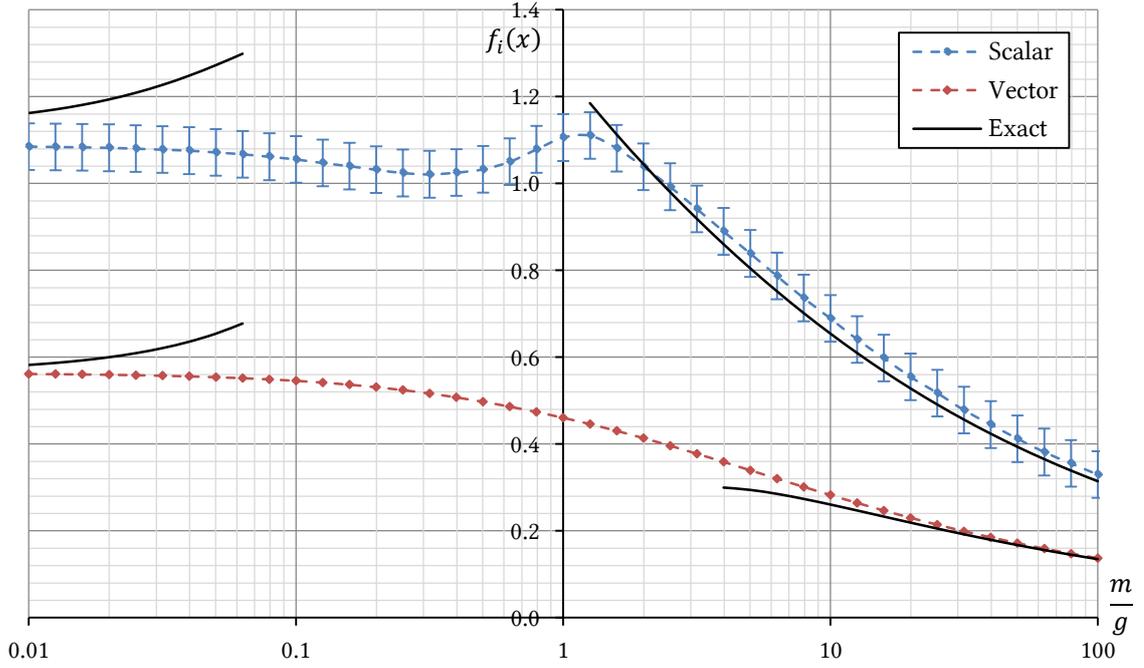

Figure 47. Scalar and vector binding energies (bulk and continuum limit estimates) for different fermion masses $m$. Errorbars for $f_-(x)$ are so small that they are omitted on the plot.

We can see that the obtained dependence is very close to the exact relations in the limit $m \to \infty$. However, the binding energies in the limit $m \to 0$ were not recreated properly. It was expected for the scalar state, since the estimated value for the massless Schwinger model in the continuum limit was clearly too low – see Table 1 and Figure 43. As for the vector state, we suspect one has to go to lower values of $m/g$ to see that the results are consistent with predictions. Besides, the error in determining $f_+(x)$ is much higher than the one of $f_-(x)$ – we conclude that the vector binding energy is generally more stable when going to the continuum limit.

Function $f_-(x)$ changes very smoothly from the non-relativistic to the relativistic regions, however the scalar state shows a peak in the vicinity of $x = 1$. We expect this to be due to the finite-size effects, and in the true bulk limit, the pike should disappear.

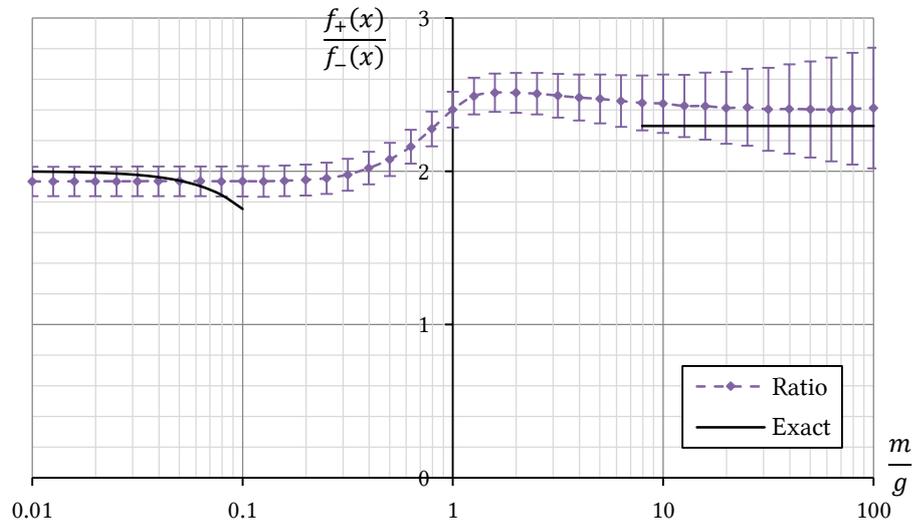

Figure 48. Ratio of the scalar and vector binding energies against the fermion mass $m$.

Ratio $f_+(x)/f_-(x)$ for different masses $m$ was shown in Figure 48. In both limits $m \to 0$ and $m \to \infty$ the obtained values are consistent with the theory, mostly due to very large errors.

### 4.2.3 The Schwinger model in the background electric field

The strong coupling expansion was used to simulate the Hamiltonian of the Schwinger model with background field, given by (1.3.25). We have studied the range $1/\sqrt{x} \in \{0.02; 0.04; \ldots; 0.98; 1.0\}$. Firstly, the objective was to determine the ground state energy $f_0(x)$ for different masses $m$ and for $\alpha = 0.5$. To do this, we have used systems $[N, M] \in \{[10,12], [12,14], [14,16]\}$. The results for $[N, M] = [14,16]$ only are presented in Figure 49, since the plots for different system sizes were almost exactly overlapping. We have also included the results for $\alpha = 0$, for comparison.

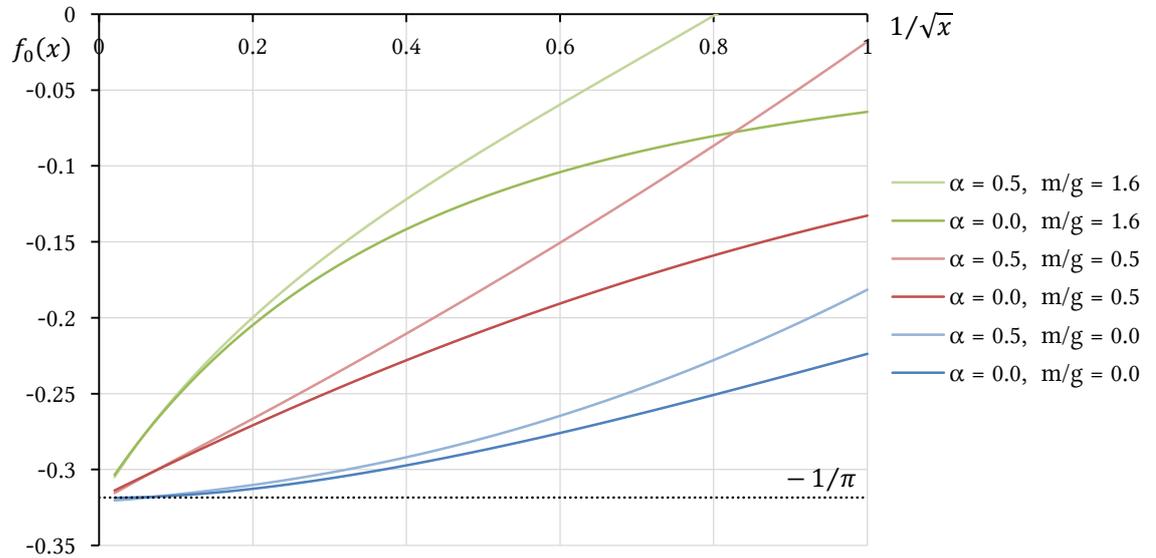

Figure 49. Ground state energy for the Schwinger model with various values of $\alpha$ and $m$.

The results indicate that the ground state energy in the continuum and bulk limits is independent of both $\alpha$ and $m$, which is consistent with theoretical predictions [2].

String tension for $\alpha = 0.5$ was also measured, using equation (2.2.20). We have investigated various system sizes with $N = M - 2$. Figure 50 shows the example results for $m/g \in \{0.0; 0.2; 1.6\}$. The exponential finite size scaling was used to determine a bulk limit values of $1/\sqrt{x}$ away from the continuum limit. We have found that the convergence of plots for different sizes is good at higher values of $m/g$. Near and below the critical mass $(m/g)_c \approx 0.33$, the lines are not overlapping any more. This behaviour suggests that we are in the vicinity of the critical point.

The linear and quadratic extrapolants were used to estimate the continuum limit values. Therefore, we have determined the dependence of the string tension on the fermion mass – see Figure 51. At $m/g = 0$ the string tension disappears, which is in agreement with chiral symmetry prediction – the background field ("confining" potential) should be completely screened at $m = 0$ [2]. If $m \neq 0$, the string tension is non-zero and thus we have a confinement of test charges (we can also imagine them as a capacitor, as discussed in Ch. 1.3.5 and in [22]). For small values of $m$, the results are consistent with the theoretical prediction – equation (1.3.30).

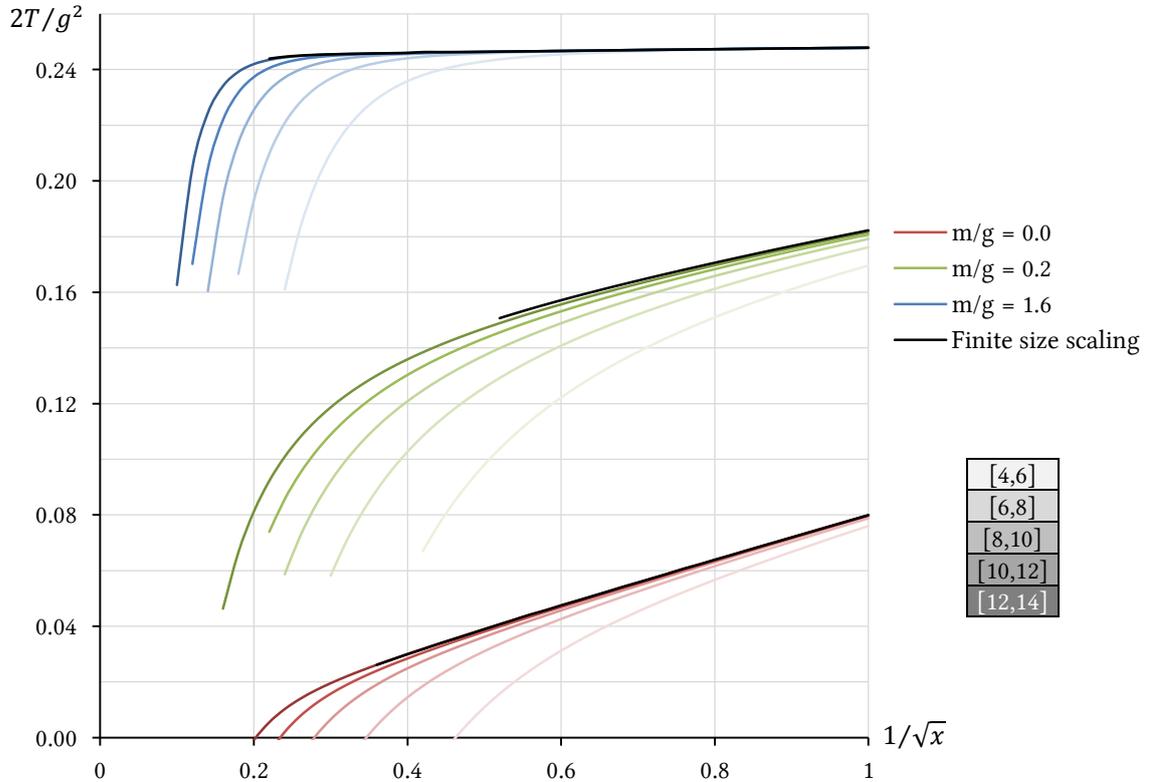

Figure 50. String tension for $\alpha = 0.5$ for different $[N, M]$ and masses.
Black solid line is the finite size scaling.

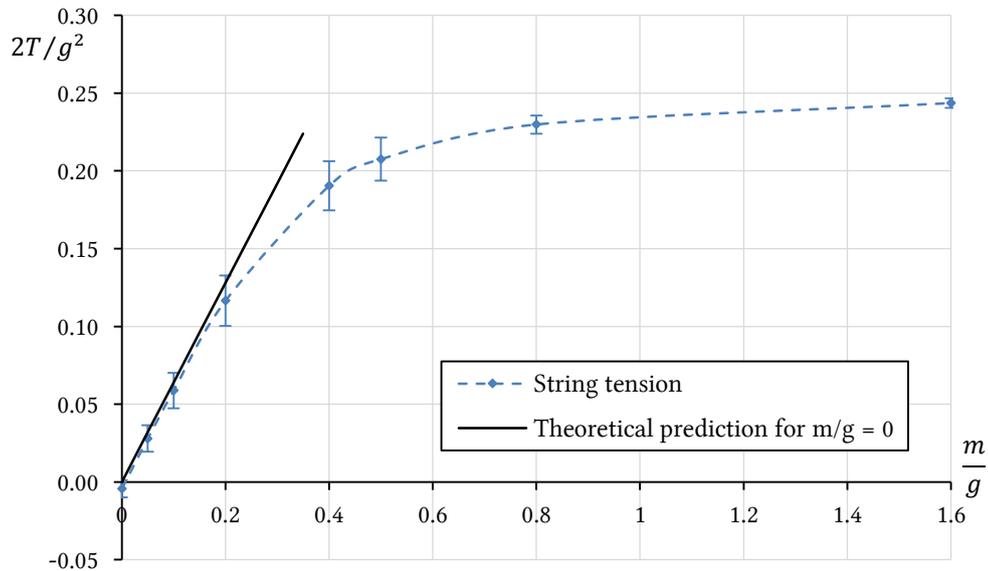

Figure 51. Dependence of the string tension $2T/g^2$ on the fermion mass $m/g$.

Another quantity we were interested in, was the chiral order parameter given by equation (2.2.21). We have studied different system sizes with $N = M - 2$ in order to see if the numbers will converge and if we can use the extrapolants method to estimate the continuum values.

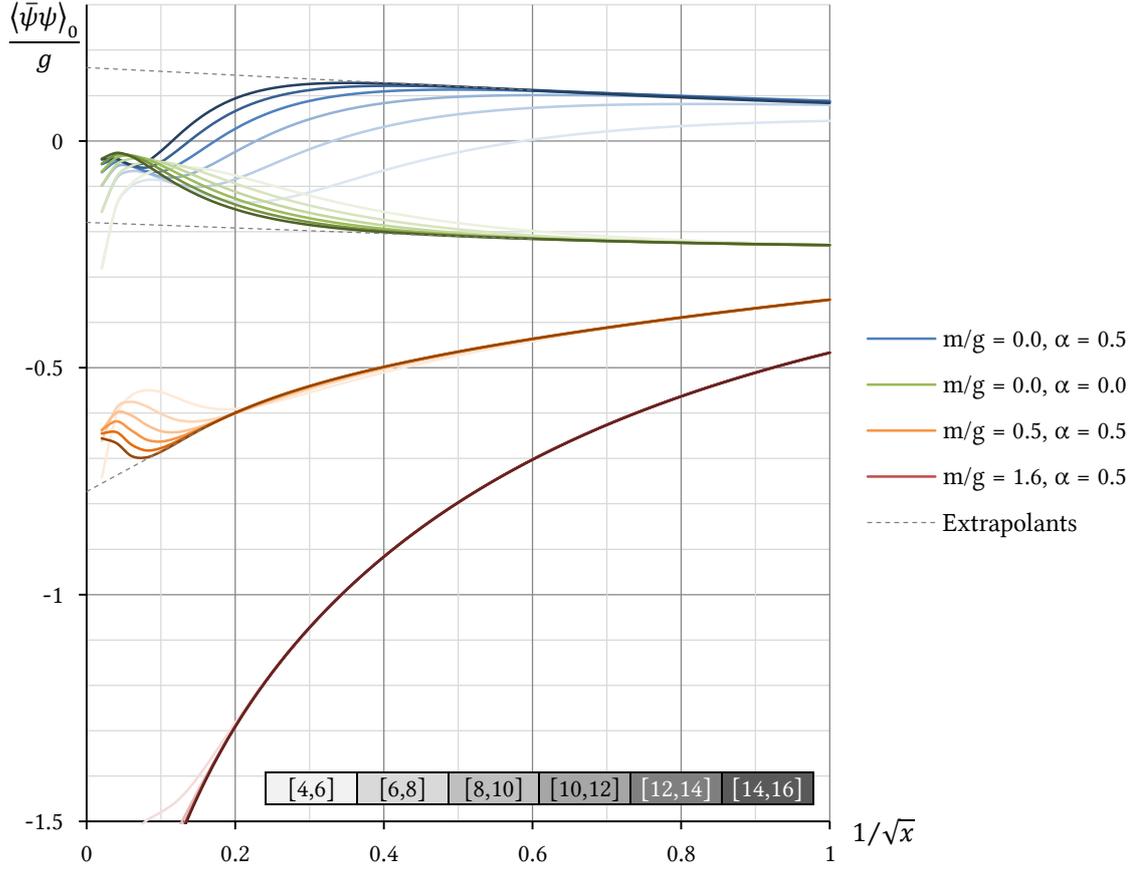

Figure 52. The chiral order parameter for various $m$ and $\alpha$.

Figure 52 presents the results. One can see that the curves for given $\{m, \alpha\}$ come together and we can estimate the continuum limit value by making the linear and quadratic extrapolants for small values of $1/\sqrt{x}$ where the plots are still overlapping. Examples of the linear extrapolants are marked by dashed black lines.

Using this method, relation between $\langle \bar{\psi}\psi \rangle_0$ and $m$ in the continuum limit was studied for $\alpha \in \{0.0; 0.5\}, m/g \in \{0.0; 0.05; ...; 0.95; 1.0\}$ and is shown in Figure 53. Errors are approximated by the difference between the linear and quadratic extrapolants. One can see that:

- for $\alpha = 0.0$, the dependence between the chiral order parameter and the fermion mass is roughly linear,
- for $\alpha = 0.5$, the relation is almost similar for masses greater than the critical mass (given by (1.3.26), marked on the plot by dashed line), and below the critical mass the values of $\langle \bar{\psi}\psi \rangle_0$ are higher than for $\alpha = 0.0$,
- the values at $m/g = 0$ are consistent within the errorbars to the theoretical prediction described by equation (1.3.32) and marked on plot by dotted lines.

From the overall shape of $\langle \bar{\psi}\psi \rangle_0$ at $\alpha = 0.5$, we conclude that there is indeed a phase transition at about $m/g \approx 0.35$. For $m/g$ below the critical value, the charge conjugation symmetry is unbroken and we have one vacuum, while above $(m/g)_c$, system experiences spontaneous symmetry breakdown and we have two vacua [22].

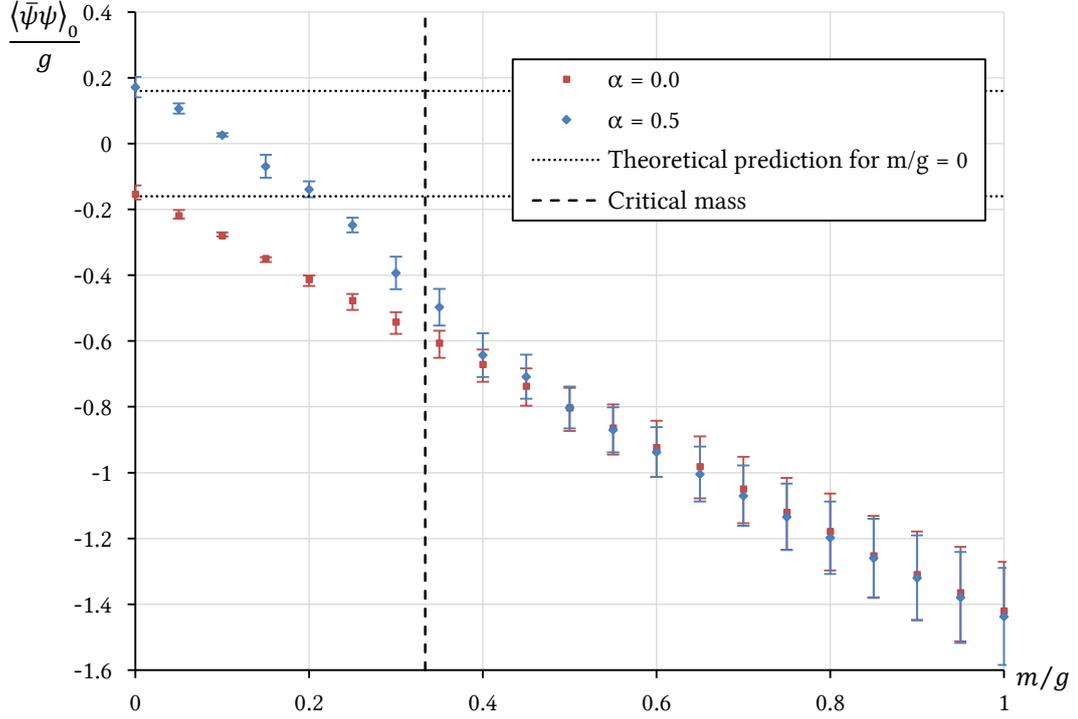

Figure 53. Dependence of the chiral order parameter on the fermion mass $m$.

Similar approach was used to inspect the average electric field $\Gamma^\alpha$ and the axial fermion density $\Gamma^5$, given by equations (1.3.27) and (1.3.28). We have simulated the systems with $m/g \in \{0.0; 0.5; 1.0\}$ to see if the numbers converge. The results are shown in Figure 54 and Figure 55.

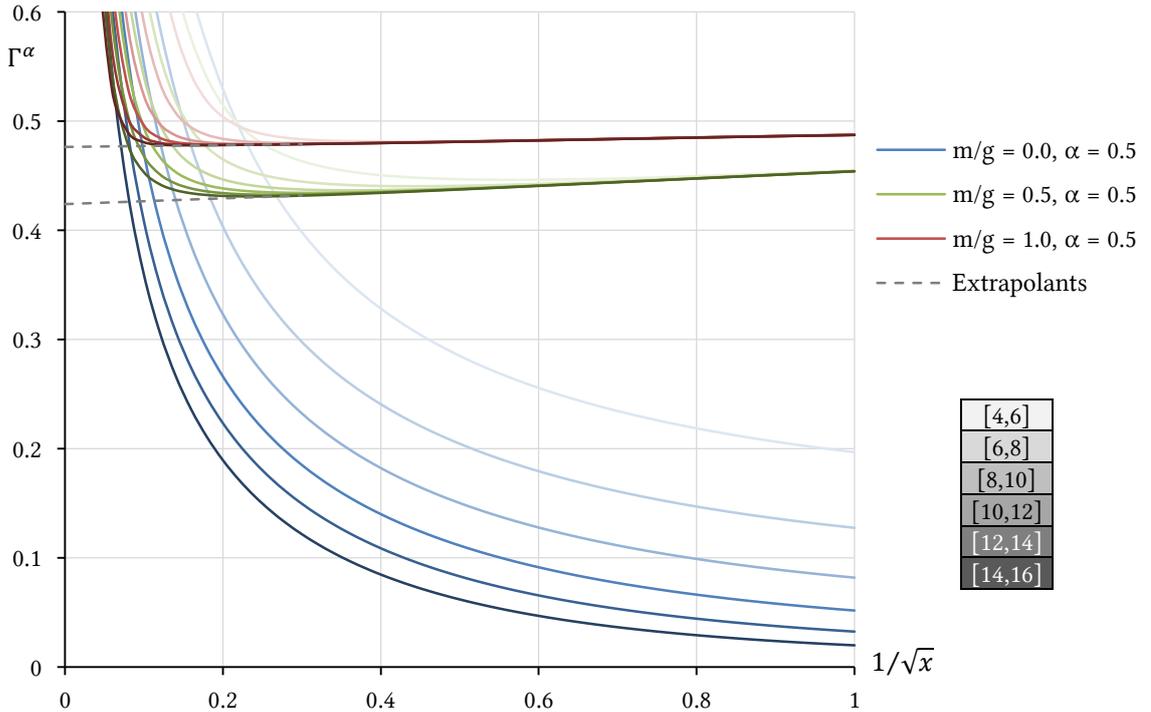

Figure 54. The average electric field $\Gamma^\alpha$ for different system sizes and fermion masses $m$.

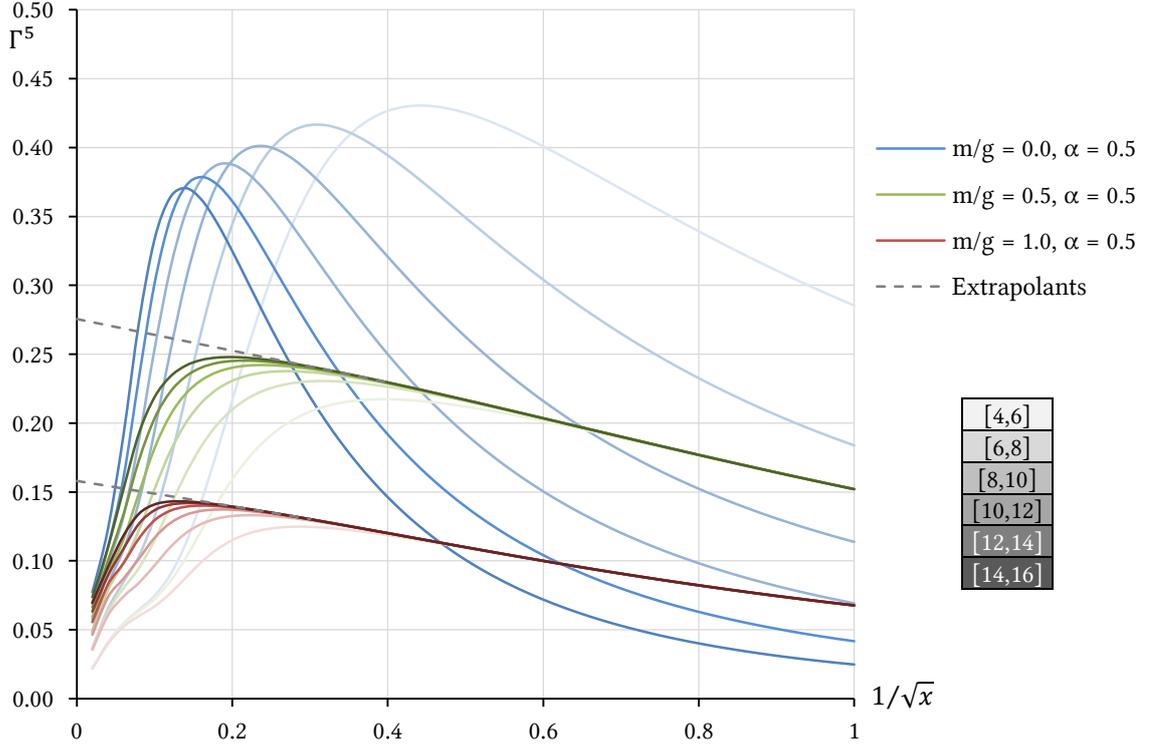

Figure 55. The axial fermion density $\Gamma^5$ for different system sizes and fermion masses $m$.

The findings indeed converge for $m/g = 0.5$ and $1.0$ and we have employed the extrapolants method to approximate the continuum values. However, for $m/g = 0.0$, we can see that the plots do not come together. For such cases, we were unable to determine the true values of $\Gamma^\alpha$ and $\Gamma^5$ in the $x \to \infty$ limit.

The dependence of $\Gamma^\alpha$ and $\Gamma^5$ on the fermion mass $m$ was investigated using this technique. The results are shown in Figure 56. We have not calculated the estimates for $m/g \in \{0.0; 0.05; ...; 0.25; 0.3\}$ since the plots were not converging in this section.

We can see now, that we were not able to determine the estimates below the critical mass (dashed line), while the results were converging for $m/g > (m/g)_c$. In the heavy mass limit $m/g \to \infty$, the average electric field is found to be approaching $+0.5$, while the axial fermion density is smoothly decreasing. This is consistent with the findings of [4]. Besides, very near the critical point, the errorbars become bigger and it gets very hard to estimate the true continuum values in this region.

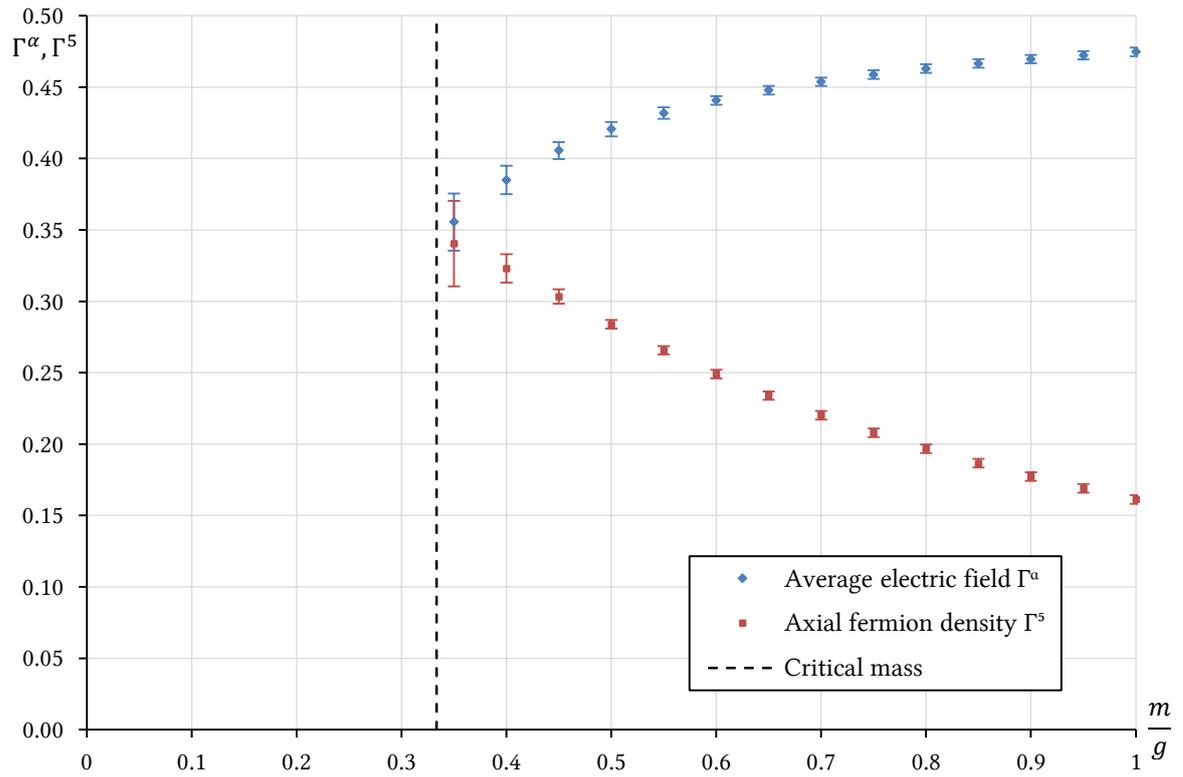

Figure 56. Average electric field $\Gamma^\alpha$ and axial fermion density $\Gamma^5$ dependence on $m/g$ for the Schwinger model with background electric field $\alpha = 0.5$.

# Chapter 5
# Summary, conclusions and outlook

The study of the quantum spin models and the Schwinger model, which was conducted in this project, allowed us to obtain various interesting results. Using computer simulations, we managed to reproduce the theoretical results and even improve some previously performed research.

The low-lying energy levels in the bulk limit of the XY model and the Heisenberg model were successfully determined using the exact diagonalisation method. However, this approach was found to use a very large amount of computer memory to simulate quite small system sizes. This can be improved by using various symmetries of the system, and as an example, we have successfully shown that when we implement the magnetisation conservation, larger systems can be simulated. The numerical renormalisation group method was also used on mentioned models, with various results. The conclusion is that the NRG technique is suitable for outlining the qualitative features of investigated systems, but further investigation is needed to determine the precise values. The DMRG method was very effective in this area, with the quantitative results that were consistent with the exact values and that had a very good numerical precision. We can see that with this method we can effectively achieve large system sizes, without using as much memory as in the exact diagonalisation.

The $J - J'$ model was also studied for several ratios $J'/J$. Using the ED method, we were able to show the low-lying energy levels and to estimate the critical point to be:

$$(J'/J)_c = 0.2419 \pm 0.0005 \tag{5.1}$$

While the NRG approach did bring moderate results for the ground state energy, the DMRG was very fruitful in this case. Therefore, we have determined the dependence between the ground state energy and the interaction constants ratio in the bulk limit.

To conclude, the simulations of the quantum spin models were a success. We have acknowledged how effective the density matrix renormalisation group is, and how it can simulate very large systems without using so much the computer memory resources. However, we recommend using larger initial systems, to achieve even more precise results.

The free massive Schwinger model was also studied, using the ED method. However, it was rather obvious that much bigger systems were needed to be simulated, in order to determine the ground state energy in the bulk limit for various fermion masses $m$.

Therefore, we have employed the strong coupling expansion with the intension of achieving better results. To determine the bulk limit values we were using two approaches:

- for various $[N, M]$ with $N = M - 2$, we kept those results where the plots were converging,
- for various $[N, M]$ with very large $N$, we have used the exponential finite size scaling, proposed by Hamer et al.

The latter one was found to be ineffective in the very high $x \to \infty$, so we have mainly used the first technique. Afterwards, we were estimating the continuum values by the extrapolants method.

First of all, we have simulated the massless Schwinger model, where we have found the ground state energy function $f_0(x)|_{x \to \infty}$ to be consistent with the exact result. The vector particle mass $f_-(x)$ was also determined correctly in the continuum limit, but the scalar particle mass $f_+(x)$ estimate had a lower value than the expected one. This shows the limitations of the extrapolants method – the fits are made from too low values of $x$. Therefore, a better approach is needed to investigate the continuum limit in the future.

Secondly, the massive Schwinger model was studied. The vacuum energy in the continuum limit was obtained to be independent of the fermion mass $m$, which is coherent with the expectations. The dependence of vector particle mass $f_-(x)$ on the mass $m$ was found to be approximating the theoretical predictions in both limits of very heavy and very light $m$. However, for the scalar particle mass, the result is consistent in the $m \to \infty$ limit, but inconsistent in $m \to 0$ limit. We deduce that this also shows that the extrapolants method is sometimes inaccurate. Also, we found out that there is a peak for $f_+(x)$ at about $m \approx 1$, but it is smoothing if we go to larger systems, so one can suspect that it is a finite-size effect, but further investigation is needed.

The Schwinger model with the background electric field was also investigated and we have found that the vacuum energy is independent of the field $\alpha$, which is also consistent with exact calculations. Chiral order parameter, average electric field and axial order parameter were studied in systems with different masses $m$ and $\alpha = 0.5$ and were found to be consistent with the theoretical predictions and the findings of Hamer et al. [2] and Byrnes et al. [4] Nevertheless, after closer inspection of $\Gamma^\alpha$ and $\Gamma^5$ behaviour, it is clear that the used method gives quite big errors near the critical mass $(m/g)_c \approx 0.33$ and is completely unable to determine the values below it. Hence, for the analysis of those order parameters, the DMRG approach would be invaluable.

Summarising, the results of this work are consistent with the theoretical values and findings of [2,4,3]. We have also simulated larger systems and have used denser sampling, which led us to derive finer conclusions. The path that we should follow in the future is outlined, so that we can improve the methods of exploring the Schwinger model and quantum spin models and use them to look into some other, more complicated theories better describing our physical reality.

The final goal of the Hamiltonian approach used in this work is to tackle QCD with such a method. However, the road to this achievement is still quite long. There are many problems one encounters when changing the dimensionality of the model from $1+1$ to $2+1$, so QCD with $3+1$ dimensions is a hard goal to reach. Besides, another challenge rises when one wants to study non-abelian $SU(3)$ models, instead of abelian $U(1)$ theories, such as the Schwinger model. Therefore, an interesting next step would be to investigate a gauge theory with the $SU(3)$ symmetry in $1+1$ dimensions.

# Acknowledgements

This study would not have been possible without the support of many people. First and foremost, I would like to express my gratitude to my supervisor, prof. UAM dr hab. Piotr Tomczak, for his invaluable knowledge and many stimulating discussions. I sincerely thank my project supervisor, dr Krzysztof Cichy, who shared his enormous amount of information and skills with me. His irreplaceable assistance, support and guidance were essential for the success of my work.

Special thanks to all my lecturers, who has taught me almost everything I know and who were always ready to answer my questions. I would also like to thank my high school teacher, mgr Bożena Moldenhawer, to whom I own my passion of physics, and who has inspired me to become a physicist.

Finally, I wish to express my love and gratitude to my dear family and friends. For their endless understanding and love they provided me throughout the duration of my studies.